\documentclass{aa}
\usepackage{graphicx}
\usepackage{txfonts}
\usepackage{todonotes}
\usepackage[normalem]{ulem}
\usepackage{amstext}
\usepackage[breaklinks=true]{hyperref}
\newcommand{\orcit}[1]{\protect\href{https://orcid.org/#1}{\protect\includegraphics[width=8pt]{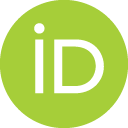}}}

\newcommand{\gaia}{\textit{Gaia}\xspace}
\newcommand{\gband}{$G$--band\xspace}
\newcommand{\gbp}{\ensuremath{G_{\rm BP}}\xspace}
\newcommand{\grp}{\ensuremath{G_{\rm RP}}\xspace}
\newcommand{\bprp}{\ensuremath{G_{\rm BP}-G_{\rm RP}}\xspace}
\newcommand{\gflux}{\ensuremath{I_{\rm G}}\xspace}
\newcommand{\bpflux}{\ensuremath{I_{\rm BP}}\xspace}
\newcommand{\rpflux}{\ensuremath{I_{\rm RP}}\xspace}
\newcommand{\cxs}{\ensuremath{C^{\ast}}\xspace}
\newcommand{\PP}{\textsf{PhotPipe}}

\newcommand{\dro}{\gaia~DR1\xspace}
\newcommand{\drt}{\gaia~DR2\xspace}
\newcommand{\edr}{\gaia~EDR3\xspace}
\newcommand{\drf}{\gaia~DR4\xspace}
\newcommand{\drthree}{\gaia~DR3\xspace}
\newcommand{\xp}{BP and RP\xspace}
\newcommand{\wrt}{with respect to}
\newcommand{\nueff}{\ensuremath{\nu_\text{eff}}\xspace}

\newcommand{\webref}[1]{\href{#1}{#1}}
\newcommand{\citeip}[1]{#1~et~al. (in preparation)\xspace}
\newcommand{\citeipt}[1]{#1~et~al., in preparation\xspace}
\newcommand{\citeipp}[1]{(\citeipt{#1})\xspace}
\newcommand{\secname}{Sect.}
\newcommand{\equref}[1]{Eq.~\ref{eq:#1}}
\newcommand{\secref}[1]{\secname~\ref{sec:#1}}
\newcommand{\asecref}[1]{Section~\ref{sec:#1}}
\newcommand{\appref}[1]{Appendix~\ref{sec:#1}}
\newcommand{\figref}[1]{Fig.~\ref{fig:#1}}
\newcommand{\afigref}[1]{Figure~\ref{fig:#1}}
\newcommand{\tabref}[1]{Table~\ref{tab:#1}}
\newcommand{\instref}[1]{\inst{\ref{inst:#1}}}
\newcommand{\widefig}[1]{\includegraphics[width=17cm]{#1}}
\newcommand{\colfig}[1]{\resizebox{\hsize}{!}{\includegraphics{#1}}}

\authorrunning{M.~Riello~et~al.}

\begin{document} 

\title{Gaia Early Data Release 3: Photometric content and validation}

\author{
M.~Riello \orcit{0000-0002-3134-0935}\instref{ioa}\fnmsep\thanks{Corresponding author: M.~Riello\newline
e-mail: \href{mailto:mriello@ast.cam.ac.uk}{\tt mriello@ast.cam.ac.uk}}
\and
F.~De~Angeli \orcit{0000-0003-1879-0488}\instref{ioa}
\and
D.~W.~Evans \orcit{0000-0002-6685-5998}\instref{ioa}
\and
P.~Montegriffo\instref{oabo}
\and
J.~M.~Carrasco \orcit{0000-0002-3029-5853}\instref{ub}
\and
G.~Busso \orcit{0000-0003-0937-9849}\instref{ioa}
\and
L.~Palaversa\inst{\ref{inst:zag},\ref{inst:ioa}}
\and
P.~W.~Burgess\instref{ioa}
\and
C.~Diener\instref{ioa}
\and
M.~Davidson\instref{ifa}
\and
N.~Rowell\instref{ifa}
\and
C.~Fabricius \orcit{0000-0003-2639-1372}\instref{ub}
\and
C.~Jordi \orcit{0000-0001-5495-9602}\instref{ub}
\and
M.~Bellazzini \orcit{0000-0001-8200-810X}\instref{oabo}
\and
E.~Pancino~\orcit{0000-0003-0788-5879}\inst{\ref{inst:oafi},\ref{inst:asi}}
\and
D.~L.~Harrison \orcit{0000-0001-8687-6588}\instref{ioa}
\and
C.~Cacciari \orcit{0000-0001-5174-3179}\instref{oabo}
\and
F.~van~Leeuwen\instref{ioa}
\and
N.~C.~Hambly \orcit{0000-0002-9901-9064}\instref{ifa}
\and
S.~T.~Hodgkin \orcit{0000-0002-5470-3962}\instref{ioa}
\and
P.~J.~Osborne\instref{ioa}
\and
G.~Altavilla~\orcit{0000-0002-9934-1352}\inst{\ref{inst:oaroma},\ref{inst:asi}}
\and
M.~A.~Barstow \orcit{0000-0002-7116-3259}\instref{lei}
\and
A.~G.~A.~Brown \orcit{0000-0002-7419-9679}\instref{leiden}
\and
M.~Castellani \orcit{0000-0002-7650-7428}\instref{oaroma}
\and
S.~Cowell\instref{ioa}
\and
F.~De~Luise~\orcit{0000-0002-6570-8208}\instref{oate}
\and
G.~Gilmore~\orcit{0000-0003-4632-0213}\instref{ioa}
\and
G.~Giuffrida\instref{oaroma}
\and
S.~Hidalgo \orcit{0000-0002-0002-9298}\instref{iac}
\and
G.~Holland\instref{ioa}
\and
S.~Marinoni \orcit{0000-0001-7990-6849}\inst{\ref{inst:oaroma},\ref{inst:asi}}
\and
C.~Pagani\instref{lei}
\and
A.~M.~Piersimoni \orcit{0000-0002-8019-3708}\instref{oate}
\and
L.~Pulone \orcit{0000-0002-5285-998X}\instref{oaroma}
\and
S.~Ragaini\instref{oabo}
\and
M.~Rainer \orcit{0000-0002-8786-2572}\instref{oafi}
\and
P.~J.~Richards\instref{stfc}
\and
N.~Sanna\instref{oafi}
\and
N.~A.~Walton \orcit{0000-0003-3983-8778}\instref{ioa}
\and
M.~Weiler\instref{ub}
\and
A.~Yoldas\instref{ioa}
}


\institute{
Institute of Astronomy, University of Cambridge, Madingley Road, Cambridge CB3 0HA, UK\label{inst:ioa}
\and
INAF -- Osservatorio di Astrofisica e Scienza dello Spazio di Bologna, via Gobetti 93/3, 40129 Bologna, Italy
\label{inst:oabo}
\and
Institut de Ci\`encies del Cosmos (ICC), Universitat de Barcelona (IEEC-UB), c/ Mart\'{\i} i Franqu\`es, 1, 08028 Barcelona, Spain
\label{inst:ub}
\and
Ruđer Bo\v{s}kovi\'c Institute, Bijeni\v{c}ka cesta 54, Zagreb, Croatia\label{inst:zag}
\and
Institute for Astronomy, School of Physics and Astronomy, University of Edinburgh, Royal Observatory, Blackford Hill, Edinburgh, EH9~3HJ, UK
\label{inst:ifa}
\and
INAF -- Osservatorio Astrofisico di Arcetri, Largo E. Fermi, 5, 50125 Firenze, Italy\label{inst:oafi}
\and
Space Science Data Center - ASI, Via del Politecnico SNC, 00133 Roma, Italy\label{inst:asi}
\and
INAF -- Osservatorio Astronomico di Roma, via Frascati 33, 00078 Monte Porzio Catone (Roma), Italy\label{inst:oaroma}
\and
School of Physics \& Astronomy, University of Leicester, Leicester LE9 1UP, UK\label{inst:lei}
\and
Leiden Observatory, Leiden University, Niels Bohrweg 2, 2333 CA Leiden, The Netherlands\label{inst:leiden}
\and
INAF - Osservatorio Astronomico d'Abruzzo, Via Mentore Maggini, 64100 Teramo, Italy\label{inst:oate}
\and
IAC - Instituto de Astrofisica de Canarias, Via L\'{a}ctea s/n, 38200 La Laguna S.C., Tenerife, Spain\label{inst:iac}
\and
STFC, Rutherford Appleton Laboratory, Harwell, Didcot, OX11 0QX, United Kingdom\label{inst:stfc}
}

\date{Received date; Accepted date}

\abstract
{\gaia\ Early Data Release 3 (\edr) contains astrometry and photometry results for about $1.8$ billion sources based on observations collected by the European Space Agency \gaia\ satellite during the first 34 months of its operational phase.}
{In this paper, we focus on the photometric content, describing the input data, the algorithms, the processing, and the validation of the results. Particular attention is given to the quality of the data and to a number of features that users may need to take into account to make the best use of the \edr catalogue.}
{The processing broadly followed the same procedure as for \drt, but with significant improvements in several aspects of the blue and red photometer (\xp) preprocessing and in the photometric calibration process. In particular, the treatment of the \xp background has been updated to include a better estimation of the local background, and the detection of crowding effects has been used to exclude affected data from the calibrations. The photometric calibration models have also been updated to account for flux loss over the whole magnitude range. Significant improvements in the modelling and calibration of the \gaia point and line spread functions have also helped to reduce a number of instrumental effects that were still present in DR2.}
{\edr contains 1.806 billion sources with \gband photometry and 1.540 billion sources with \gbp and \grp photometry. The median uncertainty in the \gband photometry, as measured from the standard deviation of the internally calibrated mean photometry for a given source, is 0.2 mmag at magnitude $G=10$ to 14, 0.8 mmag at $G\approx17$, and 2.6 mmag at $G\approx19$.
The significant magnitude term found in the \drt photometry is no longer visible, and overall there are no trends larger than 1 mmag/mag. Using one passband over the whole colour and magnitude range leaves no systematics above the $1\%$ level in magnitude in any of the bands, and a larger systematic is present for a very small sample of bright and blue sources. A detailed description of the residual systematic effects is provided.
Overall the quality of the calibrated mean photometry in \edr is superior with respect to DR2 for all bands.}
{}


\maketitle
%

\section{Introduction}\label{sec:intro}

\gaia Early Data Release 3 \citep[EDR3,][]{DPACP-130} is based on data collected during the first 34 months of the nominal mission \citep{GaiaRef} and provides an astrometric and photometric catalogue for more than 1.5 billion sources. \gaia DR3, planned for the first half of 2022\footnote{See \webref{https://www.cosmos.esa.int/web/gaia/release} for updates.}, will be based on the \edr astrometry and photometry but will provide a much more comprehensive set of data, including mean \xp spectra, radial velocities, detailed information on many different classes of variable sources, complementary astrometric information on extended and non-single sources, and classification and astrophysical parameters for different subsets of sources.
Although the number of sources in the \edr catalogue is only slightly larger than that of \drt, the cyclic nature of the \gaia Data Processing Analysis Consortium (DPAC) processing means that the new release is based on a complete reprocessing of the mission data, allowing it to benefit from substantial improvements in the various charge-coupled device (CCD) calibrations, instrument models, and photometric and astrometric calibrations. Additionally, the inclusion of one additional year of mission data with respect to \drt has allowed for the further reduction of the uncertainties on the source photometry and astrometry.

This paper provides an overview of the photometric processing that contributed to the \edr catalogue, focusing on the improvements that were introduced for this data release. A comprehensive view of the photometric processing and its evolution over \gaia data releases is given by, in addition to this paper, the set of papers published for \dro \citep{PhotCore, PhotValDR1, DR1_PhotTop}, \drt \citep{Proc_DR2, Phot_DR2}, and the companion online documentation of the \gaia archive\footnote{
\href{https://archives.esac.esa.int/gaia}{\tt https://archives.esac.esa.int/gaia}}. Since the main focus is on the \gaia photometry, this paper provides only a summary of the \xp spectra pre-processing; the principles of the internal calibration of the \xp spectra will be provided in \citeip{Carrasco}, the spectroscopic content of \gaia DR3 will be presented in \citeip{De~Angeli}, and the external calibration process will be discussed in \citeip{Montegriffo}.
Finally, in this paper we discuss the quality of the \edr photometric data, providing guidelines for making the best use of the catalogue and describing some known issues that the end users should be aware of to avoid problems in their own analysis.


\section{Data used}\label{sec:data}

\edr is based on 34 months of observations that started on 25 July 2014 (10:30 UTC) and ended on 28 May 2017 (8:45 UTC), which corresponds to 1037.9 days. In the paper, mission events are reported in the on-board mission timeline (OBMT), expressed in units of satellite revolutions (21600 s) from an arbitrary origin. A formula to convert OBMT to barycentric coordinate time (TCB) is provided by Eq. (3) in  \cite{GaiaRef}. The time covered includes the range used for \drt with an additional one year of observations, providing a $54\%$ increase in time coverage.

A detailed description of the \gaia instruments is provided in \cite{GaiaRef} and a summary of the main characteristics relevant to the photometric processing can be found in \cite{Proc_DR2}, \cite{Phot_DR2}, and \cite{PhotCore}. 
The main events in the time range covered by \drt are listed in \cite{Proc_DR2}. In the additional year of observations included in \edr, one more decontamination campaign was carried out. At the end of this last decontamination campaign, the satellite focus had not degraded and therefore it was not necessary to refocus the instruments. A list of the time ranges covered by the various events and a description of additional gaps in the data are presented in \appref{gaps}.

The key input used by the photometric and low-resolution spectra processing system \PP\ for the measurement of \gband fluxes are the results of the Image Parameter Determination (IPD) process performed by the Intermediate Data Update (IDU) system. This task estimates the observation time, across-scan position (for 2D windows) and instrumental flux of the source in each SM and AF window, along with their associated formal uncertainties. The modelling of the window contents is a complex process involving many calibrations, from the electronic bias through to the point-spread function (PSF, for 2D windows) or line-spread function (LSF, for 1D windows). Significant improvements have been made to these calibrations between \drt and \edr and hence to the fitted \gband fluxes. Foremost is the quality of the PSF/LSF; in \drt a single library with very limited parameterisation was used, whereas all of the major dependencies are activated in \edr. In this release the variation of the PSF and LSF with time due to changes in focus and contamination level is tracked. The wave number from \drt was used, when available, to properly represent the colour dependence of the profiles, and the smearing effect of the across-scan motion is included, along with the variation across each charge-coupled device (CCD). To better model the \gaia PSF a shapelets-based scheme has superseded the product of the along scan (AL) and across scan (AC) LSFs used in \drt. A detailed description of the PSF and LSF modelling is provided in \cite{EDR3_PSF}. The other calibrations used in IPD, such as the electronic bias and non-uniformity, dark signal, charge injection and release \citep{BIAS} have all been redetermined in IDU to improve their self-consistency and resilience to data gaps. Enhancements have been made to the masking of saturated samples, and to remove suspected secondary sources within a window. Finally, a local background has been fitted for the majority of windows, allowing a much better tracking of the extreme straylight features.

For \gbp and \grp, \PP\ starts from the reconstructed satellite telemetry, which collates all acquisition information for the \xp instrument into a single record for each transit and deals with the generation and application of the various calibrations required to produce bias and background corrected epoch spectra, which are then geometrically calibrated, removing the optical distortions and CCD geometric effects. The bias and non-uniformity mitigation \citep{BIAS} is based on a set of calibrations produced by IDU. Two key improvements have been introduced for \edr: the determination of the local background for each \xp observation, including both straylight and astrophysical background contributions; and an assessment of the crowding status of each observation based on the predictions of observations on the focal plane for all objects in the source catalogue covering the entire time range spanned by the processing. More information on these aspects of the \xp pre-processing is provided in \secref{bprp}.

As already described in \cite{Proc_DR2}, one critical piece of input information used by \PP\ is the cross match produced by IDU. The purpose of the cross match is to identify transits belonging to the same astrophysical source and to exclude spurious detections of artefacts around bright sources. A detailed description of this key process is provided in \cite{EDR3_XM}. It is critical for the end user of the \edr catalogue to realise that a \gaia source and all its properties are defined by the set of transits that have been associated with it by the cross match process. Direct comparisons of individual sources between the \drt and \edr catalogues should take into account that: the fact that the source identifier is the same in the two data releases does not imply that the corresponding astrophysical source is the same; even when the astrometry is consistent it is still possible that a significant fraction of the transits that were associated with that source in the \drt catalogue are not any longer in the \edr catalogue. We therefore strongly discourage the end user from drawing conclusions based on comparisons source by source between \drt and \edr. We instead suggest to perform statistical comparisons of similarly selected datasets from both archives (e.g. comparing colour--magnitude diagrams for particular sky regions). 

In the paper we will often make use of a sample of nearby sources with good photometry and astrometry\footnote{The \gaia archive query on the \texttt{gaia\_source} table required \texttt{parallax}$>3$, \texttt{parallax\_error}$<1$ and \texttt{phot\_proc\_mode}$=0$.}. We will refer to this selection as `nearby source dataset'; if additional selection criteria were applied they will be explicitly mentioned.

\section{\xp spectra processing}\label{sec:bprp}
 
Several aspects of the \xp (pre-)processing have not changed with respect to \drt. Here we focus on a few important improvements and additions that were introduced in the latest processing, in particular on aspects that are relevant for the generation of the photometric catalogue.

\subsection{Crowding evaluation}\label{sec:crowding}

The crowding evaluation process is an assessment of the crowding status of a transit based on the pre-computed `scene'. This is defined as the predicted observation time and AC coordinate for all objects in the source catalogue computed projecting their known astrometric coordinates onto the focal plane given the satellite attitude and geometry. The scene covers the entire time range covered by the data. These predictions are used to assess whether a given transit happened to be affected by crowding. Here we distinguish two cases: transits can be either contaminated by a nearby source (that may or may not have had a window assigned in that specific satellite scan of that region of the sky) or blended when some additional source was captured by the same window. Transits were flagged as blended also when the non-target source was just outside the window (within 5 TDI periods in the AL direction and 2 pixel in the AC direction). The crowding assessment of course takes into account also accidental contamination or blend from the other field of view (FoV). The result of the crowding assessment for a given CCD observation is an indication of its status as contaminated, blended or not affected by crowding.

\begin{figure}
    \centering
    \colfig{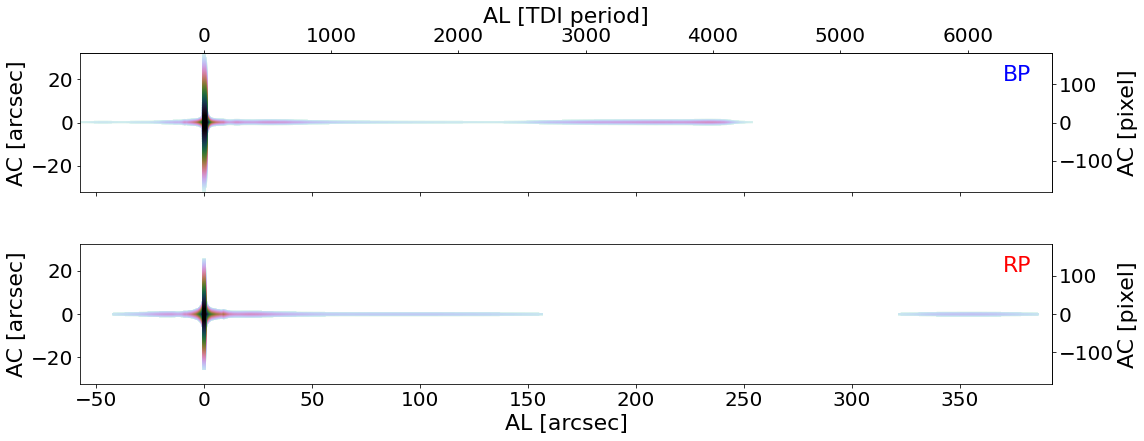}
    \caption{Reconstructed contamination due to an object of magnitude six placed at the coordinate origin. The 2D map is the result of evaluating the AL and AC contamination profiles. A full 2D mapping will be done in the next release. BP contamination is shown in the top panel, RP in the bottom panel. For each location in the plot the colour corresponds to the contaminating flux expected in a window centred at that coordinate converted to magnitude for ease of interpretation.}
    \label{fig:al_ac_profile}
\end{figure}

While for the assessment of blends the simple knowledge of the relative positions of window and scene sources is sufficient, for the contamination evaluation an estimate of the AL and AC LSFs well beyond the boundary of the window is required. For the processing leading to \edr, the contamination surrounding a bright object has been characterised using black-listed transits \citep{EDR3_XM}. These are transits that were not cross-matched to any existing source and did not trigger the generation of a new source because they were considered to be spurious detections caused by diffraction spikes around bright objects. In the AL direction the contamination profiles were described using splines.
In the AC direction a simpler approach was taken interpolating linearly in magnitude space between the central value and the distance at which the brightness level was below the typical background. This distance was estimated from the analysis of the black-listed transits as a function of the magnitude of the central source. Figure \ref{fig:al_ac_profile} shows a typical 2D reconstruction of the contamination around an object of magnitude six for BP in the top panel and RP in the bottom panel. This can then be scaled according to the magnitude of the contaminating source. This simplified approach can only reproduce diffraction spikes aligned with the AC and AL directions. There are however indications of diagonal features at a much smaller level. These will only be accounted for when a full 2D mapping of the contamination will be implemented for the next release. The map shown in the plot corresponds to an area of 7 arcmin AL by 1 arcmin AC. The second peaks in the AL profile at about 3500 TDI period in BP and 6000 TDI period in RP from the contaminating source is probably due to inner/outer reflection on the side faces of the \xp prisms.

\begin{figure}
    \centering
    \colfig{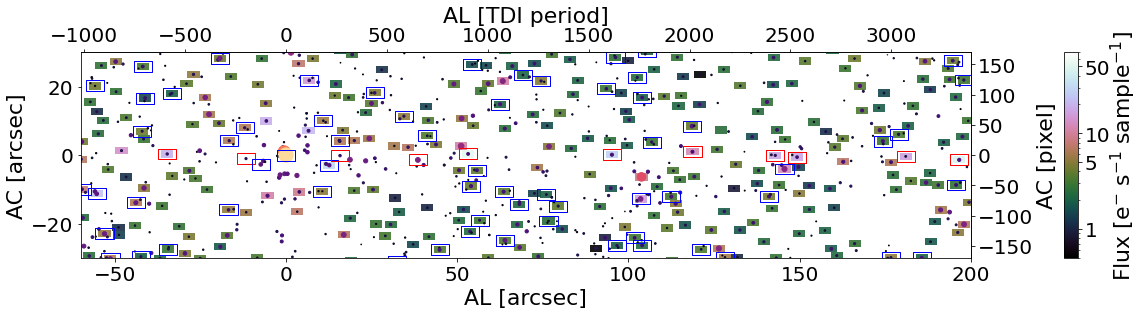}
    \colfig{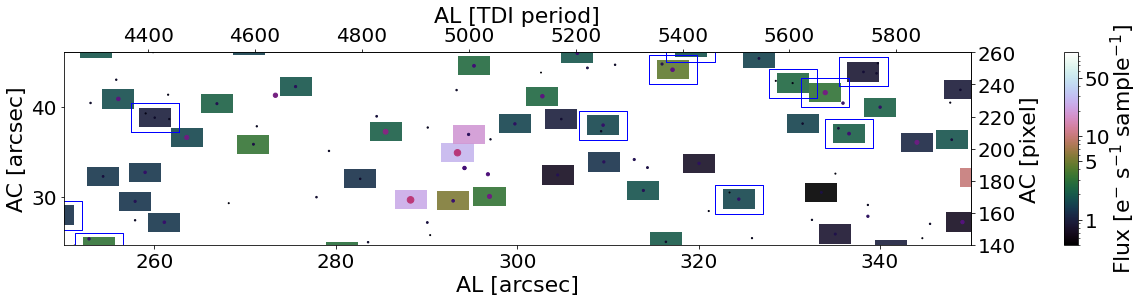}
    \caption{Example of the computed scene for a short time range. \textit{Top panel}: Scene and transits for a small stretch of data ($\approx12$ seconds or $\approx4.5$ arcmin AL and $\approx1$ arcmin AC). \textit{Bottom panel}: Zoom-in of a small group of scene objects and transits around two sources of magnitude close to 13. In both panels, the colour coding of the rectangular symbols representing the actual observations is by residual background flux measured in ${\rm e}^- {\rm s}^{-1} {\rm sample}^{-1}$. The sources in the scene are represented by filled circles with size and colour proportional to their magnitude. The colour bar in the plot refers to the colour coding used for the observations (see the text for more details). }
    \label{fig:scene}
\end{figure}

The top panel in \figref{scene} shows a small stretch of computed scene, the corresponding observations and crowding evaluation results covering about 4.5 arcmin in the AL scan direction and about 1 arcmin in the AC direction close to a source of magnitude 5.4 (located at the origin of the coordinates). In this time range, one of the two FoVs was observing a high-density region near the Galactic centre. The scene objects are shown with filled circles with size and colour proportional to the source brightness (the brightest source observed in this time range appears in yellow at coordinate (0,0), while fainter objects are represented with small dark symbols) while the transits are shown with coloured rectangles of size proportional to the size of the \xp windows. The transit symbols are colour coded according to the residual background (i.e. the background level evaluated from the edge samples of the BP spectrum after the application of the background calibration, lighter colours correspond to larger residual background values).
Larger red and blue rectangles mark transits that have been assessed as contaminated and blended respectively.
From this example it is clear that blending affects a large fraction of transits, while contamination is mostly relevant for transits at the same AC coordinate as the bright object. \afigref{scene} also shows that not all sources in the catalogue can be assigned a window during all scans (visible as filled circles with no filled rectangle around them) and that in the case of bright sources, some of the light coming from the target object is present in the edge samples thus affecting our measurement of the residual background (visible as larger and brighter filled circles with light-coloured filled rectangles). 
This is even more evident in the bottom panel, which shows a zoom-in on a few transits around a couple of sources of magnitude close to 13. The symbol sizes and colours have the same meaning as in the top panel. The two brightest sources (visible as red circles at coordinates 285-295 arcsec AL and 30-36 arcsec AC) have the highest residual background (shown by the lightest-coloured filled rectangles in the same plot).

Even though no attempt was made in the processing leading to \edr to correct the spectra for the effects of crowding, the results of the crowding evaluation were fundamental in cleaning the inputs used in all the following calibration procedures from affected data: observations flagged as contaminated or blended were not used in the computation of all calibrations. A dedicated procedure to remove the effect of crowding from the actual spectra is being developed and will be in place in the processing leading to \drf. Crowded observations were not filtered when computing mean spectra or mean source photometry as this would have caused much reduced completeness in dense sky regions, however the \edr catalogue contains for each source contamination and blends counters (in the columns \texttt{phot\_bp/rp\_n\_contaminated\_transits} and \texttt{phot\_bp/rp\_n\_blended\_transits} in the \texttt{gaia\_source} table) which can be used to detect problematic cases. See also \secref{quality:crowding} for some additional considerations for the end user.

\subsection{Background calibration}\label{sec:overproc:bkg}

The two main components of the background in the \xp spectra are the straylight caused by  diffraction from loose fibres in the sunshield \citep{Preproc_DR1} and the astrophysical background (e.g. non-resolved sources, diffuse light from nearby objects, zodiacal light).
In the processing for \drt the background calibration was optimised to remove the straylight component by accumulating background measurements (from empty windows known as Virtual Objects) over periods of approximately 8 satellite revolutions \citep{Proc_DR2}. This process generated 2D maps of resolution 1 degree in the AL direction and 100 pixels in the AC direction (corresponding to approximately 17.7 arcsec).
While this was appropriate for the smooth behaviour of the straylight in most devices, it was clearly not sufficient to characterise the small-scale variations due to the astrophysical background. The validation of the \drt data showed clear indications that significant residual background was affecting the performances in crowded regions and in areas in the sky where the level of diffuse light is expected to be higher.

The resolution of the background calibration is constrained by the amount of background measurements available. In the latest processing, in order to increase the resolution of the 2D maps, science windows assigned to sources fainter than $G=~18.95$ were used to provide additional background measurements from the edge samples in the window. This enabled increasing the resolution to $\approx0.5$ degree in the AL direction and $~8$ arcsec in the AC direction. 
Finally, to be able to characterise the local astrophysical background, a k-nearest neighbour approach was applied to the map residuals. The median of the 30 closest background measurements (with a maximum distance of 25.6 arcsec) was taken as the estimate of the local background for each observation.

\begin{figure}
    \centering
    \colfig{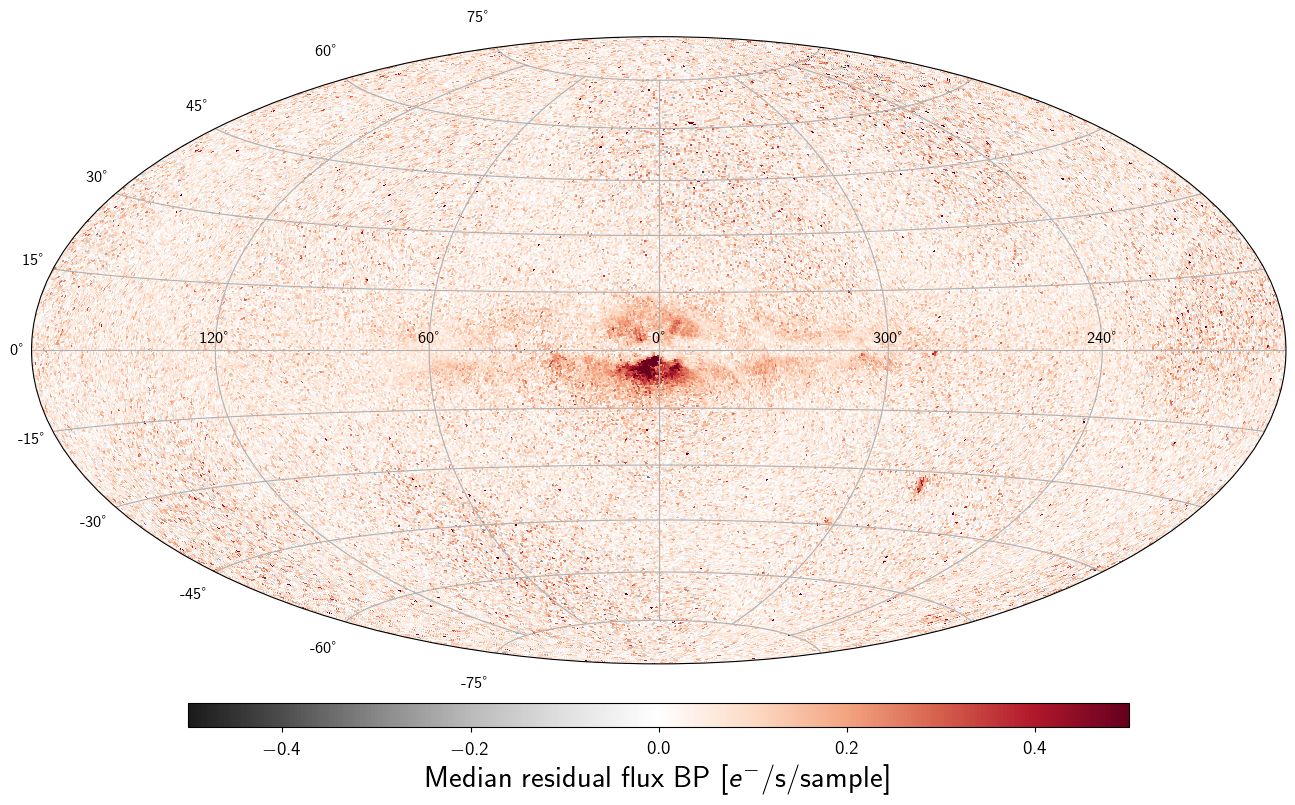}
    \caption{Sky distribution of the source median residual background as measured from the BP spectra. The residual background measurement is obtained from the edge samples of all calibrated epoch spectra for a given source and is given in units of e$^-$ s$^{-1}$ sample$^{-1}$.}
    \label{fig:bkg_sky}
\end{figure}

To show the performance of the background calibration, we defined a quantity called residual background which is computed for each transit as the median of the flux values in the edge samples of a spectrum. In the following we present the results of the analysis on BP data as representative of both BP and RP.
\afigref{bkg_sky} shows the distribution in the sky of the median residual background in BP spectra for the nearby source dataset with the additional magnitude cut $G>17$.
Signatures of the Galactic Plane and other crowded regions are still visible, but the background flux residuals are limited to the range $[-0.5, 1.0]$ in e$^-$ s$^{-1}$ sample$^{-1}$. This is equivalent to an effect at the mmag level for a source of magnitude 15, a hundredth of a magnitude for a source of magnitude 17 and a tenth of a magnitude level for a source of magnitude 19 (a more detailed estimate is provided in \tabref{bkg_flux_to_mag}. The table is provided to help readers understand the significance of the features shown in \figref{bkg_sky}. No correction for this effect is suggested here. As indicated by the sky map, the size of this effect is not constant over the sky and a correction would have to be applied at the epoch level to take into account the satellite scanning law and the overlapping fields of view.

\begin{table}
\caption{Conversion between the residual flux level given in ${\rm e}^-{\rm s}^{-1}{\rm sample}^{-1}$ and a magnitude difference at three different magnitudes (15, 17, and 19). The $\Delta$ mag  values listed in the table for a given level of residual background indicate the corresponding expected change in magnitude. We note that since this is a magnitude difference, the values provided are applicable to both \gbp and \grp.}\label{tab:bkg_flux_to_mag}
\centering
\begin{tabular}{r|rrr}
    \hline\hline
    \multicolumn{1}{c}{residual flux} & \multicolumn{3}{|c}{$\Delta$ mag} \\
    \multicolumn{1}{c}{$[$e$^-$ s$^{-1}$ sample$^{-1}$$]$}& \multicolumn{1}{|c}{15} & \multicolumn{1}{c}{17} & \multicolumn{1}{c}{19} \\
    \hline
    -0.5 & -0.002 & -0.015 & -0.097 \\
    0.5 & 0.002 & 0.015 & 0.099 \\
    1.0 & 0.005 & 0.030 & 0.209 \\
    \hline
\end{tabular}
\end{table}

\begin{figure}
    \centering
    \colfig{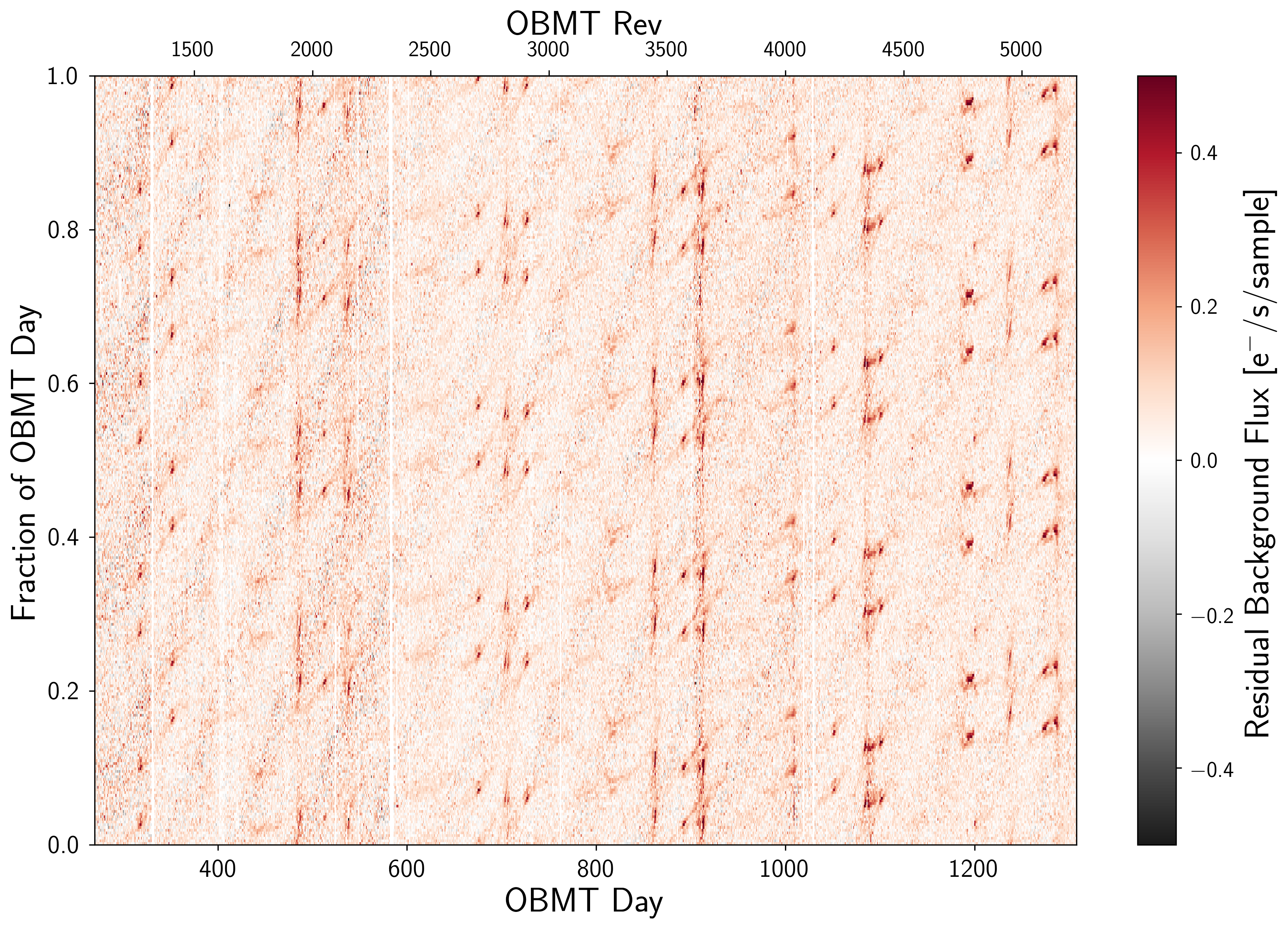}
    \caption{Temporal distribution of the residual background measurements in BP spectra. Each column in the heatmap shows the measurements within a given OBMT day for each OBMT day. The OBMT revolution is shown on the top abscissa axis for ease of interpretation. The high–residual features are the Galactic Plane crossing the two FoVs either in the Galaxy inner or outer direction (see the text for more details). The gaps related to major events such as decontamination and refocussing are visible. Other small gaps are due to telemetry data that could not be included in the processing for various reasons.}
    \label{fig:bkg_time}
\end{figure}
\afigref{bkg_time} shows the variation in time of the median residual background in BP spectra observed by \gaia in the time range covered by \edr (abscissa) with intra–day resolution (ordinate). For a given abscissa position (i.e. one OBMT day), the ordinate shows the residual background variation within the four OBMT revolutions of that day thus allowing a much higher level of detail to be visible compared to a standard histogram. The 16 daily Galactic Plane (GP) crossings are clearly visible: eight in the inner and eight in the outer direction of the Galaxy, four for each FoV. The GP features are seen becoming progressively steeper in the plot as a result of the spacecraft spin axis becoming perpendicular to the GP itself and leading to a Galactic Plane Scan (GPS) when both Gaia FoVs are effectively scanning the GP continuously for several days (e.g. at $\approx$ 1945 rev and then again at $\approx$ 2120 rev, etc.). The eight thin streaks visible before 1200 rev are due to the LMC crossing the two FoVs at each revolution during the ecliptic poles scanning mode (see below). After that the LMC is still visible as increased density spots at periodic intervals. The larger gaps are related to decontamination and refocus events. Other minor gaps are due to outages in the daily processing pipelines or genuine spacecraft events.

It is important to remark that the measure of the residual background from the edge samples of a spectrum will unavoidably be affected by the presence of other sources in the window (i.e. blending) and by contamination from nearby bright sources. We could have used the results of the crowding evaluation to filter out these cases, but this would have prevented us from creating a full sky map as particularly dense regions would have had very little data left.
By using the median values for validation we have mitigated problems with occasional blend or contamination coming from the other FoV, but we should remember that in crowded regions our results will be biassed towards larger positive residuals.

\subsection{Flux and LSF calibration and mean spectra}\label{sec:meanspectra}

Mean source spectra will be released for the first time in \drthree and a detailed description of the processing that lead to the generation of calibrated spectra will be provided for that release (\citeipt{Carrasco} , \citeipt{Montegriffo}).
In this subsection we will only give a very brief overview of this process considering that the reference colour information used in the photometric processing was extracted from calibrated mean source spectra.

The general flow of the flux and LSF calibration of the \xp spectra is very similar to the one in place for the photometry: also in this case the calibration is divided into an internal calibration using a large number of sources to constrain the calibration of all different instrument configurations to a single homogeneous system, and an external calibration which relies on a small set of sources with high accuracy external data to tie the internal system to the absolute one. As for the photometry, no external data is used in the internal calibration of the spectra implying that the process needs to be iterated between a step creating a reference catalogue of spectra for all calibrators and a step updating the calibrations. 
Once the reference catalogue is established, a single run over all observations will generate the final set of calibrations.

During the internal calibration, the spectra are first converted to an internal wavelength scale, called `pseudo-wavelength' applying the calibrated differential dispersion function.
The calibration model for each calibration unit is then defined as a kernel function describing the flux contribution at each pseudo-wavelength from a range in pseudo-wavelength thus characterising changes in response and LSF between different observing conditions and across the wavelength range covered by the BP and RP instruments.
The calibration is defined as a forward model, meaning a model that when applied to the mean source spectrum predicts an observed spectrum for a given time, CCD, FoV, window class, and gate.

The process of generating the mean source spectrum collects all epoch spectra for a given source and fits a function that offers the best predictions in the least squares sense when the calibration is applied to it. This function is defined as a superposition of Hermite polynomials and it is continuous over the pseudo-wavelength range covered by the BP and RP instruments. Integrals of this function over specific wavelength ranges provide the colour information used in the photometric processing, see \secref{calmodel}.

\section{Photometric processing}\label{sec:photsys}

The principles of the photometric calibration have been outlined in \cite{PhotCore} while \cite{Proc_DR2} and \cite{Phot_DR2} provided additional information on how the calibration process was implemented for \drt. This section provides a summary of the changes that were introduced in the photometric processing for \edr.
The main differences \wrt\ \drt are: 1) the OBMT time range; 2) the set of sources used to establish the photometric system; 3) the large-scale (LS) calibration model and the type of colour information used. The small-scale (SS) calibration model is the same as in \drt, a simple zero point \citep[see][]{Proc_DR2}. The following sections provide detailed information about these aspects.

\subsection{Time range}

The period used to establish the photometric system (`INIT period' hereafter) is composed of two time ranges:
$\approx2574.7$ to $\approx2811.7$ OBMT rev and 
$\approx4121.4$ to $\approx5230.1$ (i.e. the end of the period covered by \edr). 
\begin{figure}
    \centering
    \colfig{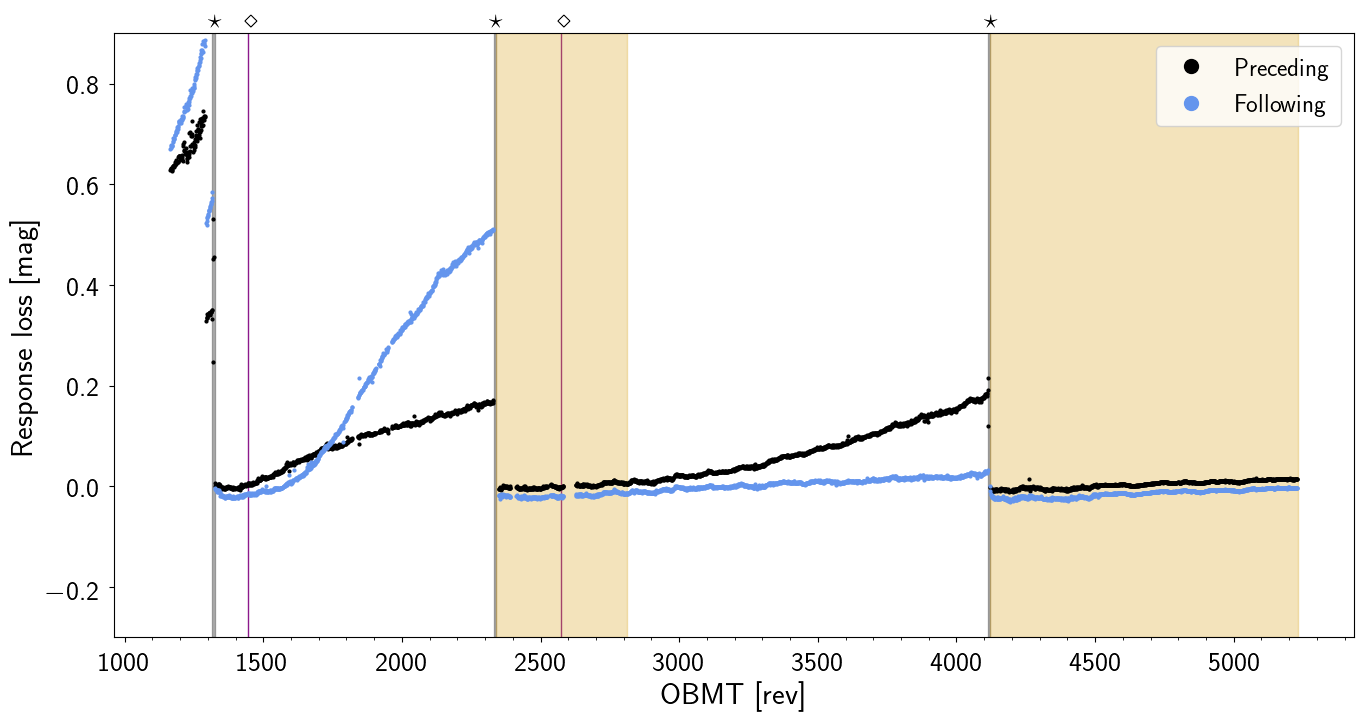}
    \caption{Temporal evolution (expressed in units of satellite revolutions) of the response loss in BP due to contamination during the period covered by \edr. The preceding FoV is shown in black, the following FoV is shown in light blue. The very thin grey shaded areas (marked with $\star$) are the three decontamination campaigns; the two vertical purple lines (marked with $\diamond$) are the two refocussing events; the shaded orange areas show the two time ranges that constitute the INIT period.}
    \label{fig:calperiods}
\end{figure}
The two periods were selected because they have both the lowest and most stable contamination level \citep[see][]{Proc_DR2}. Additionally, the two periods together cover $\approx 1345$ OBMT rev corresponding to $\approx 336$ days and together they provide almost twice the full sky coverage.
\afigref{calperiods} shows the response loss due to contamination measured by the FirstLook instrument health monitor system. The orange shaded areas show the two time ranges that constitute the INIT period. Both time ranges start after a decontamination campaign; the duration of the first time range was selected in order to avoid large variations in the response: in the time range between the last two decontamination campaigns it is clear that the preceding FoV is still affected by contamination which builds up during the time range reaching a response loss of $\approx 0.2$ mag before the last decontamination campaign. The second time range used for the INIT period is instead very well behaved with a nearly constant response level.

\subsection{Algorithm overview}\label{sec:algorithm}

The photometric system is established using a set of calibrators that were selected as described in \secref{standards}. The iterative calibration process follows the same principles used for \drt: an initial set of reference source fluxes is produced by accumulating the uncalibrated epoch photometry from the INIT period and then used to derive a set of LS calibrations. The calibrations are then applied to produce calibrated epochs that are accumulated for each source to produce an updated set of reference fluxes \citep[see][for more details]{Proc_DR2, PhotCore}. Having an explicit time dependency in the calibration model is not very practical due to the irregular time evolution which is both smooth, during most periods, and discontinuous, during decontamination and refocussing campaigns. Instead the LS calibrations are solved independently over short periods of $\approx4$ rev for each time range and instrument configuration \citep[see][for more information]{Proc_DR2}. 
The shorter gate configurations (Gate04 and Gate07) for the AF CCDs are particularly difficult to calibrate due to the very few observations acquired with these configurations. To improve the statistics, the LS calibrations for these configuration were solved over extended time ranges of $\approx20$ OBMT revolutions. The same approach was taken for the shorter gate configurations (Gate05 and Gate07) for the BP CCDs. For RP instead the Gate05 and Gate07 configurations were calibrated using the Gate09 calibrations because the CCD response did not show any additional major feature in the longer Gate09 configuration. 
LS calibration solutions could not be derived for non-nominal calibration units (i.e. gated observations for window class 1 and window class 2) even when further extending the time ranges to $\approx60$ OBMT rev because of the lack of a sufficient number of observations. These instrument configurations were therefore calibrated using the corresponding ungated calibrations.
A total of 20 iterations were performed (see \secref{convergence} for a discussion of the convergence criteria).

Using the reference fluxes from the last iteration, the SS and LS were then solved iteratively in the same way as for \drt \citep{Proc_DR2}. The resulting LS and SS calibrated mean photometry for the calibrators represents the final set of reference fluxes used to then derive the LS and SS calibrations for the full \edr time range.

\subsection{Selection of calibrators}\label{sec:standards}

A set of calibrators were selected among all sources observed in the INIT period. The main purpose of this selection is to provide a more compact dataset to use for the iterative initialisation of the photometric system. The selection was designed to provide a wide colour range and a uniform sky coverage in both magnitude and colour. The main reason to require uniform sky coverage is to ensure that each one of the time ranges for which the LS calibrations are solved would have an adequate set of calibrators observed regardless of the satellite scan direction.  To be selected, sources were required to have \drt photometry in $G$, \gbp and \grp (so that they could be assigned to a colour-magnitude bin) and to have at least $5$ available \xp observations in the INIT period. The \drt photometry was only required for the selection stage and was not used in the \edr calibration process. The magnitude range was restricted to $5.0\leq G\leq19.0$ and the colour to $-1.0\leq\bprp\leq 6.0$. In the regime $G\leq13.5$ sources will normally be assigned a 2D window and gating will be used on board to minimise the effects of saturation. In order to have enough calibrators to solve for daily calibrations for these instrument configurations, all sources brighter than $G=13.5$ were automatically included by the selection process. At fainter magnitudes instead, for each level $k=6$ HEALPix \citep{HPIX} pixel a grid of 70 colour bins and 140 magnitude bins (in the ranges specified above) was created and the first four sources to be assigned to each bin were selected. The order of the sources in each bin was randomised (in a reproducible way) before the selection started. Three additional conditions were added to this selection process: 1) all spectro-photometric standard stars  \citep[SPSS,][]{SPSS} and sources from the passband validation library (PVL) were automatically included in the selection (see \appref{extds} for detailed information on these datasets); 2) all sources that had epochs acquired with a gate configuration were automatically included in the selection to ensure proper linking between the various gated configurations \citep[see][for more details]{Phot_DR2}; 3) all sources with $G-\grp\leq0.5(\gbp-G) + 1.2$ were excluded because contamination from extragalactic objects is quite high in that region of the $G-\grp$ versus $\gbp-G$ colour--colour diagram.
This process produced a selection of $\approx100$ million sources.

\subsection{Calibration models}\label{sec:calmodel}

\begin{figure*}[!ht]
    \centering
    \includegraphics[width=6cm]{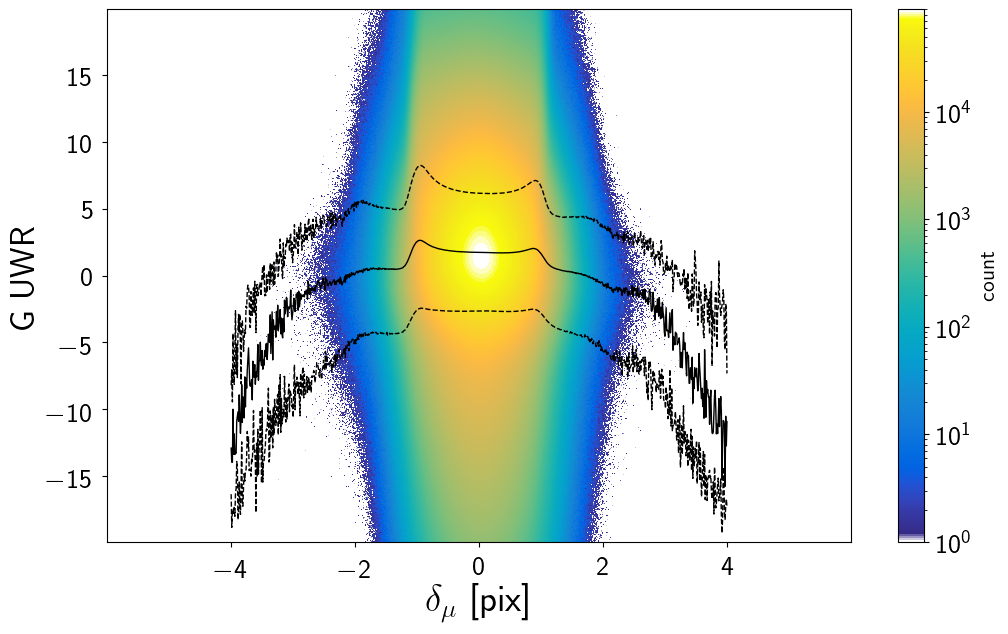}
    \includegraphics[width=6cm]{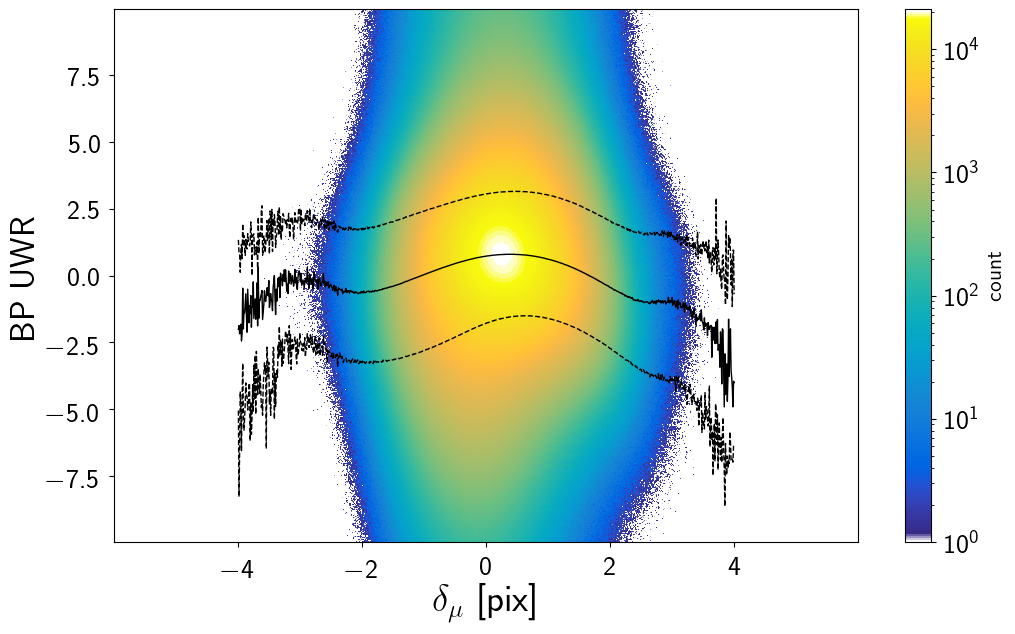}
    \includegraphics[width=6cm]{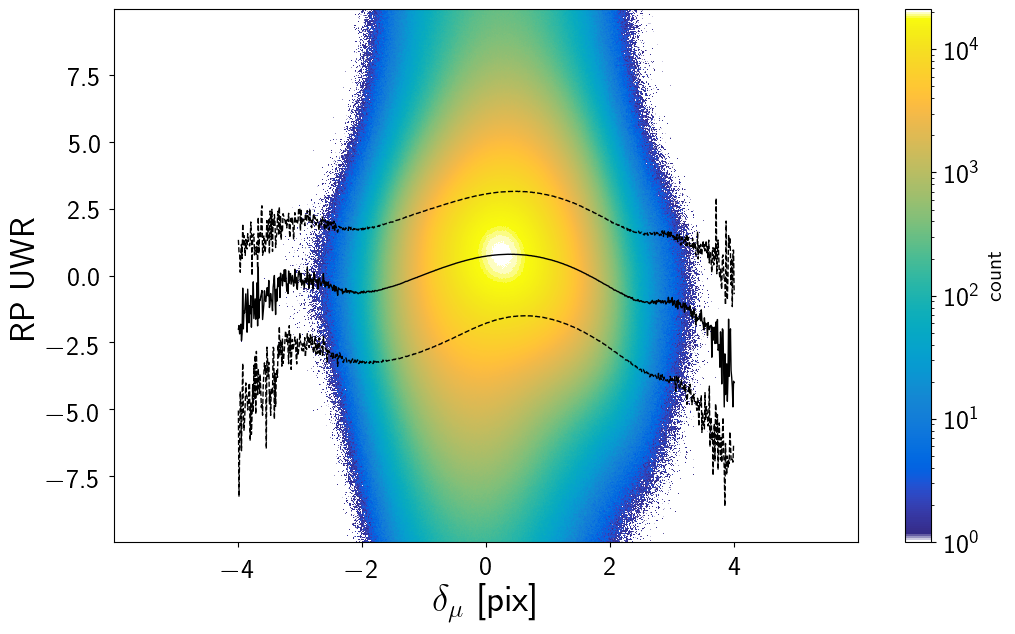}
    \caption{Raw unit weight residuals of the photometry vs. the centring error $\delta_\mu$. The left panel shows $G$ (all AF CCDs), the central panel shows \gbp and the right panel shows \grp. The solid black line shows the median of the distribution, the dashed lines show the $\pm1\sigma$ of the distribution. The centring error is defined as the difference between the source predicted AC position at the observing time and the nominal window centre. The residuals were produced only for the set of calibrators and only using data from the INIT period.}
    \label{fig:fluxloss}
\end{figure*}

The large scale (LS) calibration model describes features that vary smoothly across the focal plane and that might change smoothly with time over timescales of several satellite revolutions. The main changes to the LS model for \edr are: 1) the colour information is represented by spectral shape coefficients (SSC) computed from the internally calibrated source mean spectra; 2) the SSCs are no longer normalised as in \drt \citep[see Appendix A in][]{Proc_DR2} but are used to form flux ratios (see below); 3) additional terms have been included to model the flux loss caused by the on-board window acquisition process. It is important to notice that because of point 1 above, the colour information is now fixed all the way through the calibration process: in \drt instead the reference colour information was updated at every iteration making the overall process less stable. In \drt it was not possible to use SSCs derived from the mean spectra since the internal calibration process of the spectra was not considered to be mature enough for use in the photometric calibration.
For all bands the LS model is a polynomial with a zero point $z_0$, quadratic dependence from the across scan (AC) position of the observation $\mu$ and a quadratic dependence from the centring error $\delta_{\mu}$, defined as the difference between the predicted AC position at the observing time\footnote{The predicted position is computed from the astrometric source parameters, the reconstructed satellite attitude and the geometric calibration for \gband and \xp.} and the nominal window centre. The raw residuals (meaning with respect to the identity model) in \figref{fluxloss} show that there are no observations with a centring error larger than $\pm2$ pix for the \gband and for BP as well. For this reason only observations with a centring error in the range $\pm2$ pix were used to solve for the calibration; the centring error was then clamped to $\pm2$ pix when applying the calibrations to avoid problems caused by extrapolation. For RP the situation is more complex due to the optical design of the instrument: a wider range of $\pm4$ pix was required for the calibration solution and for clamping when applying the calibration (additional information is available in \appref{rpzooming}).

The colour dependencies are modelled in terms of the SSCs fluxes computed from the internally calibrated mean spectra (\secref{meanspectra}). The SSCs fluxes are used to form different ratios to provide pseudo-colours. For \edr we defined four SSC flux ratios:

\begin{eqnarray}
r_1=\dfrac{s_0}{s_1+s_2}&\quad& r_2=\dfrac{s_3}{s_1+s_2}\label{eq:sscbp}\\
r_3=\dfrac{s_4}{s_5+s_6}&\quad& r_4=\dfrac{s_7}{s_5+s_6},\label{eq:sscrp}
\end{eqnarray}
where $s_0$ to $s_3$ are the four SSCs fluxes computed from the BP mean spectrum and $s_4$ to $s_7$ are the four SSCs fluxes computed from the RP mean spectrum. The wavelength ranges defining each SSC are the same as used in \drt \citep[see Table 5 in][]{PhotCore}. The \gband LS model includes a linear dependence from all the ratios defined by \equref{sscbp} and \equref{sscrp}. The LS model for BP includes only the two ratios defined by \equref{sscbp} while the model for RP includes only the two ratios defined by \equref{sscrp}.
The small scale (SS) calibration models the column-level CCD sensitivity. The same model (a simple zero point for each 4 pixel wide AC bin) was used as in \drt. The LS calibration models described above are summarised by \equref{lsmodel} with \equref{ls_colour_model} providing the three different colour dependencies for \gband, BP and RP:
\begin{eqnarray}
\dfrac{f}{I_s}&=&z_0+\sum_{i=1}^2{a_i\mu^i}+\sum_{j=1}^2{b_j\delta_{\mu}^j}+C(r)\label{eq:lsmodel}\\
C(r)&=&\left\{
\begin{array}{ll}
     c_1r_1 + c_2r_2+c_3r_3+c_4r_4& \text{for }G\text{--band},\\
     c_1r_1 + c_2r_2& \text{for BP},\\
     c_1r_3 + c_2r_4& \text{for RP},
\end{array}\right.\label{eq:ls_colour_model}
\end{eqnarray}
where $f$ is the uncalibrated flux of a given CCD observation for a source $s$, $I_s$ is the reference source flux, $r_k$ are defined by Eqs. \ref{eq:sscbp} and \ref{eq:sscrp} and $z_0$, $a_i$, $b_j$ and $c_k$ are the model coefficients to be fitted in the calibration procedure. We define as `calibration factor' the right hand side of \equref{lsmodel} evaluated for a given CCD observation of a given source. The ratio of the raw epoch flux $f$ and the calibration factor provides the calibrated epoch flux.

The time link calibration (mitigating the effect of contamination) that was introduced in \drt was not required for \edr because the throughput in the INIT period was more stable and less affected by contamination than the one used for \drt. For \drt an additional calibration was introduced to help with the mixing between the different instrument calibrations: this was not used for \edr since the use of a more compact set of calibrators allowed to perform more iterations for the initialisation of the photometric system leading to a better mixing between the different instrument configurations.

\subsection{Validation of the LS and SS calibrations}

\begin{figure*}
    \centering
    \widefig{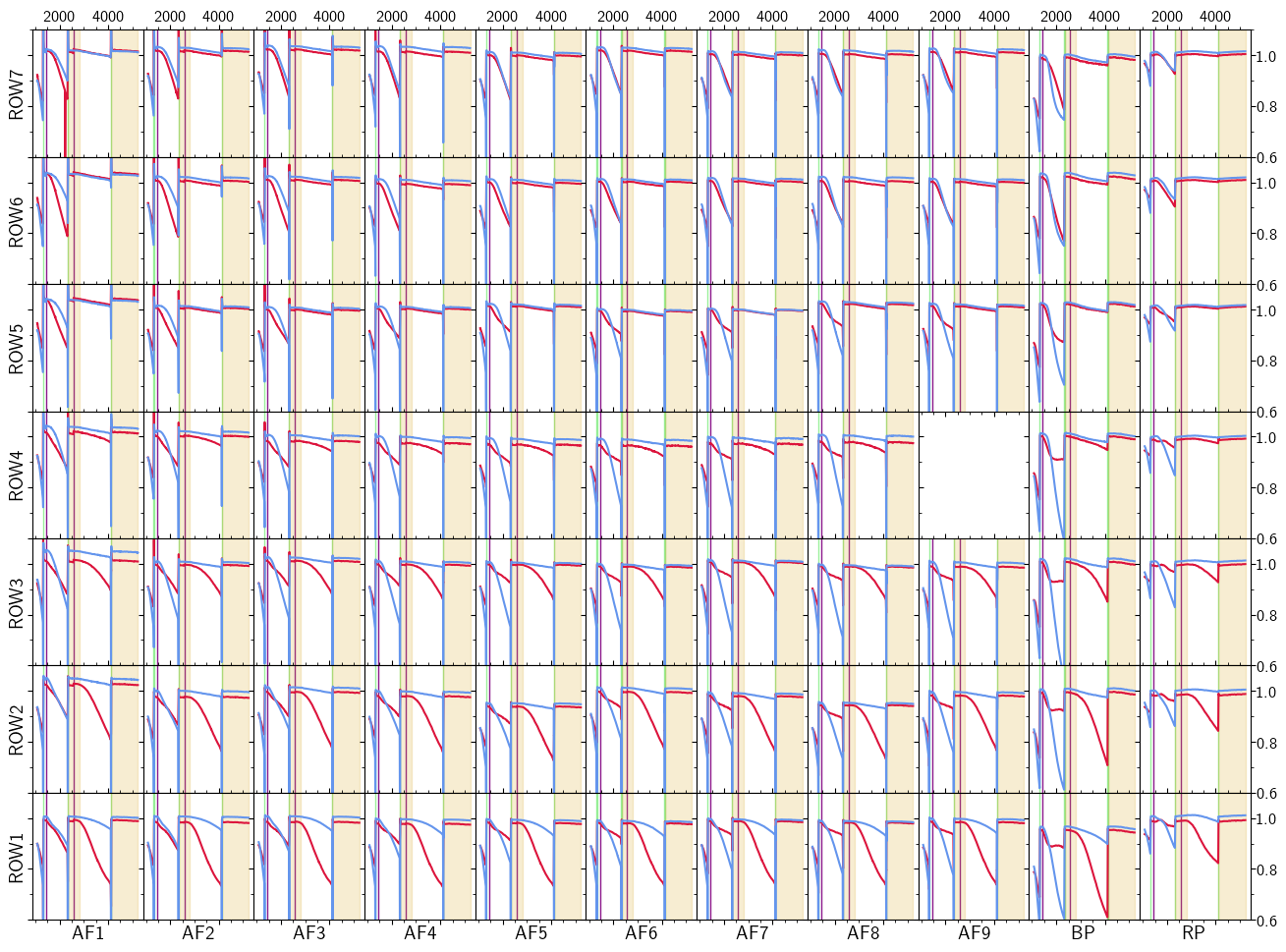}
    \caption{
    Time dependence of the calibration factor for the whole focal plane showing the AF, BP, and RP CCDs for each row. This plot covers the Window Class 1 and 2, ungated configurations. The blue line shows the preceding FoV and the red line shows the following FoV. The shaded area shows the INIT period. The vertical green lines show the decontamination events and the purple vertical lines the refocussing events.
    }
    \label{fig:focalPlanePlot}
\end{figure*}

\afigref{focalPlanePlot} shows the time dependency of the calibration factor for the whole focal plane. This effectively shows the response loss mainly caused by the contamination which affected the early stages of the mission more severely \citep{GaiaRef}. It is interesting to note that the rate of contamination changed behaviour between the two FoVs following the three decontamination events, possibly indicating that the deposition of the contaminant (assumed to be water ice) flipped from one FoV mirror system to the other. This can be seen in the bottom right part of these diagrams. It is also noticeable that the behaviour is very different at different locations of the focal plane.

\begin{figure}
    \centering
    \colfig{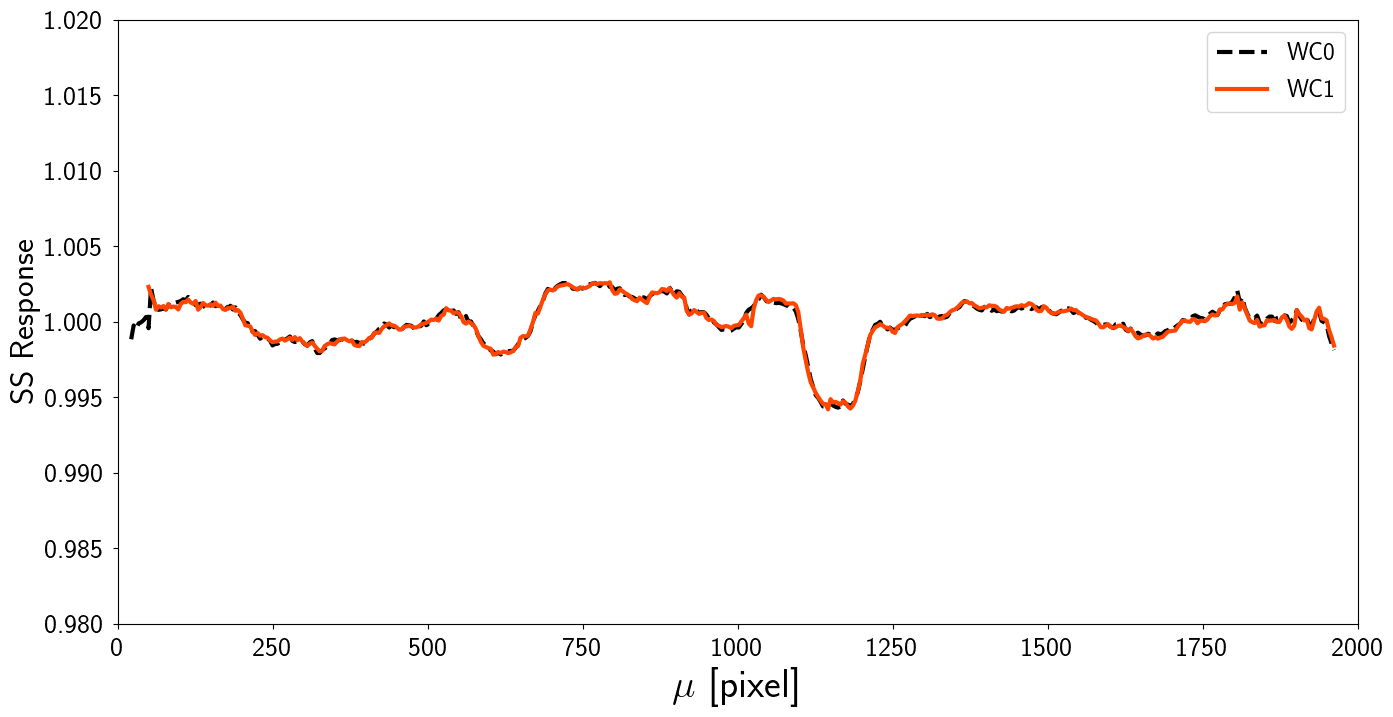}
    \caption{
    Response as measured by the SS calibrations for the BP row 2 CCD in the Following FoV. Two sets of calibrations are shown corresponding to the 1D window configuration (labelled WC1 and shown in dark orange) and the 2D window configuration (labelled WC0 and shown in black). 
    }
    \label{fig:SsDiffBp}
\end{figure}

An example of the quality of the SS calibrations (equivalent to a 1D flat field) is shown in \figref{SsDiffBp}. Here two sets of calibrations are shown for a particular CCD corresponding to 1D (shown in red) and 2D (shown in black) configurations which cover different magnitude ranges. The fact that the two calibrations overlap almost perfectly, even though they are produced using completely independent datasets, confirms that even the smallest features visible in the calibrations are indeed real and not noise. We can therefore conclude that the SS calibration is measuring the CCD response to better than the mmag level.

\begin{figure}
    \centering
    \colfig{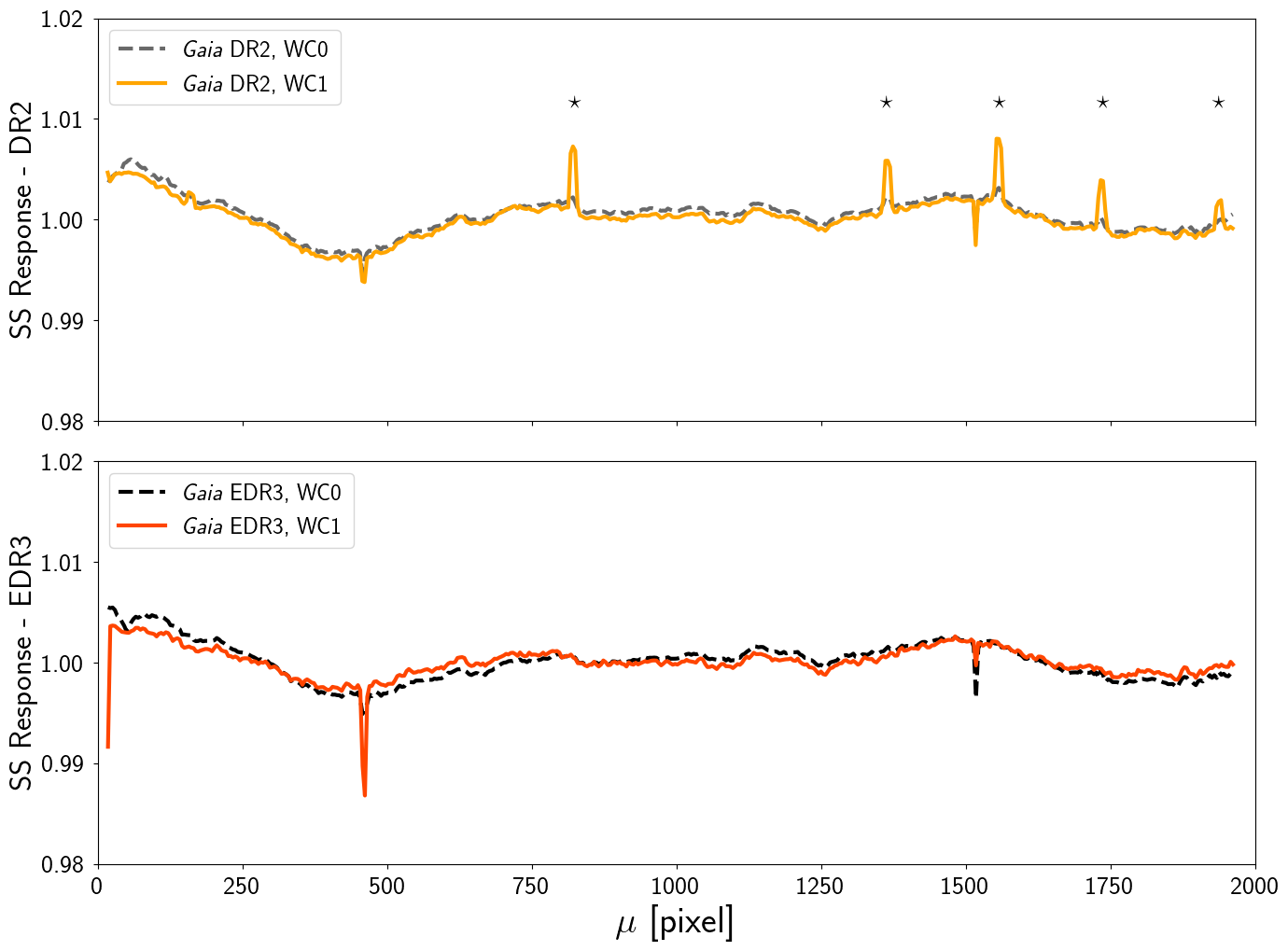}
    \caption{
    Responses as measured by the SS calibrations for the AF5 Row 1 CCD in the Preceding FoV for the 1D and 2D window configurations, labelled WC1 and WC0 respectively.
    The top panel shows the \drt calibration and the bottom one the \edr ones. The $\star$ symbols in the top panel show the location of the hot columns which are no longer visible in the \edr calibration due to improvements in the IPD.
    }
    \label{fig:SsDiffAfBadCol}
\end{figure}

One of the improvements made in the IPD processing leading to \edr is a better handling of hot columns. Before the LSF/PSF fit is carried out, samples corresponding to identified hot columns are masked. In \drt, the effect of hot columns was partially accounted for by the SS calibrations. This is shown in \figref{SsDiffAfBadCol} where for \drt (upper plot) the hot columns can be seen as five narrow peaks. In \edr, these peaks are absent showing that the hot columns have been dealt with correctly. The CCD shown has a particularly large number of defective columns with anomalous response.

\begin{figure}
    \centering
    \colfig{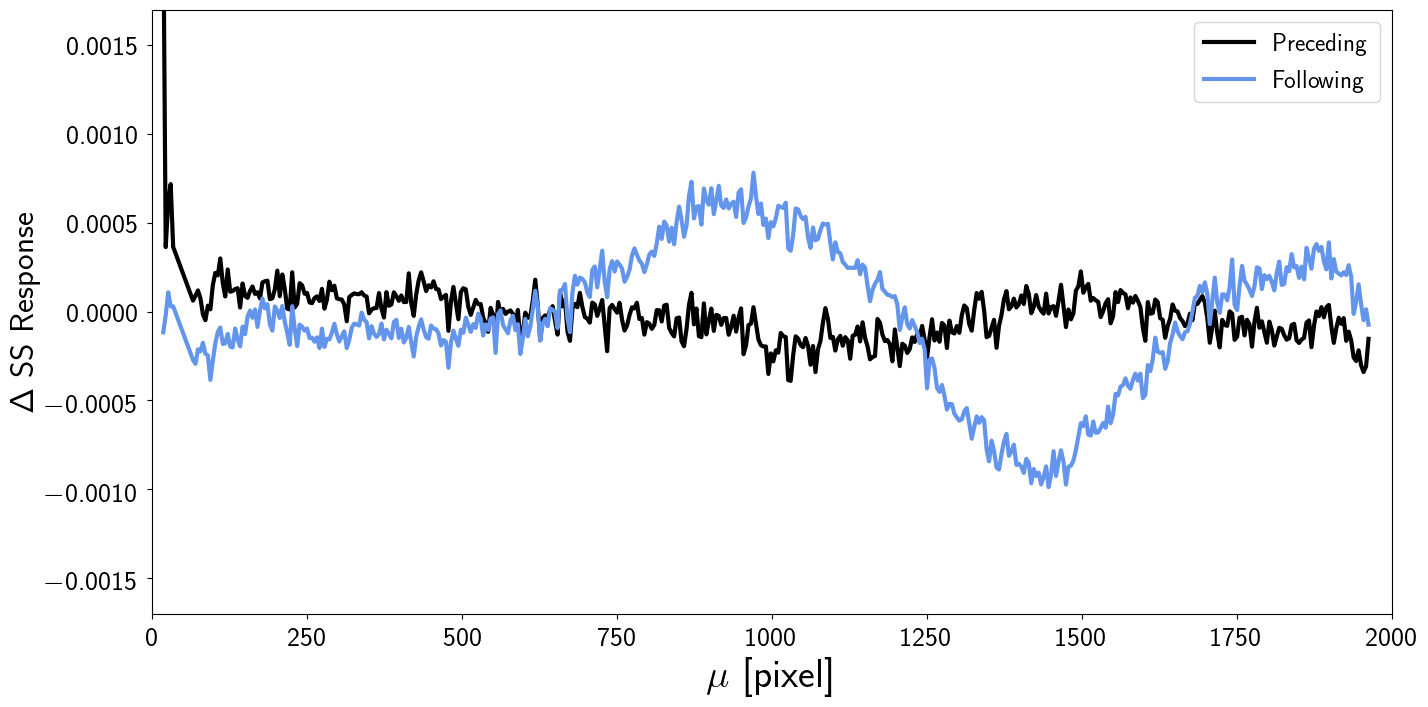}
    \caption{
    Difference in the SS calibration response between two time periods for the AF9 row 1 CCD for the preceding (black) and following (blue) FoV.
    }
    \label{fig:SsDiffTimeScopes}
\end{figure}

The SS calibrations have been calculated for \edr in three separate time periods. This is to provide a crude form of time functionality to the calibration model. Long time periods are needed to ensure that enough data is present for the calibrations, especially for the gated observations. Comparisons between the calibrations obtained for different time periods confirm that the instrument response at the small--scale level does not vary significantly. \afigref{SsDiffTimeScopes} shows an example of one of the largest variations between SS calibrations covering different time ranges. Even in this case the differences are smaller than 1 mmag.
The typical difference between these two time periods for all CCDS is 0.12 mmag.

\subsection{Convergence of the reference system}\label{sec:convergence}
\begin{figure}
    \centering
    \colfig{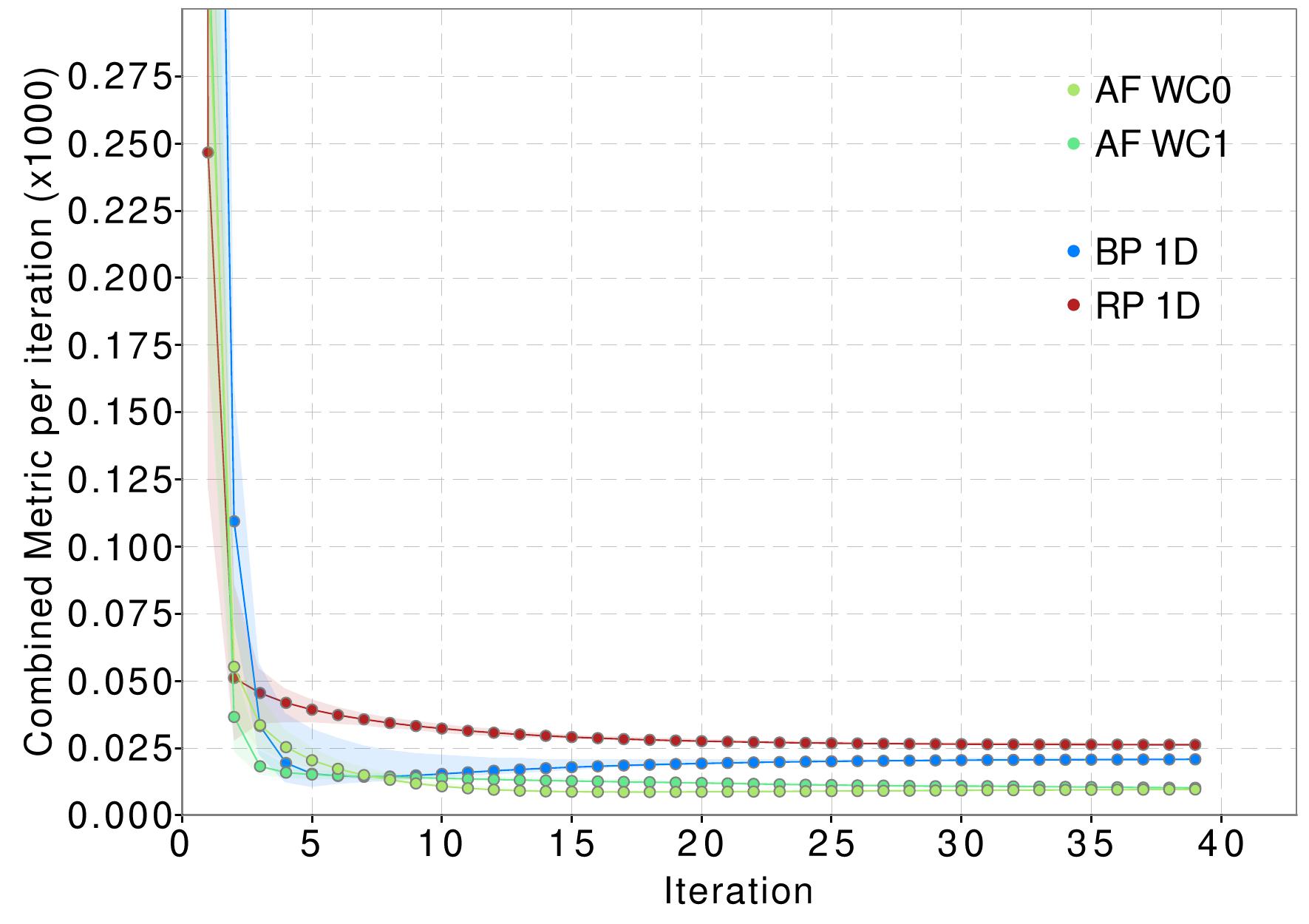}
    \caption{
    The L1 Norm convergence metric as a function of iteration for four major configuration groupings: AF Window Classes 0 and 1, 1D BP and RP observations.
    }
    \label{fig:convergencePlot}
\end{figure}

The main method used to assess the convergence of the photometric system is the same one as used in \dro and DR2 and is described in \cite{PhotValDR1}. This uses the L1 Norm metric to determine the typical change in photometry using the calibration coefficients. \afigref{convergencePlot} shows this metric for four major configuration groupings. While this is much better than seen in \drt \citep{Phot_DR2}, the metric does not converge to zero. Using these plots it was decided to terminate the iterations at the 20th iteration.

Another set of metrics to analyse as a function of iteration number is the calibration coefficients. In an ideal system, when the photometry has converged, the coefficients will remain the same between iterations. For \edr, while this is true for the coefficients involving the AC position and the centring error and also the overall calibration factor and standard deviation of the solution, it is not true for the colour coefficients and these values can change by up to 0.01 over 20 iterations. The reason for this is that there is a strong correlation between the colour coefficients and this causes a partial degeneracy in the calibration model. However, the overall calibration factor is stable at the sub-mmag level and the only implication of this is that these coefficients cannot be used in additional validation analyses for example plotting them as a function of time.

\begin{figure}
    \centering
    \colfig{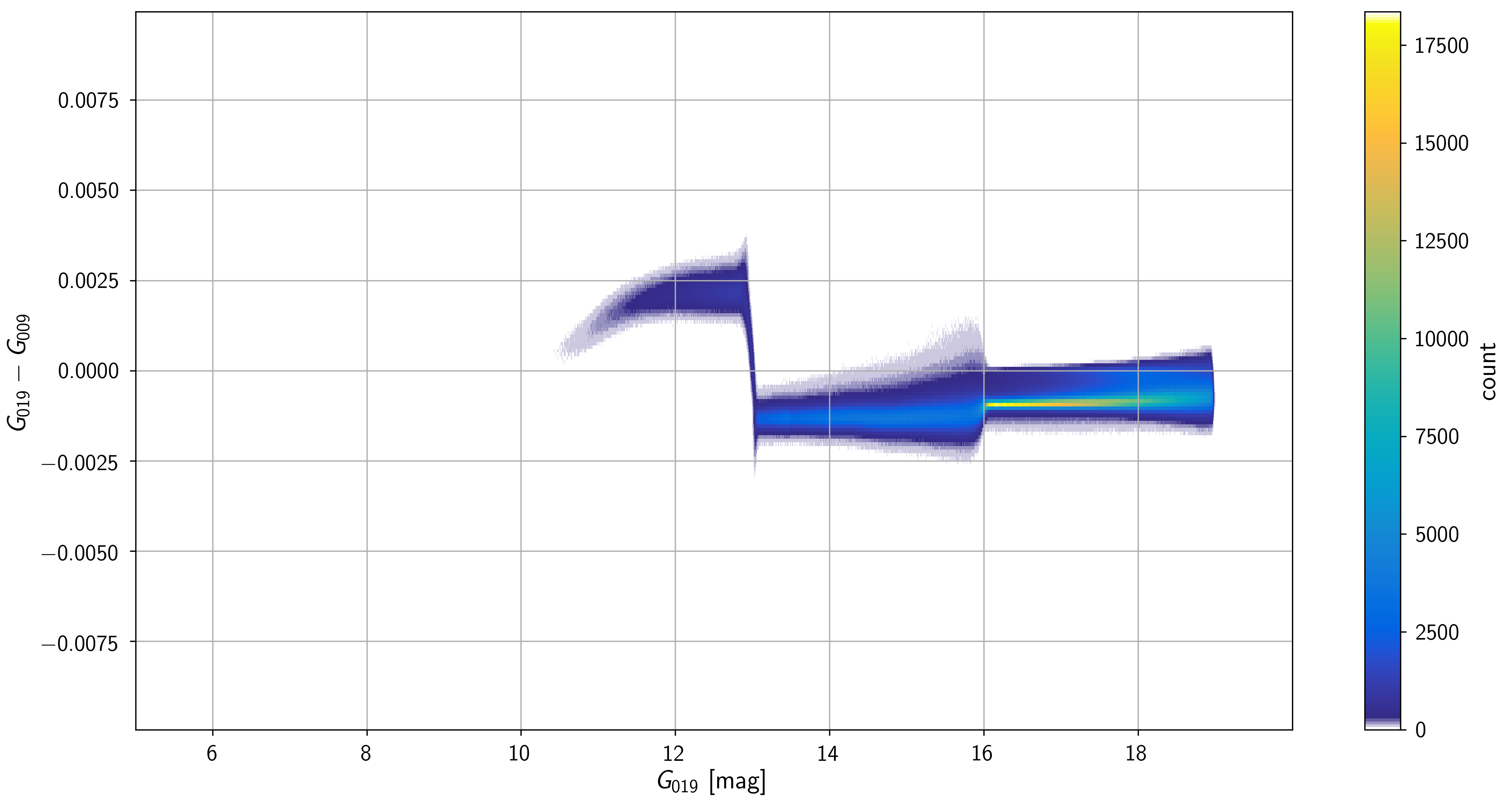}
    \colfig{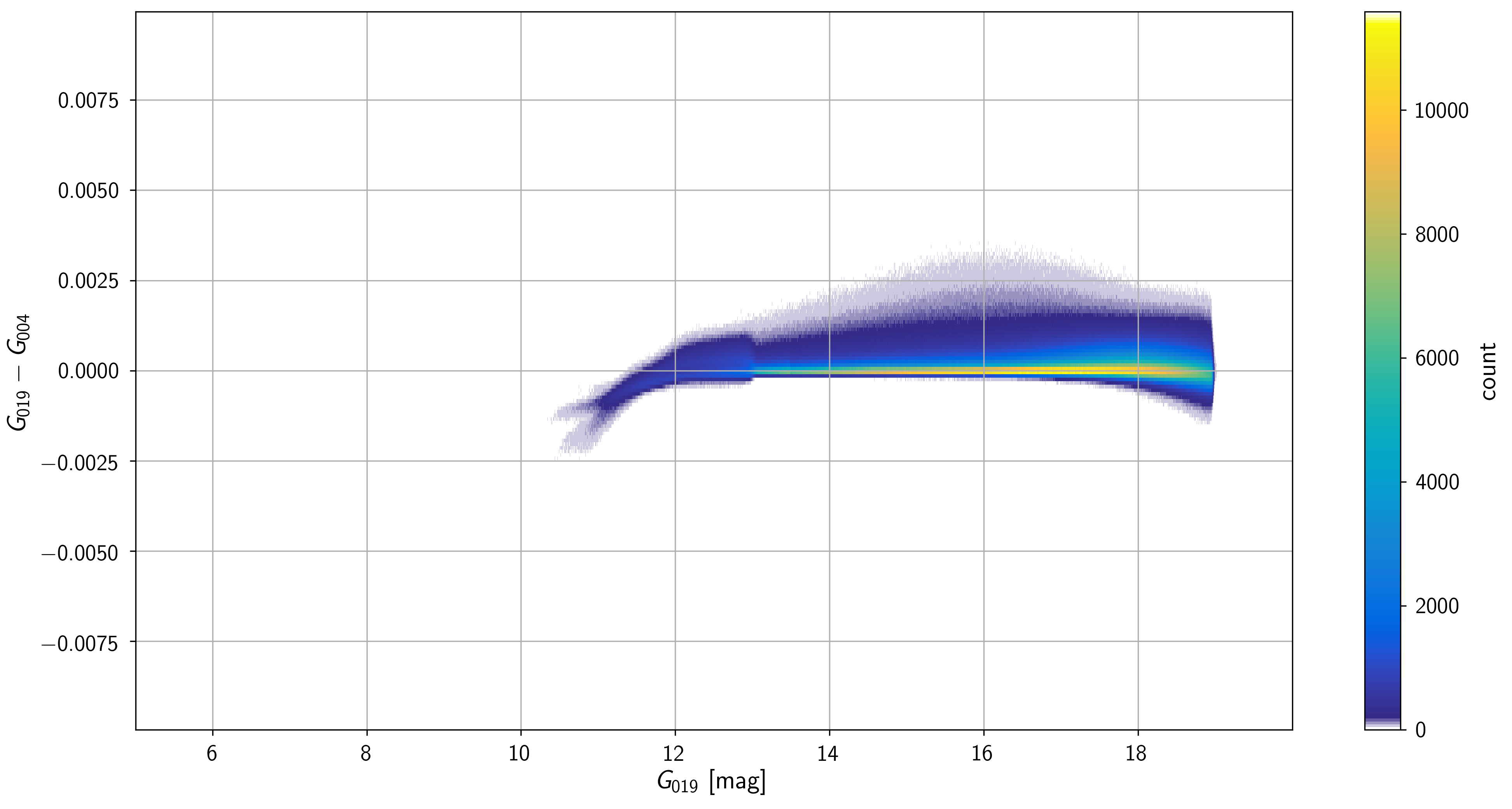}
    \caption{
    The difference between the \gband photometry in magnitudes between different iterations as a function of magnitude for a test (\textit{upper panel}) and the final (\textit{lower panel}) calibration model.
    }
    \label{fig:linkIteration}
\end{figure}

A single overall photometric system will form when there is good mixing between the different configurations, meaning that each configuration is calibrated with many calibrators and each calibrator is observed under many different configurations. Problems can arise when configurations are limited to certain magnitude ranges and if systematic effects are present in the data that are not accounted for by the calibration model. For the different gate configurations this is not a problem as the magnitude ranges of activation of each gate are small and the uncertainty of the on-board magnitude determination is large in comparison. However, for the window class configurations (with boundaries at $G=13$ and 16) the on-board magnitude accuracy is good (about 1\%) which means that the number of sources that are observed in more than one window class is small. The effect of this is that the ability of the iterations to create a consistent photometric system across all configurations is limited. This can be seen in \figref{linkIteration} which shows the difference in photometry between subsequent iterations as a function of magnitude for a test calibration model (top panel) and the final one used (bottom panel). In the test calibration model, it can be seen that there are discontinuities occurring in correspondence to window class configuration changes. 
While there is a physical reason for the discontinuity occurring at $G=13$ due to different flux loss effects in 2D and 1D windows, the jump at $G=16$ can only be due to a problem in the convergence to a consistent system across different configurations. The convergence process occurs very slowly due to the poor mixing between these magnitude ranges. For the final model an offset was introduced between the window class configurations separated at $G=13$ to speed up the convergence and the two faintest window class configurations were combined into a single one. The improvement can be seen in the lower panel of \figref{linkIteration}. The lack of discontinuities at $G=13$ and $G=16$ can also be seen in \figref{extcatcomp}.
We note however, that while this strategy will ensure that the final photometric system does not show discontinuities due to poor mixing between different instrument configurations, it is still possible for unmodelled systematics to cause small inhomogeneities in the internal photometric system that might require further treatment (see also \secref{extcal}).

\section{Mean source photometry}\label{sec:sourcephot}

The generation of the mean \gband, \gbp, and \grp source photometry follows the same process used for \drt and described in \cite{Proc_DR2}: epochs are calibrated by applying the appropriate LS and SS calibrations and the resulting calibrated epoch flux is accumulated, for each band, to produce a mean source flux as the weighted mean of the valid contributions (with the weight defined as the inverse of the variance on the calibrated epoch). 
The uncertainty on the weighted mean flux, accounting for any intrinsic scatter that might exist within the data, will be given by
\begin{equation}
   \sigma_{\overline{f}}=
       \sqrt{\sum f_i^2w_i - \overline{f}^2\sum w_i \over {n-1}}
       {1 \over \sqrt{\sum w_i}}.
\end{equation}
It can be shown that if the underlying distribution is Gaussian, then an inverse variance weighted mean is the maximum likelihood estimator for the mean of the distribution \citep{lupton1993statistics}. 
Furthermore, this type of weighting always guarantees the maximum signal to noise in the answer.

An epoch contribution to this weighted mean flux is considered valid when both the LS and SS calibrations have been successfully applied and the calibrated flux is at least 1 ${\rm e}^-{\rm s}^{-1}$. This minimum flux threshold was introduced in \drt to mitigate the impact of extreme outliers \citep{Proc_DR2}: the impact of this flux threshold for \edr\ is discussed in \secref{fluxlimit}. Calibrated epochs could also be excluded a priori from contributing to the mean photometry in a given band depending on quality metrics based on acquisition and processing flags. AF observations were excluded from the mean photometry when any of the following criteria were met: AC trimmed windows acquired around 2230 OBMT revolutions as part of a set of tests that were performed to assess the impact of reducing the AC size of AF windows; windows affected by a charge injection; windows that had some of the samples removed because of inter-FoV truncation; windows for AF2 ROW5 with a reference AC position larger than 1203 pixels (the data is severely affected by a deep trap in the serial register); windows for which the IPD was flagged as not successful; windows for which the source predicted AC position was not available; AF observations in the periods immediately following a decontamination campaign have also been excluded due to large variations in the system response caused by the focal plane having not reached thermal equilibrium yet (see \appref{gaps} for more information).
BP and RP observations were excluded from the mean photometry when any of the following criteria were met: truncated windows; windows affected by a charge injection; windows acquired with multiple gates; windows for which the source predicted AC and AL positions were not available (this information is required for the pre-processing of the epoch spectrum from which the raw epoch flux is produced); windows affected by bad columns.
To apply the LS calibration to the epoch observations of a given source, it is necessary to use the source SSCs derived from the internally calibrated mean spectra. Depending on the availability of the SSCs (see \secref{calmodel}), there are three different calibration procedures: `gold' -- when all eight SSCs are available; `silver' -- when for either BP or RP some or all SSCs are missing; `bronze' -- when SSCs are missing or incomplete for both BP and RP or if the silver processing failed (see below). Since the calibration model involves ratios of SSC fluxes (see \equref{sscbp} and \equref{sscrp}) the set of BP SSCs is considered complete when all four SSCs are present and $s_1+s_2>0$ and analogously for RP but with $s_5+s_6>0$.
It is important to stress that the `grade' of a source is determined solely by the availability of the SSCs and has no implications about the availability of mean photometry in the various bands. In particular, it is possible for a gold source to be missing the photometry in any of its bands or for a bronze source to have photometry in any of the bands. This is a consequence of the independence between the spectral processing (leading to the generation of the mean spectrum and SSCs) and the calibration of the integrated photometry.

In order to calibrate non-gold sources it is necessary to produce an estimate of the missing SSCs. For bronze sources, a set of default SSCs are used for every source: this is analogous to how bronze sources were calibrated in \drt. For silver sources, the missing SSCs are estimated from the \gband and the available \xp band using empirical relationships derived using a set of gold sources. For silver sources an iterative process is used to generate the mean photometry: an initial estimate of the source photometry is derived using the default SSCs; this initial guess is then used to obtain an updated set of SSCs for the missing band using the empirical relationships described in \appref{sscs}; the resulting set of estimated SSCs is then used to produce the updated mean photometry. The iterative process is considered successful when the mean $G$ flux between two consecutive iterations has changed by less than $0.05\%$ or if a maximum of 20 iterations is reached. If the mean $G$ flux fails to be produced then the iterations are stopped and the source is then handled as bronze.

\begin{figure}
    \centering
    \colfig{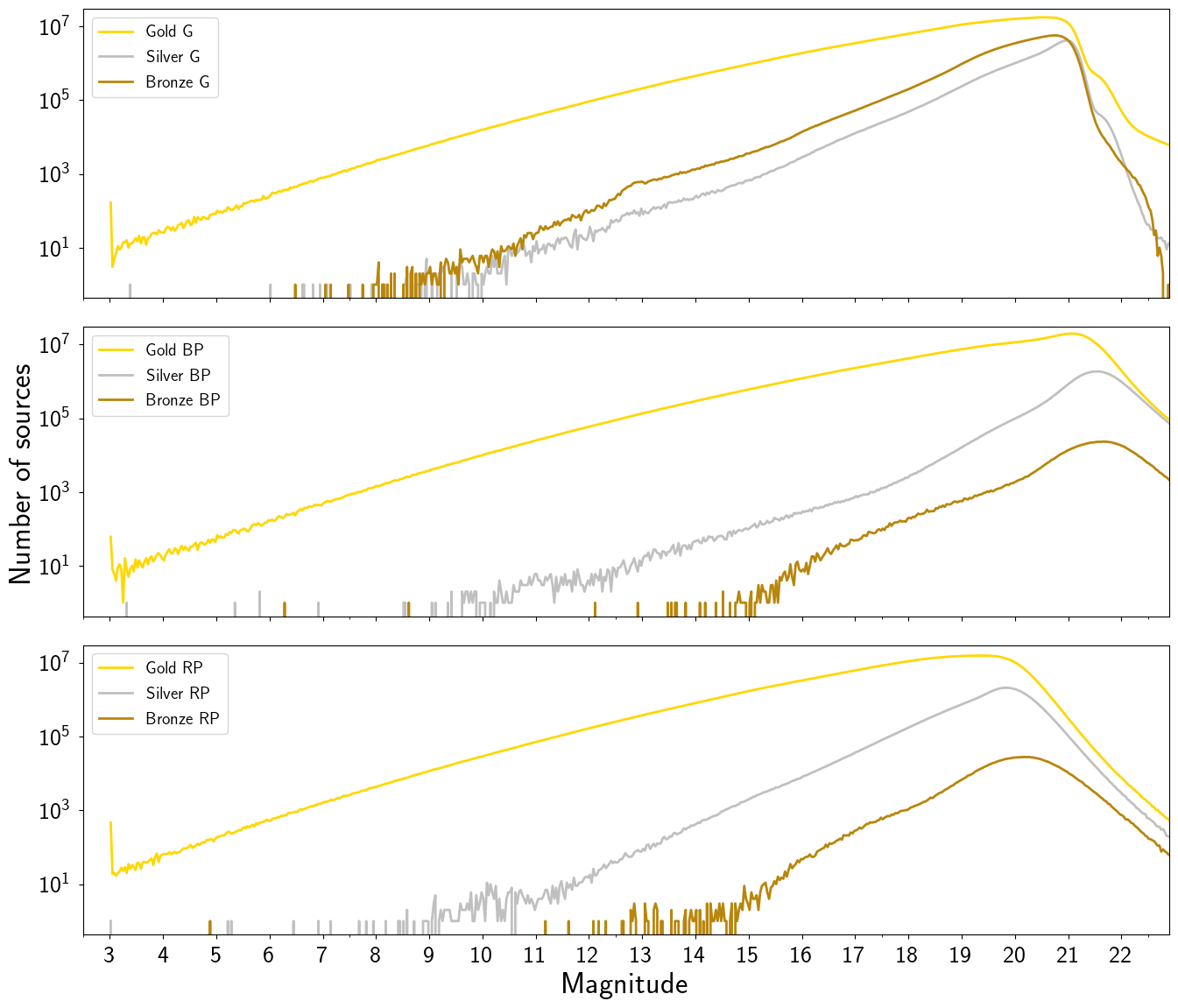}
    \caption{Number of sources with gold, silver, and bronze \gband, BP, and RP photometry as a function of $G$, \gbp, and \grp magnitude in the top, mid, and bottom panel, respectively. Although the photometric grade (gold, silver, bronze) is the same for all bands of a given source, there are sources with incomplete photometry (regardless of their grade), as discussed in the text.
    }
    \label{fig:magdist}
\end{figure}

A total of 1,602,086,411 sources were calibrated using the gold procedure, 204,074,348 sources were calibrated using the silver procedure and 746,399,821 sources were calibrated using the bronze procedure. The actual number of sources for each grade in the \edr archive will be lower because of various data quality filters applied during the catalogue preparation. The magnitude distributions of the gold, silver and bronze sources in \figref{magdist} show that silver and bronze sources are concentrated at the faint end where BP and RP spectra have lower signal-to-noise and completeness can be affected by a combination of crowding and the limitations in the resources available to the on-board video processing unit (VPU) which do not allow the allocation of a \xp window for every single observed transit.

\begin{figure}
    \centering
    \colfig{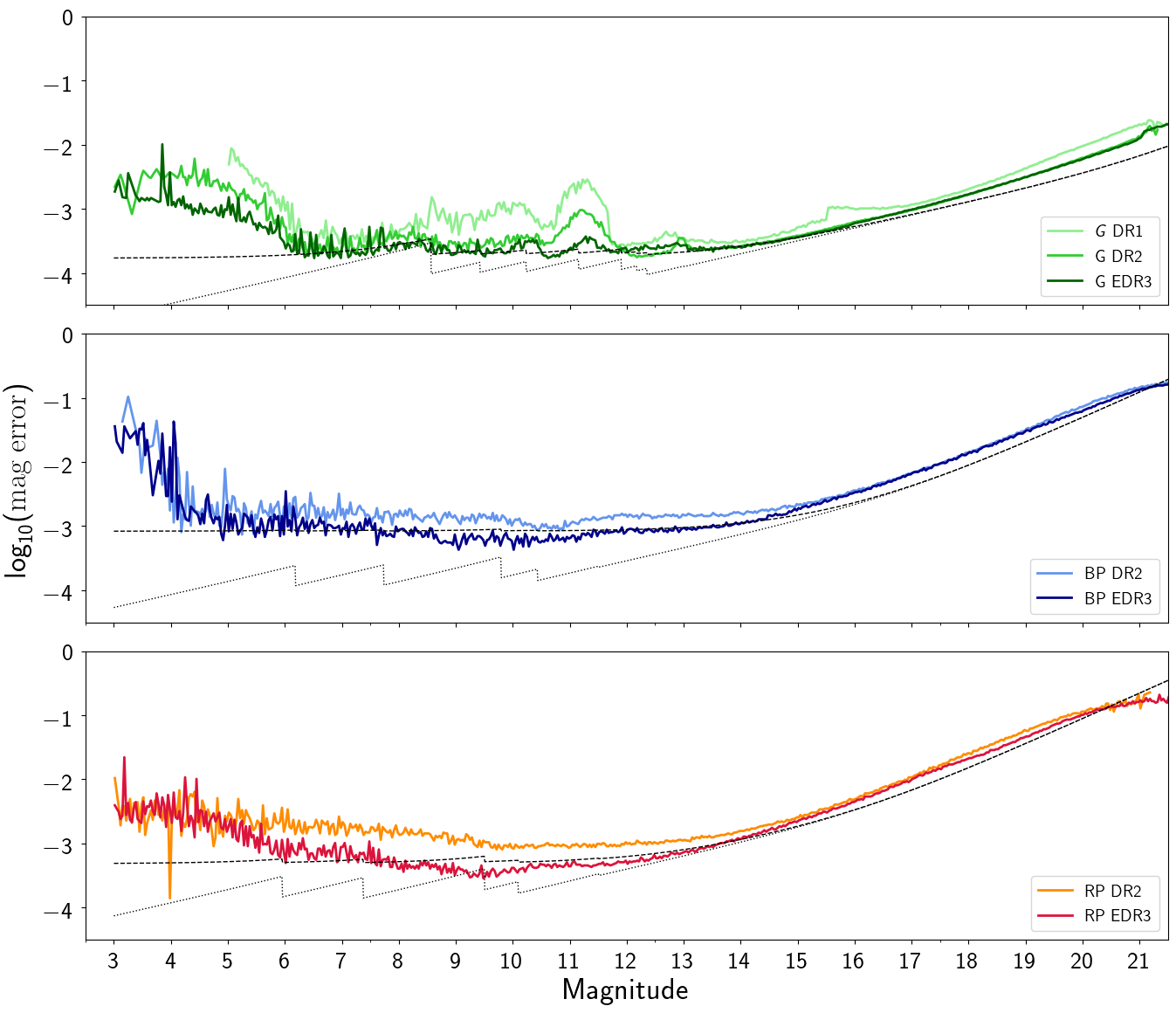}
    \caption{Distribution of the uncertainty on the weighted mean $G$ (\textit{top panel}), \gbp (\textit{central panel}), and \grp (\textit{bottom panel}) as a function of the $G$, \gbp, and \grp magnitude, respectively. Only sources with $\approx20$ transits (corresponding to $\approx200$ CCD observations in $G$) have been included in this analysis. The black dotted line shows the expected uncertainties for sources with 200 \gband (20 \gbp, \grp) contributions for a nominal mission with no calibration error. The dashed dark line shows the same expected uncertainties with an additional calibration error on the single measurement of 2.0 mmag for \gband, 3.1 mmag for \gbp and 1.8 mmag for \grp added in quadrature. The \dro and DR2 uncertainties are shown for comparison.}
    \label{fig:photerrors}
\end{figure}

\afigref{photerrors} shows the uncertainty on the weighted mean as a function of magnitude for the gold photometry. Only sources with approximately 200 \gband CCD observations (and analogously 20 in \gbp and \grp) have been included to allow comparing with the predicted uncertainties \citep{CARME}. The dotted line in each of the three panels shows the predicted uncertainty for a nominal mission and 200 CCD observations. The dashed line in each of the three panels shows the same predictions but combined with a calibration error of 2.0, 3.1, and 1.8 mmag for $G$, \gbp, and \grp, respectively. The figure includes also the \drt and DR1 uncertainties for comparison (for the latter only the \gband uncertainties are shown since the \xp photometry was not part of that release). In the \gband a large improvement can be seen in the range $10.5\lesssim G\lesssim11.5$ thanks to improvements in the handling of saturated samples in the IPD process. The \gband uncertainty can be seen to increase in the range $11.5\lesssim G\lesssim13$ to then drop again following the dark dashed line (see top panel of \figref{photerrors}). This increase in the uncertainty is due to the fact that the PSF modelling did not include the dependency from the AL rate \citep[see][]{EDR3_PSF}: the effect is expected to become more significant for longer gates which is indeed reflected by the behaviour observed for the errors. The AL rate effect on the PSF will be included in the modelling for \drf which is therefore expected to have improved errors in this magnitude range.
A significant improvement is also noticeable at the very bright end, $G\lesssim6$, which is mostly due to improvements in the handling of saturated samples in the IPD process. 
For \gbp and, to a larger extent (see \appref{rpzooming}), \grp the improvements at the brighter end are also due to the modelling of flux-loss in the photometric calibrations. At the fainter end, instead, the improvements are due to the improvements in the background mitigation, which for \edr includes an estimate of the local background (see \secref{overproc:bkg}).

\begin{figure}
    \centering
    \colfig{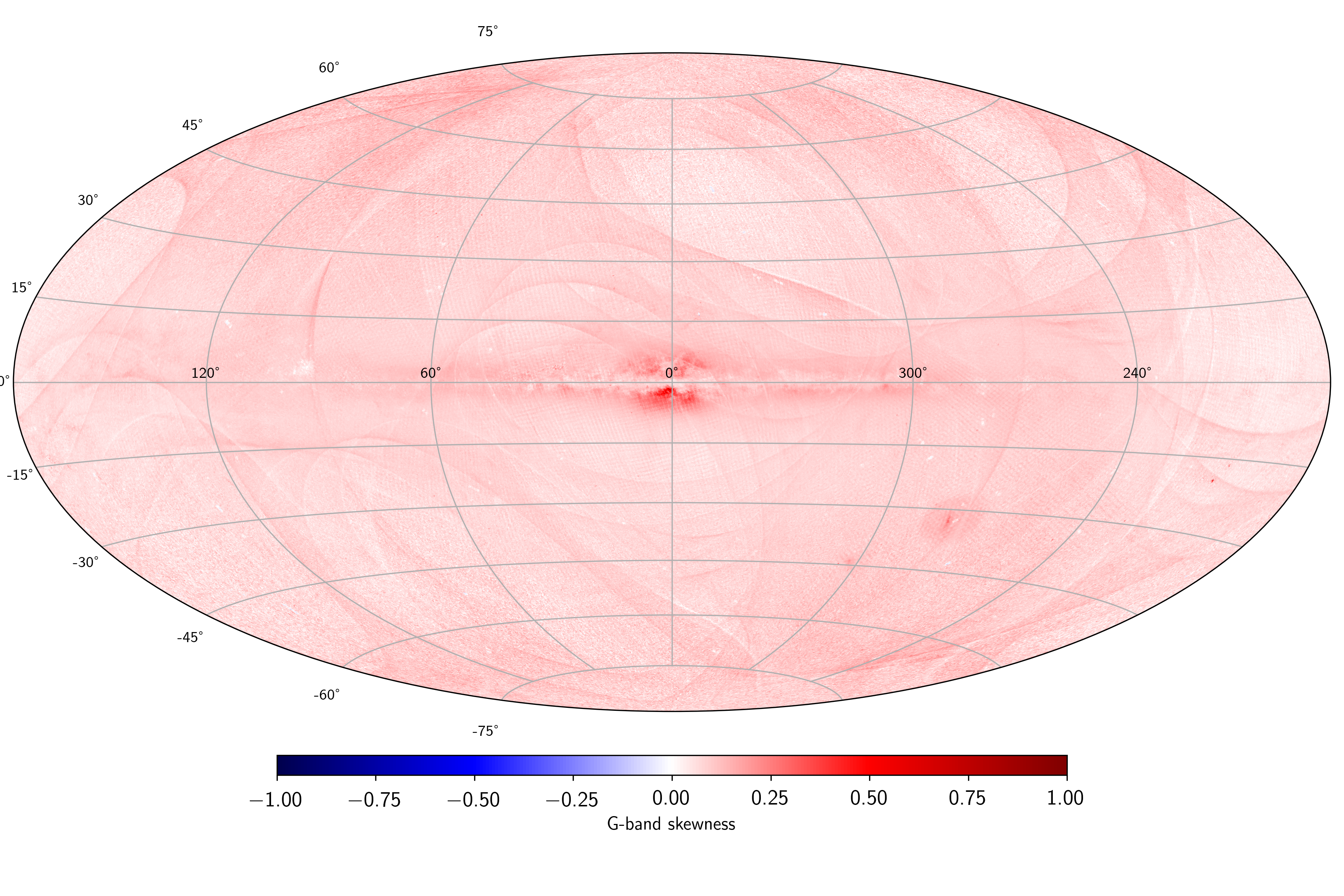}
    \caption{Sky distribution of the median skewness of the G flux. The map was produced by computing for each level $k=8$ HEALPix pixel the median $G$ flux skewness value of all gold sources.}
    \label{fig:skewnesssky}
\end{figure}

By plotting various statistics as a function of sky position it is possible to identify problems with the processing. In \drt, the skewness of the flux distribution of each source was used to identify periods where the calibration had been problematic. During these periods, for example after decontamination, the calibration had not worked well and caused observations acquired during such periods (about four days) to be poorly calibrated and become outliers for these sources. These sources would tend to have larger skewness values than normal and they would form great circles in the sky distribution of the skewness. \afigref{skewnesssky} shows the sky distribution of the source $G$ flux skewness for \edr. As can be seen, larger skewness values do not distribute along great circles but in areas of very high source density (Galactic centre and LMC) and in regions with higher scan coverage. This second effect is not fully understood yet but it is of much lower significance than the one related to the sky density.

\begin{figure}
    \centering
    \colfig{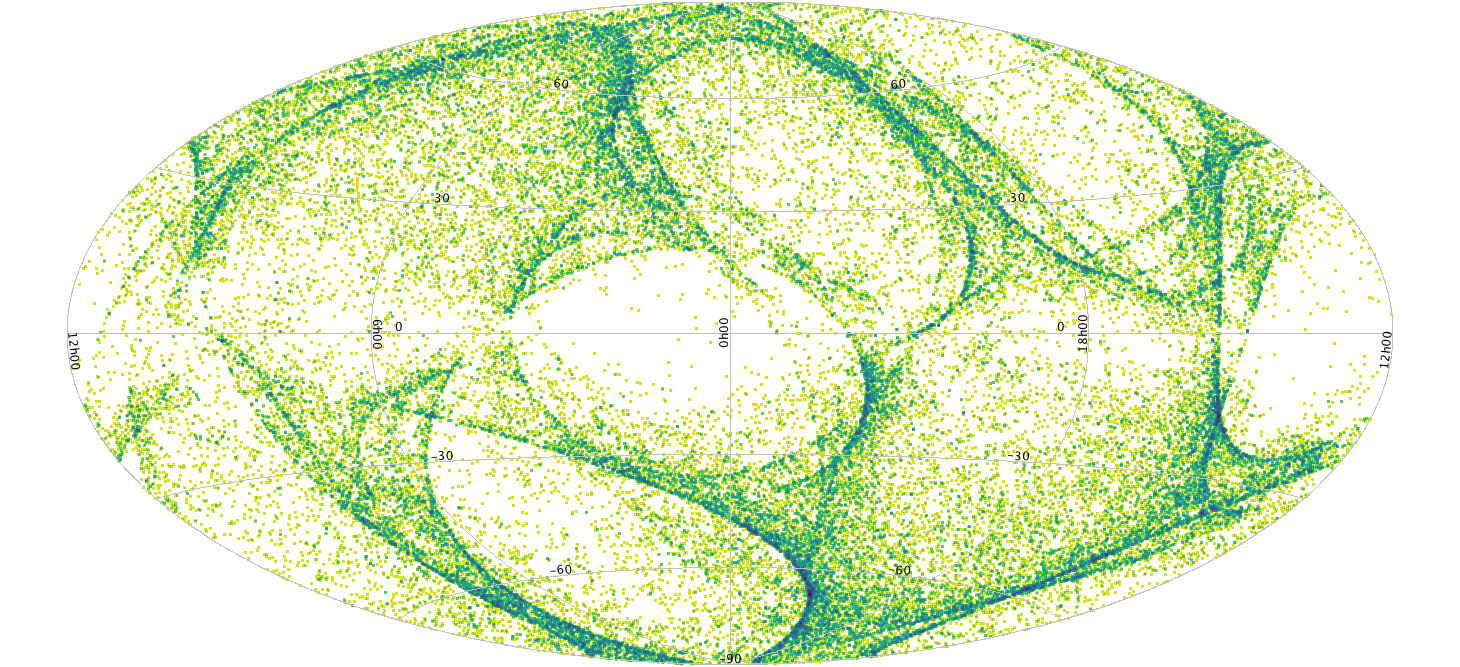}
    \caption{Sky distribution of sources with $G>22$. The only visible features are related to the \gaia scanning law.}
    \label{fig:veryfaintgsky}
\end{figure}

Looking at the sky distribution of the faintest sources can also provide useful insights on the quality of the photometry. In \drt, the distribution of sources fainter than $G=21.7$ showed a number of features in the shape of great circles therefore indicating problems with the processing \citep[see for example][]{SKEWNESS}. \afigref{veryfaintgsky} shows the sky distribution of \edr sources with $G>22$: the only visible features are linked to the scanning law and are explained by the fact that regions with higher number of observations (primarily because of more frequent scans) tend to reach a fainter magnitude limit. No other features are visible, indicating the lack of processing problems and the improved quality of \edr.

\begin{figure}
    \centering
    \colfig{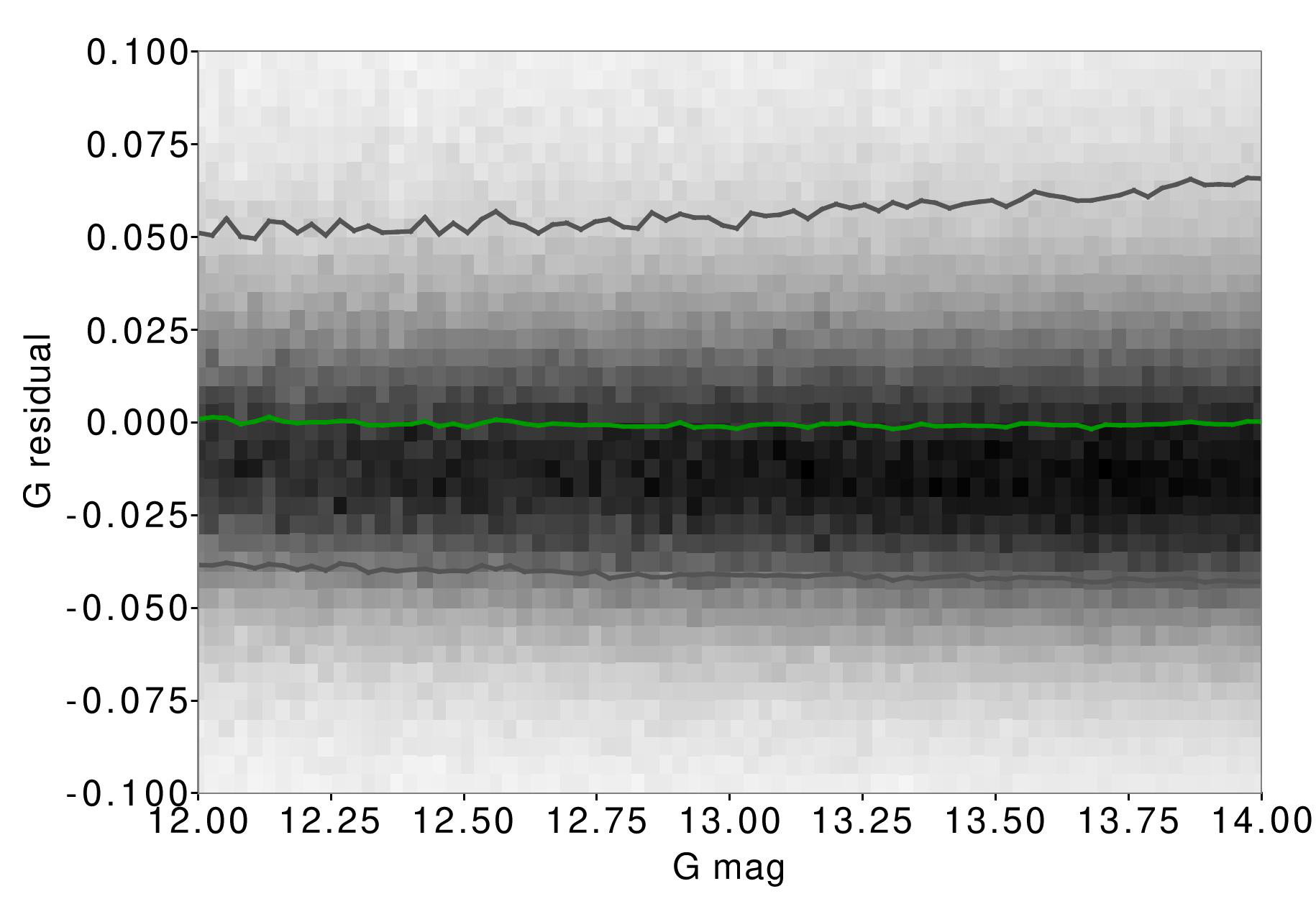}
    \colfig{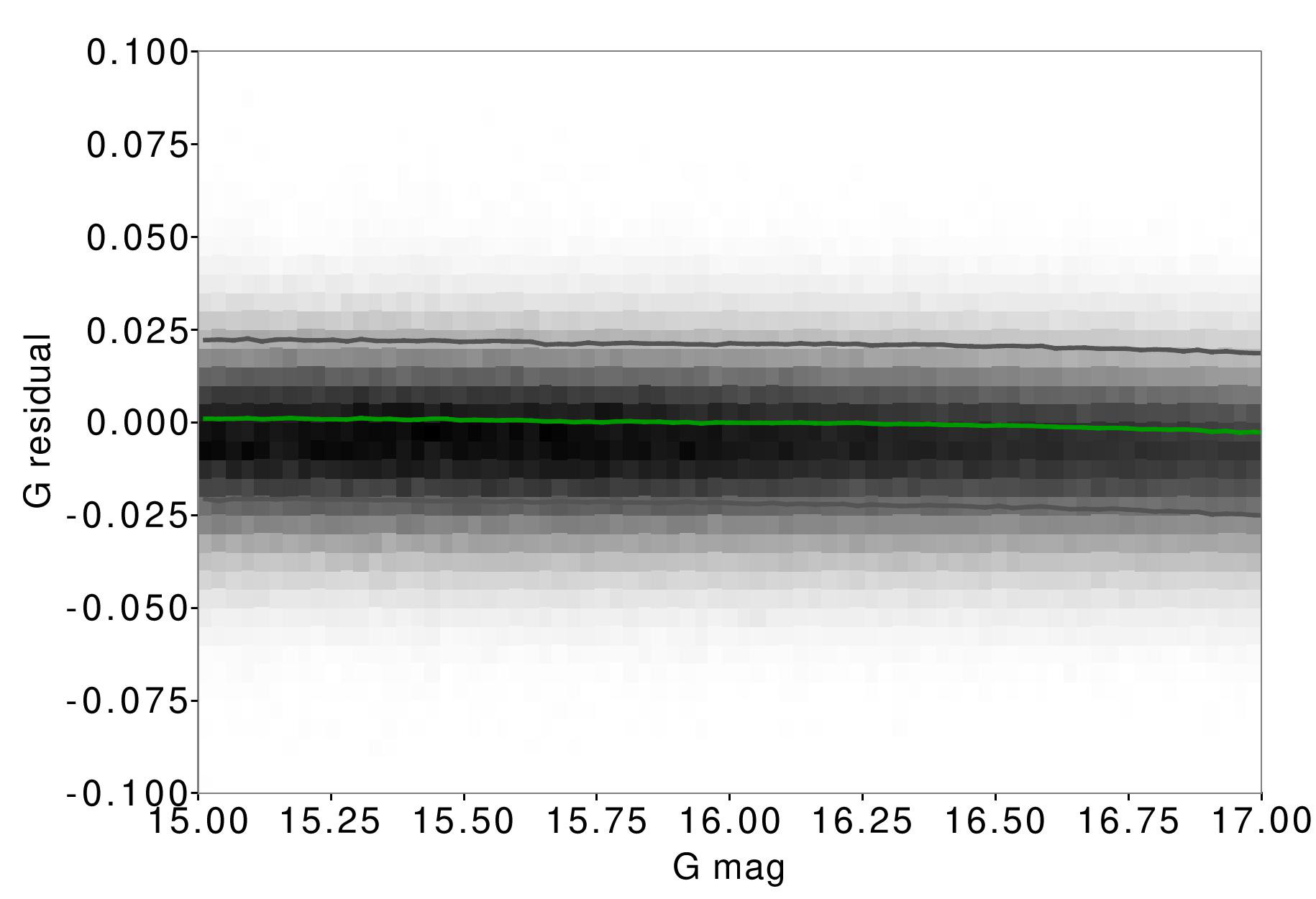}
    \caption{Comparisons of \edr  with APASS (upper plot) and SDSS DR15 (lower plot) showing that no discontinuities are detected at $G=13$ or $G=16$.}
    \label{fig:extcatcomp}
\end{figure}

Comparisons with external catalogues are usually quite difficult to carry out since they involve different passbands. Additionally, if the comparison shows a discrepancy, it can be difficult to establish whether it should be ascribed to the internal catalogue or external one. In \dro \citep{PhotValDR1} and \drt \citep{Phot_DR2}, a discontinuity was present in the comparisons with APASS at $G=13$ \citep{APASS} and with SDSS DR15 \citep{SDSSDR15} at $G=16$. Since at these magnitudes there are two important changes in the \gaia window configuration, it was reasonable to conclude that the discontinuities were a result of the \gaia processing or observation process. 
The equivalent comparisons have been carried out also for \edr (using the colour transformations given in \appref{cctran}) and are presented in \figref{extcatcomp} showing that the discontinuities are not visible anymore.

\section{BP and RP flux excess}\label{sec:xpxs}

\begin{figure}
    \centering
    \colfig{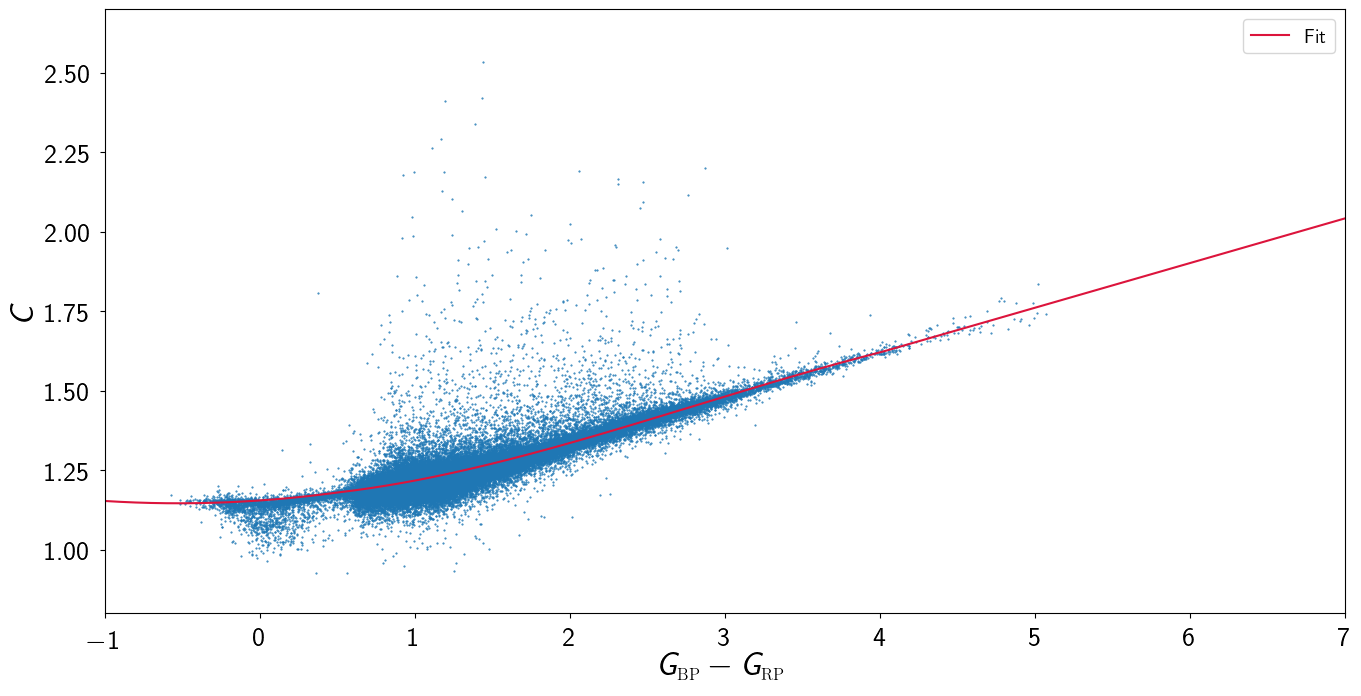}
    \colfig{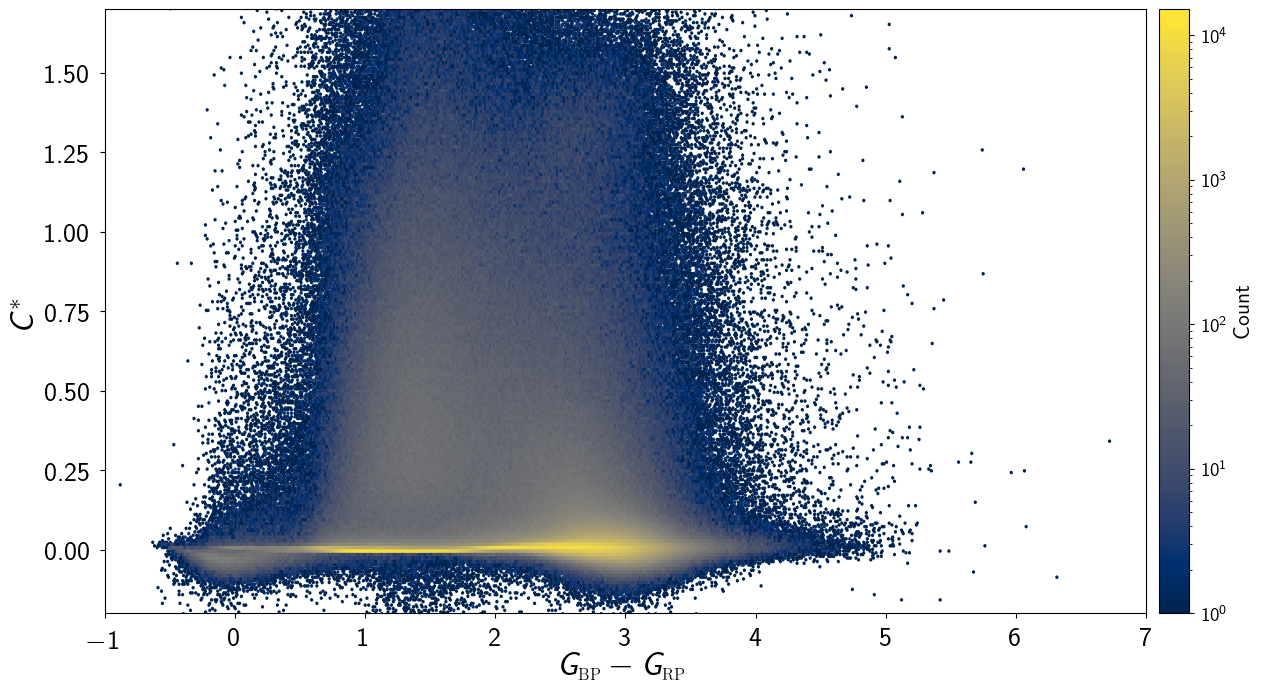}
    \caption{Determination of the \xp flux excess dependence on \bprp colour and its application to define the corrected excess factor \cxs. \textit{Top panel}: \xp flux excess vs. \bprp colour for the set of standard sources from \cite{2000PASP..112..925S} secondary standards and \cite{Ive07}. The red line represents the combined fit based on two different polynomials for the bluer-end and the central region and a linear fit for the red-end. \textit{Bottom panel}: corrected flux excess factor \cxs vs. \bprp colour for a set of nearby sources selected from the \edr archive.}
    \label{fig:xpfit}
\end{figure}

In \drt the background treatment for \xp was limited to the mitigation of the time and CCD-dependent straylight contribution \citep{Proc_DR2} and was based on maps derived from $\approx8$ revolutions. For this reason the maps were very insensitive to variations in the local background level, which therefore was still affecting the \gbp and \grp integrated photometry, especially at the faint end. \cite{Phot_DR2} introduced a quality metric, the \xp flux excess factor defined as a simple ratio between the total flux in BP and RP, and the \gband flux: $C=(I_{\textrm{BP}}+I_{\textrm{RP}})/I_G$. The motivation for $C$ as a quality metric was simply that because of the instrument passbands and response the $C$ ratio should be only slightly larger than one. The actual distribution of $C$ versus \bprp colour is more complex with the excess becoming progressively larger towards redder colour while flattening out to a constant level towards the blue end of the colour range. In \drt \cite{Phot_DR2} concluded that large values of the excess factor $C$ were caused by problems in the \gbp and/or \grp photometry and therefore recommended to filter sources with a large excess factor considering them problematic.
Because of the strong dependence on colour, using the \xp flux excess can often lead to results that are difficult to interpret. To overcome this limitation, we introduce the corrected \xp flux excess factor \cxs defined as:
\begin{equation}\label{eq:cxs}
\cxs=C-f(\bprp),
\end{equation}
where $f(\bprp)$ is a function providing the expected excess at a given colour for sources with good quality photometry. By definition \cxs is expected to be close to zero with positive values indicating that the source has more flux in BP and RP than in the \gband and vice versa for negative values. In order to derive the colour dependency $\mathbf{f}(\bprp)$ we used a sample of about 200,000 isolated and well observed sources based on a selection of the Stetson secondary standards (see Appendix \ref{app:stetson} for a description of this dataset) and a selection of the \cite{Ive07} standards. Only \edr photometry from gold sources was used in the analysis. Using a single polynomial to fit the data tends to perform poorly at the blue and red ends of the distribution. The blue end of the distribution is better described by a quadratic polynomial; the central part of the distribution is well fitted by a cubic polynomial whereas the red end can be well represented by a linear fit. The coefficients of the three polynomials and their applicable colour range are provided in \tabref{cxs_fit}.
\begin{table}[]
\caption{Coefficients of the polynominals $\mathbf{f}(x)=\sum{a_ix^i}$ fitting the \xp flux excess factor $C$ dependence on the $x=\bprp$ colour with their applicability range.}
    \label{tab:cxs_fit}
    \centering
    \begin{tabular}{l|c|c|c}
    \hline\hline
    $a_i$ & $x<0.5$ & $0.5\leq x<4.0$ & $x\geq4.0$\\
    \hline
    $a_0$ & 1.154360 & 1.162004 & 1.057572\\
    $a_1$ & 0.033772 & 0.011464 & 0.140537\\
    $a_2$ & 0.032277 & 0.049255 & N/A\\
    $a_3$ & N/A & -0.005879 & N/A\\
    \hline
    \end{tabular}
\end{table}
The resulting fit, valid in the colour range $-1.0\leq\bprp\leq7.0$, is shown in the top panel of \figref{xpfit} and was used to compute the corrected \xp flux excess \cxs for a selection of $\approx6.8$ million nearby sources which are shown in the bottom panel of \figref{xpfit}: the \cxs has a flat distribution in colour centred on zero.

\begin{figure}[!htp]
    \centering
    \colfig{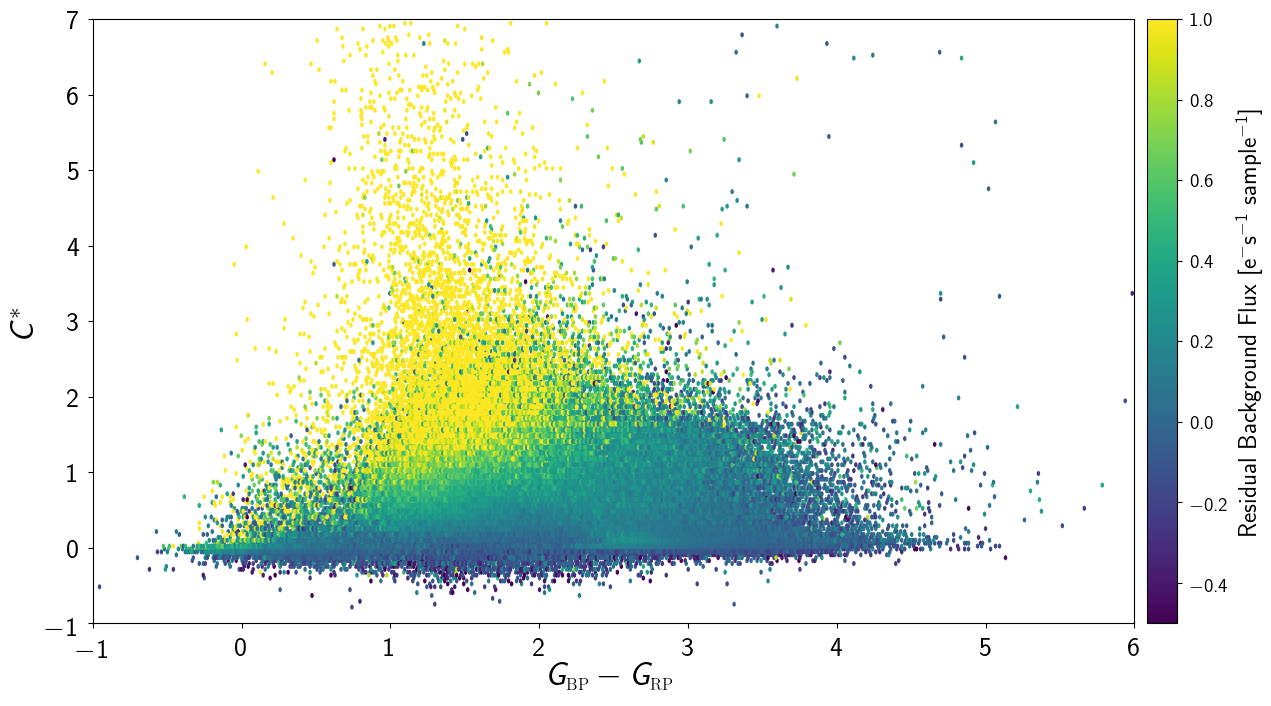}
    \colfig{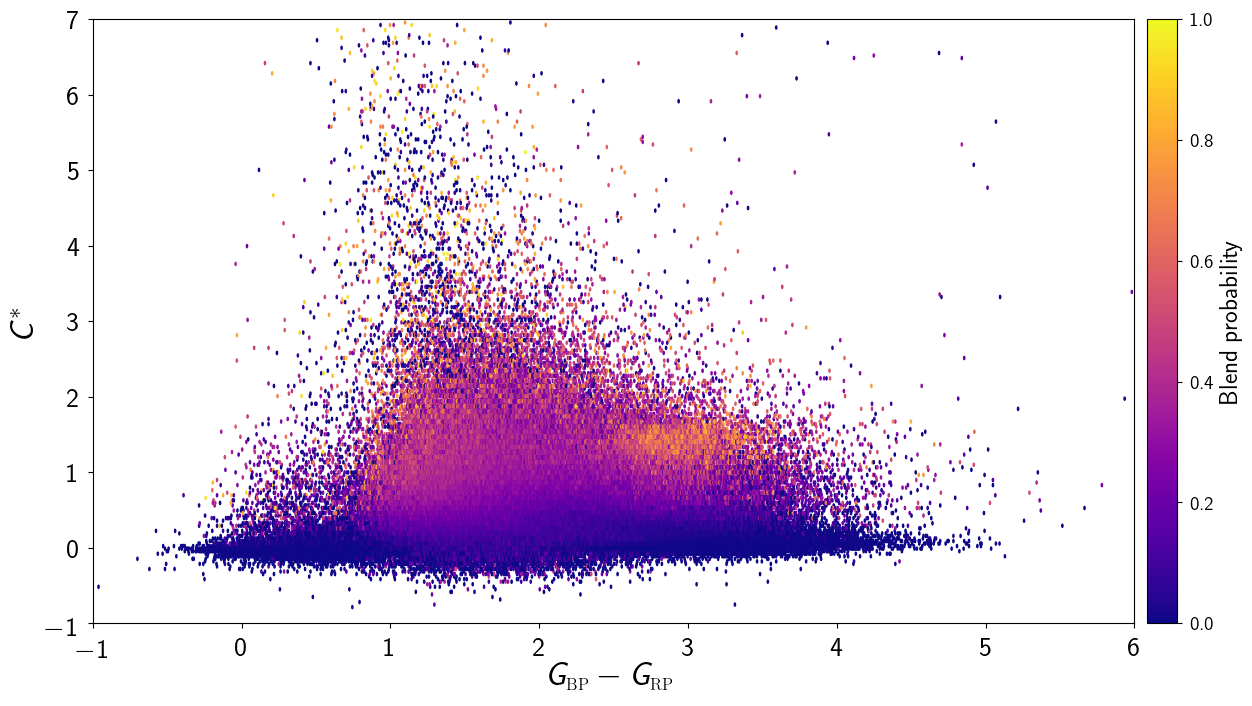}
    \caption{Effect of residual background flux and blending probability on the corrected \xp flux excess. \textit{Top panel}: Distribution of corrected \xp flux excess vs. \bprp colour for a subset of nearby sources fainter than 17 in \gband. The symbols are colour coded by the source median residual background as measured from the BP spectra. 
    \textit{Bottom panel}: Distribution of corrected \xp flux excess vs. \bprp colour for the same sources shown in the top panel but colour-coded by blend probability (see text for details).
    For these plots sources have been further selected to retain sources where the blend probability for BP and RP was in agreement to within $50\%$.}
    \label{fig:bkg_and_blend}
\end{figure}

We note here that, for DPAC-internal operational reasons, the \edr archive provides only the uncorrected \xp flux excess $C$ via the \texttt{phot\_bp\_rp\_excess\_factor} column: the user can compute the corrected \xp excess factor \cxs as defined by \equref{cxs} using the polynomials provided in \tabref{cxs_fit}.
The corrected \xp excess factor \cxs can be used to identify sources for which the \gband photometry and \xp photometry is not consistent. We will now consider a number of possible problems that might occur in the processing to try and quantify the size of their effect on \cxs. This should help understanding the possible causes for (some of) the large \cxs values seen in the \edr photometry.

\begin{figure}
    \centering
    \colfig{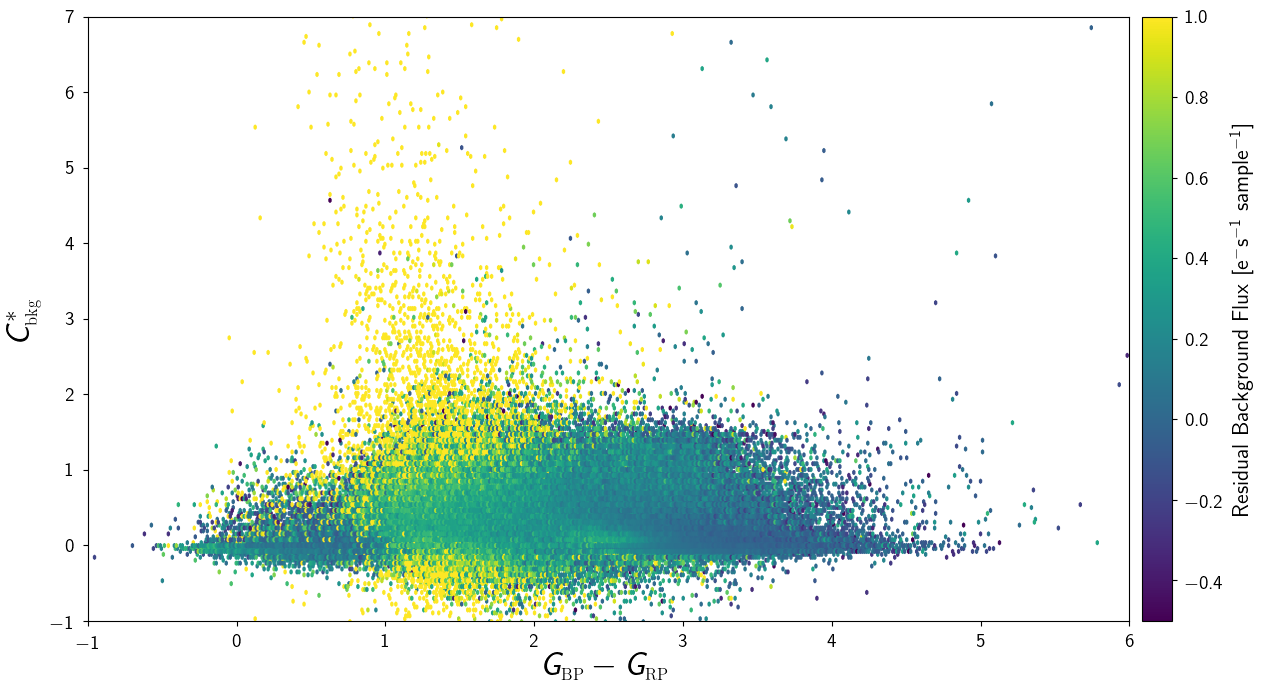}
    \caption{Corrected \xp flux excess vs. \bprp colour distribution for a selection of nearby sources with magnitude $G>17$ and more than five calibrated epoch spectra in both BP and RP. In this plot the corrected flux excess $\cxs_{\textrm{bkg}}$ has been computed after removing the median background residual flux for both BP and RP for each source . Sources are colour-coded by the median residual background as in the top panel of \figref{bkg_and_blend}. See \tabref{bkg_flux_to_mag} for an estimate of the impact of the residual background flux at different magnitudes.}
    \label{fig:bkg_corrected_excess}
\end{figure}

In \edr the background treatment for \xp has been considerably improved to deal with local variations for each individual transit (see \secref{overproc:bkg}). As we have shown, some systematic effect related to crowded regions seems to be still present in the data judging from the analysis of residual background. We have also pointed out how difficult it is to disentangle background from crowding effects when measuring the residual background using the edge samples of \xp spectra. The two panels of \figref{bkg_and_blend} show the distribution of the flux excess factor versus colour for the same selection of sources with different colour coding: in the top panel the colour of the dots indicates the median residual background in BP, while in the bottom panel the colour-coding is by an estimated `blend probability'. This latter parameter is a combination of the fraction of blended transits (as available in \edr) and an additional indicator resulting from a clustering analysis of all BP and RP epoch spectra for a given source. The number of blended transits included in the release is based on the available source catalogue. There will be cases where the blending source was not in the catalogue, this could be due to the secondary source being too close and/or too faint with respect to the primary and therefore never detected or in very crowded regions, because the priority scheme on board simply favoured brighter sources. In the blend cases, some of the epoch spectra present clearly multiple peaks showing the presence of more than one source in the window, however the position and brightness of the peaks change with the scan angle and due to the scanning law, often two groups of epoch spectra form with quite distinct features. This is what the clustering analysis is trying to detect. We have defined the metric based on the clustering analysis as the fraction of the spectrum in which the analysis clearly detected a split in the epoch spectra measurements (e.g. if only in ten out of 50 samples used for the analysis the epoch sample measurements formed two distinct groups, the metric will be equal to 0.2). In the bottom panel of \figref{bkg_and_blend} the fraction of blended transits and the newly defined metric have been multiplied to form a single blend probability.
Clearly there are large correlations between the residual background and the blend probability and sources with large flux excess tend to have large values for both parameters. From the top plot is also clear how the low flux excess values are very likely due to a slight overestimate of the background. There is however a population of sources that have high blend probability and not so large residual background. It is also important to notice that the residual background estimates obtained for these sources are often not sufficient to justify their position in the flux excess versus colour diagram. Using the residual background estimates to correct the integrated BP and RP fluxes entering the computation of the flux excess still leaves a significant fraction of sources with large flux excess. This can be seen in \figref{bkg_corrected_excess}.

\begin{figure}
    \centering
    \colfig{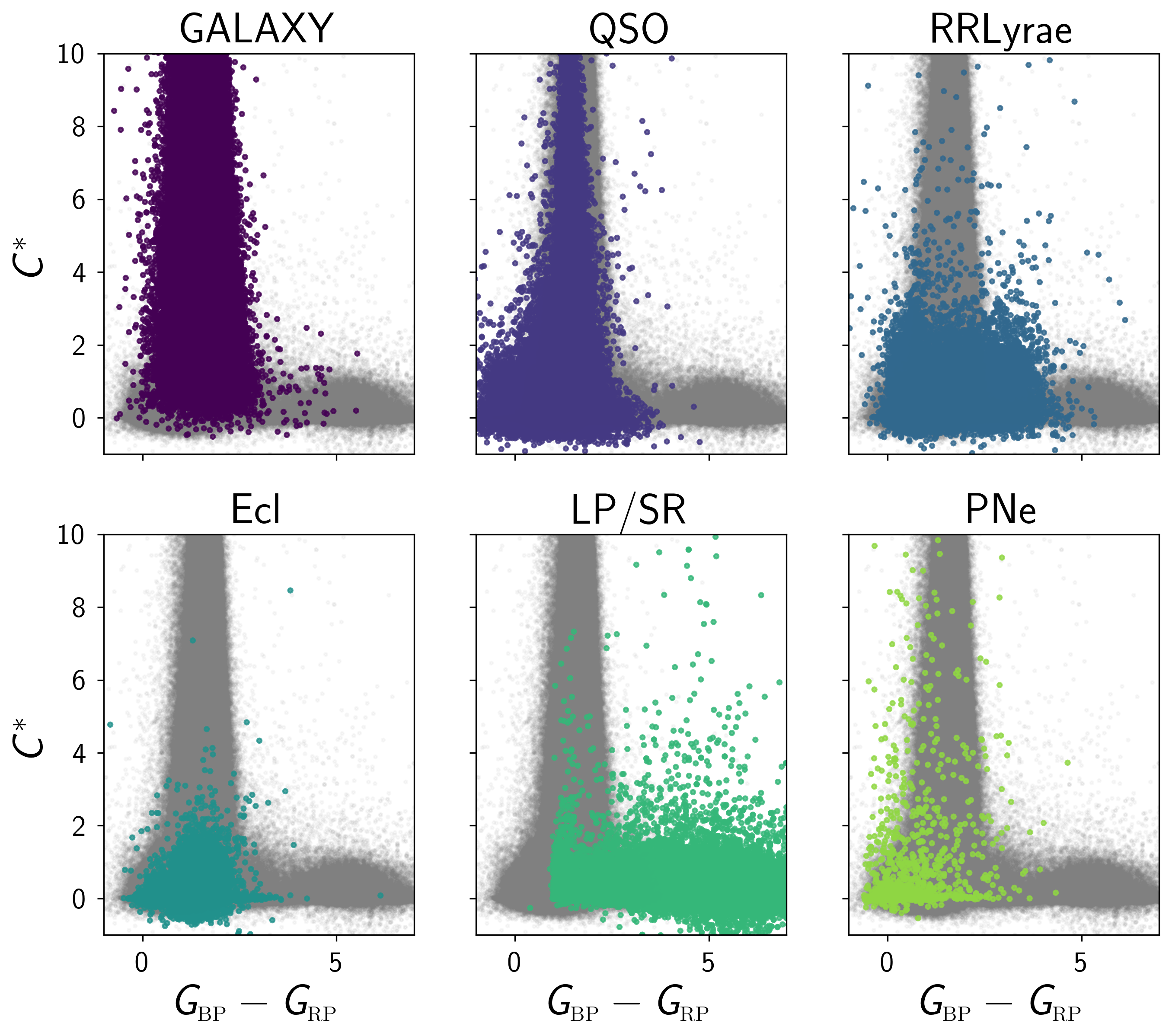}
    \caption{Corrected flux excess factor vs. \bprp colour for a selection of variable and extended sources collated from catalogues available in the literature and cross-matched with the \edr catalogue. The following types are shown: GALAXY, QSO, RRLyrae (including all subtypes), Ecl (Eclipsing variables), LP (Long Period variables) or SR (Semi-Regular variables), and PNe (Planetary Nebulae).}
    \label{fig:xsVarExt}
\end{figure}

It is therefore interesting to analyse in more detail some of the sources with very small and large corrected excess factor to assess the origin of the discrepancy between the $G$ and \xp photometry. First we considered the small number of sources in the dataset shown in the top panel of \figref{xpfit} with very low \cxs (e.g. $\cxs\leq-0.15$): analysis of the epoch spectra for these sources showed that in all cases the background had been over corrected, leading to anomalously low flux level in \xp. Looking instead at the mean spectra of the $\approx100$ sources with highest excess the situation was less clear. Sometimes there was a clear indication of variability, sometimes there was clear indication of occasional multiple sources (e.g. blends) and sometimes the spectra did not show any apparent anomaly. In all cases the background appeared to have been corrected appropriately. To explore this further, we used a catalogue of $\approx8$ million sources that was collated from the literature (and then cross-matched with the \edr catalogue) including several different types of variable stars, galaxies, quasars and planetary nebulae. \afigref{xsVarExt} shows the corrected flux excess \cxs versus \bprp colour plot for this selection colour-encoded with the source type (for a subset of those deemed to be of most interest).

\begin{figure}
    \centering
    \colfig{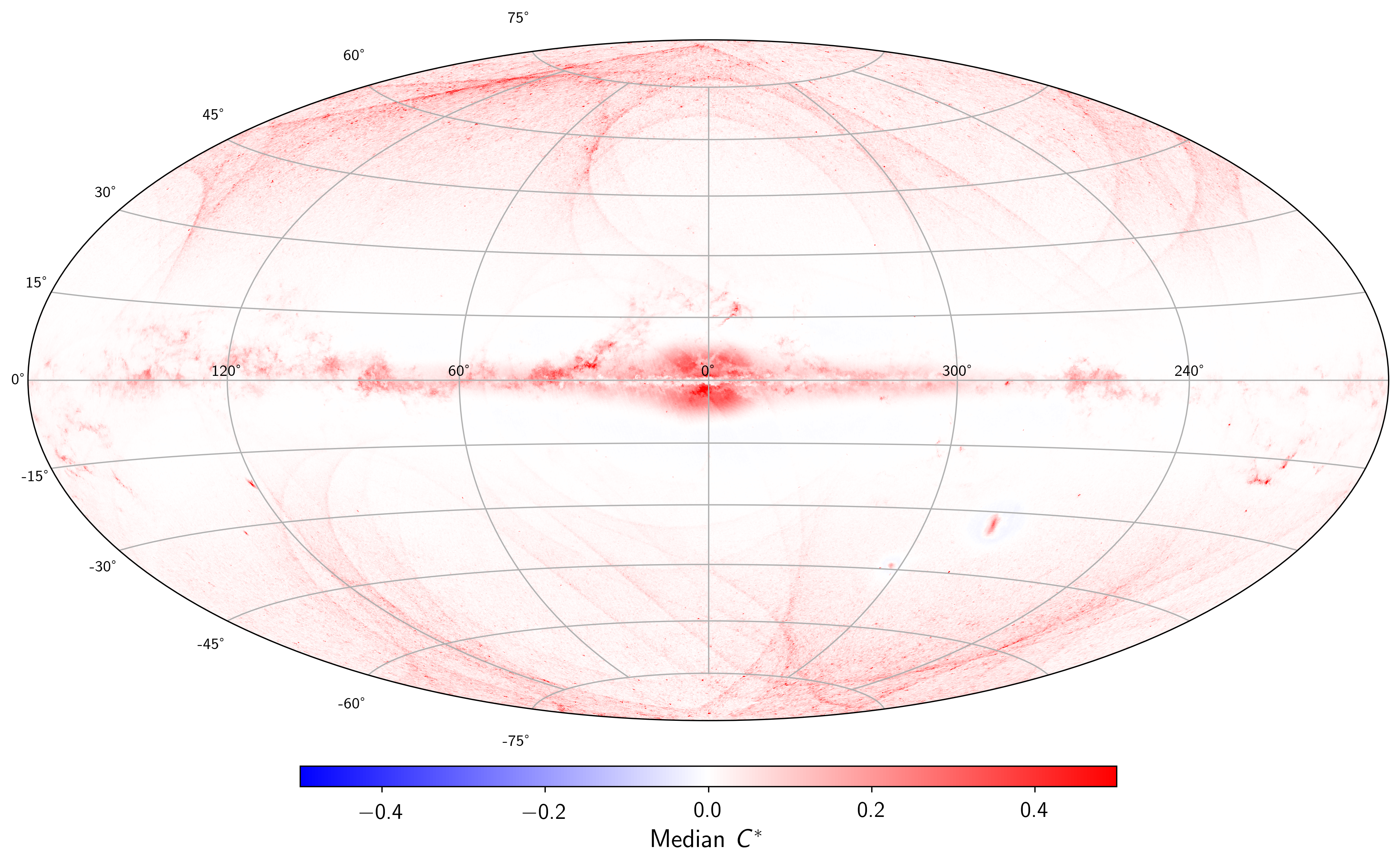}
    \caption{Sky distribution of the median corrected flux excess \cxs factor. The map was produced by computing for all gold sources the median \cxs value for each level $k=8$ HEALPix pixel.}
    \label{fig:skyCxs}
\end{figure}

One important feature revealed by this plot is that galaxies tend to have a large discrepancy between the $G$ and the \gbp, \grp fluxes. This is not surprising since the IPD and LSF/PSF modelling producing the integrated $G$ fluxes is optimised for point sources. Additionally, for extended sources the different satellite scan angle under which each epoch observation is acquired will lead to large fluctuations in the integrated $G$ flux. For \xp the window size is much larger and will therefore mitigate the effect, more so for sources with smaller angular sizes (see \appref{galaxies} for further details). Improvements in the LSF/PSF modelling and the \xp background mitigation in \edr have also led to an improvement in the \xp flux excess. In particular when comparing the sky distribution of the corrected \xp flux excess in \figref{skyCxs} with the distribution of $C$ from \drt \citep[see Fig. 18 from][]{Phot_DR2} it is clear that the ecliptic plane pattern caused by zodiacal light is no longer visible.

From \figref{xsVarExt} it is also evident that sources with anomalous SEDs, primarily those dominated by emission lines such as QSOs and PNe, also show positive values of the corrected flux excess. In the case of point-like sources such as QSOs this is primarily due to the presence of strong emission line in the wavelength range where the RP passband has a larger transmissivity with respect to the G passband. For PNe both mechanisms (the presence of an extended structure and of strong emission lines) may be at play. Similar deviations from zero of the corrected flux excess can be expected for other analogous cases such as sources in HII regions.

The analysis presented in this section is based only on sources with gold photometry. Looking at at \cxs for silver and bronze sources is only possible for the subset of these two samples with photometry available for all bands: $34\%$ and $0.0007\%$ of the whole sample, respectively. Although the distribution of \cxs versus colour is similar to that of the gold sources, some deviations can be observed. This is not surprising because of the limitations in the calibration process caused by the partial or complete lack of colour information. For these sources the most reliable photometric information is the \gband: \gbp and \grp have been provided when possible but are clearly not expected to be very reliable. Since \cxs is a quantity derived from the $G$, \gbp, and \grp fluxes it is expected that its reliability and usefulness will diminish as the quality of its components is reduced. For this reason we argue that \cxs is a metric that end users will find effective mostly for sources with gold photometry.

The recommendation for \edr is to treat the \xp flux excess $C$, or better the corrected one \cxs, purely as an indicator of consistency between the $G$ photometry and the \gbp and \grp photometry and not as a data quality indicator. In particular it is clear from the analysis presented so far that \cxs on its own it is not sufficient to discriminate between data affected by processing problems and sources that could be variable, extended or peculiar, either because of nearby sources or because of anomalous SEDs. In this sense a one-size-fits-all approach to quality filtering based on \cxs is neither possible nor desirable. Some suggestions for filtering based on \cxs are given in \secref{quality:cxs} but end users should evaluate their suitability depending on the scientific goal they are trying to achieve.

\section{External calibration}\label{sec:extcal}

The goal of external calibration is to provide for each of the filters $G$, \gbp and \grp the shape of the passbands and the corresponding zero points to allow for the transformation of internally calibrated source fluxes into meaningful magnitudes. The strategy employed to achieve this is the same one adopted for \drt: the passband is described by a parametric model whose shape is adapted to minimise the differences between observed and synthetic fluxes computed on a set of calibrators with known spectral energy distribution (SED). The mean source flux $n_p$ is given in units of ${\rm e}^- {\rm s}^{-1}$ and is related to the source photon flux distribution $n_p(\lambda)$ by the relation:
\begin{equation}
\label{ext_meanflux}
n_p = P \int\limits_0^\infty \, n_p(\lambda)\, S(\lambda) \, {\rm d}\lambda,
\end{equation}
where $P$ represents the telescope pupil area and $S(\lambda)$ is the system overall response function including the scaling factor to convert from photons to ${\rm e}^{-}$.
This function represents the system passband and is modelled as the product between a reference response function $R_*(\lambda)$ and a parametric function based on a linear combination of Legendre polynomials $P_i(\lambda_{norm})$:
\begin{equation}
\label{eq:ext_passbandModel}
S(\lambda) = R_*(\lambda) \exp\left(\sum_{i=0}^{n_R} r_i \;P_i(\lambda_{norm})\right),
\end{equation}
where $\lambda_{norm}$ is a normalised wavelength ranging in the interval $[-1, +1]$ defined as:
\begin{equation}
\label{ext_normwavelength}
\lambda_{norm} = 2 \, \frac{\lambda - \lambda_{min}}{\lambda_{max}-\lambda_{min}} - 1.
\end{equation}
The reference response $R_*(\lambda)$ for the \gband\ is equal to the nominal pre-launch response \citep{CARME}:
\begin{equation}
\label{ext_rnominal}
R_{G}(\lambda) = T_0(\lambda)\, \rho_{att}(\lambda) \,Q(\lambda),
\end{equation}
where: $T_0(\lambda)$ is the reflectivity of the telescope (mirrors); $\rho_{att}(\lambda)$ is the attenuation due to rugosity (small-scale variations in the smoothness of the surface) and molecular contamination of the mirrors; $Q(\lambda)$ is the quantum efficiency (QE) of the CCD. 
For \gbp and \grp the reference response $R_*(\lambda)$ is a cubic spline interpolation on a 1 nm fine grid lookup table derived from the \xp instrument model \citeipp{Montegriffo}. Provided that the reference response function is non negative, the exponential form of the parametric function in \equref{ext_passbandModel} guarantees the non negativity of the passband $S(\lambda)$.

\begin{figure}
\centering
\colfig{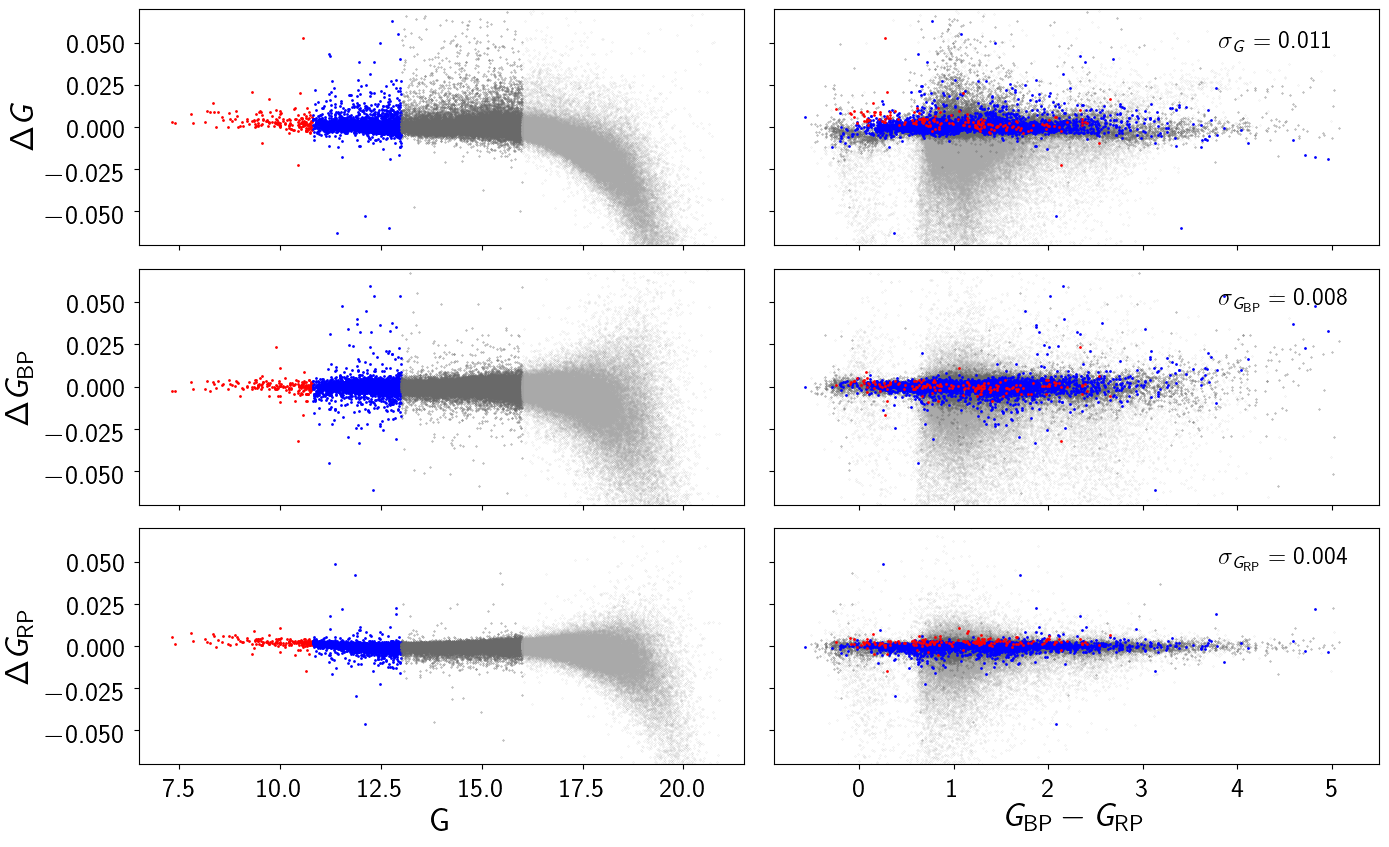}
\caption{Residuals between \edr photometry and synthetic magnitudes obtained with the final passbands and computed for a sample of $\sim100000$ SEDs obtained from externally calibrated \xp spectra for $G$ (\textit{top}), \gbp (\textit{middle}), \grp (\textit{bottom}) as function of $G$ magnitudes (\textit{left}) and \bprp (\textit{right}). The colours red, blue, dark, and light grey indicate sources with $G<10.8$, $10.8<G<13$, $13<G<16$, and $G>16$, respectively.}
\label{fig:ext_residuals}
\end{figure}

The determination of \drt passbands relied uniquely upon the set of Spectro-Photometric Standard Sources \citep[SPSS, ][see also Appendix~B]{SPSS, SPSS2, SPSS3} as calibrators; however that experience revealed the low sensitivity of the calibration strategy to the actual shape of the passbands, witnessed also by rather large number of published curves: two different sets in \cite{Phot_DR2}, others in \cite{Weiler2017} and \cite{Maizap}, all providing minimal changes in the SPSS residuals between observed and synthetic photometry. The problem is that the most reliable and stable flux calibrators, such as the SPSS, constitute necessarily a limited set of spectral distributions that can only constrain a subset of model components leaving others completely unconstrained \citep{Weiler2018}. To mitigate this limitation, for \edr we decided to employ a much larger set of auxiliary calibrators covering a wide range in stellar types: we selected a large number of sources ($N\simeq100000$) from the \cite{2000PASP..112..925S} secondary standards with $10<G<20$ (see Appendix \ref{app:stetson} for more details) and reconstructed the corresponding SEDs from externally calibrated \xp spectra. These data will be publicly available with the forthcoming \drthree release, while the complete description of the spectral instrument model calibration and methods will be provided in \citeip{Montegriffo}. The interested reader can find a summary of the approach followed for the external calibration of \xp spectra in \appref{extxpcal}. It is however worth mentioning that no external photometry from the Stetson sample has been used in the calibration process: the calibration relies entirely on \gaia data. Furthermore, the usage of a large set of \xp spectra ensures that the internal photometric system is robustly tied to the photometric system of \xp spectra: extended validation of such photometric system will be provided in \citeip{Montegriffo}. Finally, the external calibration of \xp spectra is based on the SPSS calibrators and hence the \edr flux scale is tied to the Vega flux scale to an accuracy of $1\%$. In the external calibration of the \xp instrument model, the issues related to the limited size of the SPSS dataset have been mitigated by the addition of different kind of calibrators, mostly emission line sources such as QSO or WR stars that were used to constrain specific components (LSF and absolute wavelength scale).
The comparison between \edr magnitudes and the synthetic ones computed with a preliminary set of passbands revealed two different problems affecting the internally calibrated flux scales: a colour term in the $G$ band was present for sources with $G<13$ and colour $\bprp < 1.1$; a small discontinuity in the \gbp\ residuals was visible around $G\simeq 10.8$.
Magnitude $G=13$ corresponds to the transition between window class 0 and 1 \citep{GaiaRef}, therefore the first issue has been interpreted as an interaction between a non-optimal convergence of the internal calibration between the two instrument configurations
and the known colour-dependent issues with the PSF calibration \citep{EDR3_PSF}, rather than being due to some unidentified issue affecting the calibration of \xp spectra, a hypothesis enforced also by the lack of a similar effect in the \gbp and \grp residuals.
Regarding the second effect, discontinuities in the behaviour of the \gbp photometry around this value of G magnitude were noticed also in \drt (first reports were in \citeauthor{arenou2018}, \citeyear{arenou2018}, but see also \citeauthor{Weiler2018}, \citeyear{Weiler2018}, and  \citeauthor{Maizap}, \citeyear{Maizap}). While the \edr and DR2 photometric systems should be considered independent due to the substantial differences in the processing that lead from the raw data to the final photometric catalogue and to the addition of a large amount of new data, the fact that a similar feature is still present indicates that this might be connected to a change in the instrument configuration, for example a gate activation, occurring at approximately this magnitude and that has not yet been fully calibrated out.
To minimise these effects in the final photometry a correction has been applied to the (epoch and mean) \gflux and \bpflux fluxes available from the \edr archive.
The \gflux fluxes correction has been derived as a polynomial function of the \xp flux ratio:
\begin{equation}
\label{gfluxcorr}
\gflux^* = \gflux \sum_{i=0}^{3} c_i \left(\frac{\bpflux}{\rpflux}\right)^i,
\end{equation}
with  $c = (0.9938297, ~0.0118275, -0.0019720, ~2.253619\,10^{-4})$. This correction has been applied only to sources with $G < 13$ and $-0.5 < \bprp < 1.1$: the correction was clamped below $\bprp = -0.5$ to avoid extrapolations below that limit.
Similarly, the \gbp discontinuity has been removed by correcting the corresponding fluxes:
\begin{equation}
\label{ }
\bpflux^* = \bpflux \times 10^{-0.4\; \delta_{\rm BP}},
\end{equation}
with  $\delta_{\rm BP} =  0.003763096$ for $G < 10.8$.
To avoid the creation of artefacts in the data, such as visible gaps in the colour-magnitude diagrams around the limiting magnitudes, these two corrections have been applied gradually (a linear onset $\pm10\%$ in flux around $G = 13$ and $G = 10.8$).
The final passbands were then computed using only sources in the range $13<G<16$,  $11<G<16.5$, and $G<16.5$ for $G$, \gbp, and \grp, respectively; in all cases three Legendre polynomials have been used in \equref{ext_passbandModel} to model the passbands. \afigref{ext_residuals} shows the final residuals between \edr photometry and synthetic magnitudes derived from the externally calibrated \xp spectra for the whole set of calibrators. Residuals do not show significant trends with colour and the rms ranges from $\sim0.01$ mag for $G$ to less than 5 mmag for the \grp case.
The downturn at faint magnitudes visible in all three passbands for sources fainter than $G \simeq 16.5$ is possibly caused by some bias in the background either in the integrated photometry or in the \xp spectra. Since it is not possible to confidently attribute the origin of this effect solely to the integrated photometry in each of these plots, we do not consider it appropriate to either attempt or suggest any correction for fluxes at the faint end.
The residuals in the top left plot show that the significant magnitude term that was affecting \drt photometry \citep{CasagrandeVandenBerg, Weiler2017} is not present anymore although a small trend below a few mmag/mag cannot be excluded (an exception to this general behaviour is represented by extremely blue and bright sources as discussed in \secref{issues_bright}). Moreover the \gband residuals as a function of colour (top right plot) show that the effects of the two photometric systems have been mitigated by the flux scale correction applied to the blue and bright end, leaving some residual differential colour terms below the $1\%$ level for $\bprp < 0.5$. Residuals of \grp as function of magnitude (bottom left panel) seems to indicate a small trend for magnitudes $G<13$; a linear fit reveals a trend below 0.8 mmag/mag for which no correction has been attempted.
We are aware that such inhomogeneities in the photometry may
lead to claims for the presence of multiple passbands (as was the case in \drt for BP): however, given that the overall accuracy of the absolute calibration is around $1\%$ we consider it unnecessary introducing additional complexities to account for effects that are below this level and that only concerns a small subset of the whole catalogue.

\begin{figure}
\centering
\colfig{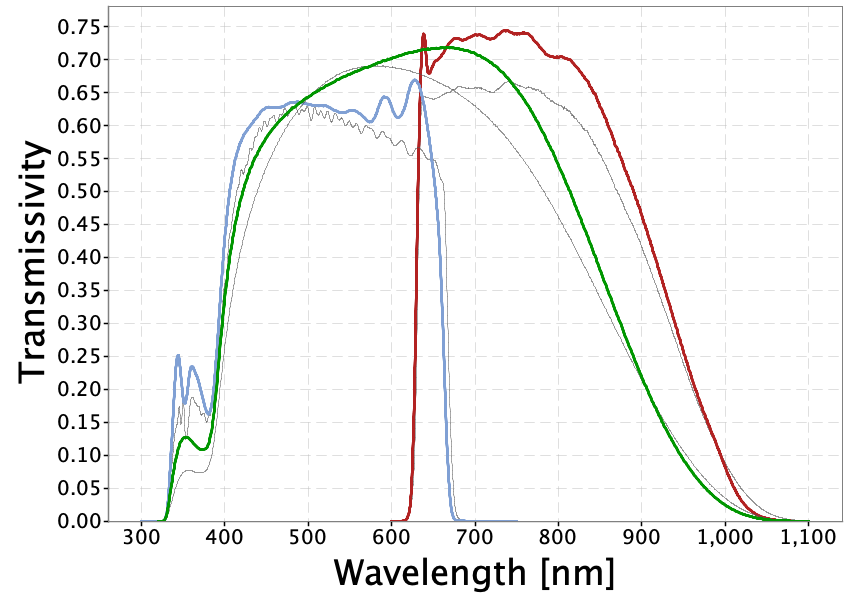}
\caption{$G$ (green), \gbp (blue) and \grp (red) passbands for the \edr photometric system; grey curves represent nominal pre-launch passbands.}
\label{fig:ext_passbands}
\end{figure}

The passbands are shown in \figref{ext_passbands} together with nominal pre-launch curves (represented in grey colour) for comparison and are distributed in electronic tabular format as part of this paper and are available via the VizieR service.
We emphasise here that each new release of \gaia data is based on a complete reprocessing of raw data (see also the considerations made at the end of \secref{data} and in \secref{dontmix}), hence the photometric system defined by the internal calibration has no relations with photometric systems of previous releases. This photometric system indeed refers to an abstract `average instrument' whose properties descend not only from the real instrument but also from all the calibration models that are layered in the long chain of the data reduction pipelines. 
For this reason it is meaningless to compare the \edr passbands with the different sets released for \drt and changes in the instrument passband between different/subsequent releases are not be interpreted as some sort of evolution of the instrument.

Once the passbands have been defined, the corresponding zero points can be evaluated in the VEGAMAG and in the AB systems following a standard procedure composed of three main steps. First the synthetic fluxes are computed for each calibrator by evaluating the mean energy per wavelength unit
\begin{equation}
\label{vegamagflux}
<f_\lambda>\, = \frac{\int\, f_\lambda(\lambda) \,S(\lambda)\,\lambda \, {\rm d}\lambda }{\int\, S(\lambda)\,\lambda \, {\rm d}\lambda },
\end{equation}
for VEGAMAG, and the mean energy flux per frequency unit
\begin{equation}
\label{abflux}
<f_\nu>\, = \frac{\int\, f_\lambda(\lambda) \,S(\lambda)\,\lambda \, {\rm d}\lambda }{\int\, S(\lambda)\,\left(c/\lambda\right) \, {\rm d}\lambda },
\end{equation}
for AB.
Then, the synthetic fluxes are converted to magnitudes by applying the relative zero point. For VEGAMAG:
\begin{equation}
\label{eq:vegamagmag}
m_{VEG} = -2.5 \log <f_\lambda> + 2.5 \log 
 \frac{\int f_\lambda^{Vg}(\lambda) S(\lambda)\lambda \, {\rm d}\lambda } {\int S(\lambda)\lambda \, {\rm d}\lambda },
\end{equation}
where  $f_{\lambda}^{Vg}(\lambda)$ is the Vega spectrum from the CALSPEC Calibration Database\footnote{Provided by the \texttt{alpha\_lyr\_mod\_002.fits} file.} rescaled to set the flux equal to $f_{550} = 3.62286\,10^{-11}$ W m$^{-2}$ nm$^{-1}$ at the wavelength $\lambda=550.0$ nm, which is assumed as the flux of an unreddened A0V star with $V=0$ (see the online documentation for a detailed description of the reference $f_{550}$ flux definition). Instead, for AB:
\begin{equation}
\label{abmag}
m_{AB} = -2.5 \, \log <f_\nu> - 56.10,
\end{equation}
where the value of the zero point corresponds to fluxes measured in units of W m$^{-2}$ Hz$^{-1}$.
Finally, the passband zero point is computed as the mean value of the ratios between synthetic and uncalibrated magnitudes for the whole set of calibrators:
\begin{equation}
\label{zp}
ZP_{X} = \left< \frac{m_X}{ -2.5 \, \log (n_p)} \right>,
\end{equation}
where $X$ stands for either VEGAMAG or AB.

The values of the zero points for both systems are reported in \tabref{ext_zptab} along with some useful passband related quantities such as the filter full width at half maximum (FWHM), the mean photon wavelength $\lambda_0$ and the pivot wavelength $\lambda_p$ (a useful parameter introduced by \citet{koornneef} that allows for an exact conversion between the broadband fluxes $<f_{\lambda}>$ and $<f_{\nu}>$).
\begin{table*}[]
\caption{Photometric zero points in the VEGAMAG  and AB systems, the FWHM, the mean photon wavelength $\lambda_{0}$, the pivot wavelength $\lambda_{p}$ for $G$, \gbp and \grp.}
\label{tab:ext_zptab}
\begin{center}
\begin{tabular}{l|c|c|c|c}
\hline\hline 
~ & $G$ & \gbp & \grp & Units\\
\hline
 $ZP_{VEG}$ & 25.6874 $\pm$ 0.0028 & 25.3385 $\pm$ 0.0028 & 24.7479 $\pm$ 0.0028 & mag\\
 $ZP_{AB}$  & 25.8010 $\pm$ 0.0028 & 25.3540 $\pm$ 0.0023 & 25.1040 $\pm$ 0.0016 & mag\\
 FWHM & 454.82 & 265.90 & 292.75 & nm\\
 $\lambda_{0}$ & 639.07 & 518.26 & 782.51 & nm\\
 $\lambda_{p}$ & 621.79 & 510.97 & 776.91 & nm\\
 \hline
\end{tabular} 
\end{center}
\end{table*}
It is important to note that these passband zero points are not suitable for synthetic magnitude evaluations for which the correct value must be computed according to \equref{vegamagmag}: \gaia fluxes are in units of e$^-$ s$^{-1}$ (not normalised by the telescope area) while synthetic fluxes refer to mean energy densities in units of W m$^{-2}$ nm$^{-1}$.

\begin{figure}
\centering
\colfig{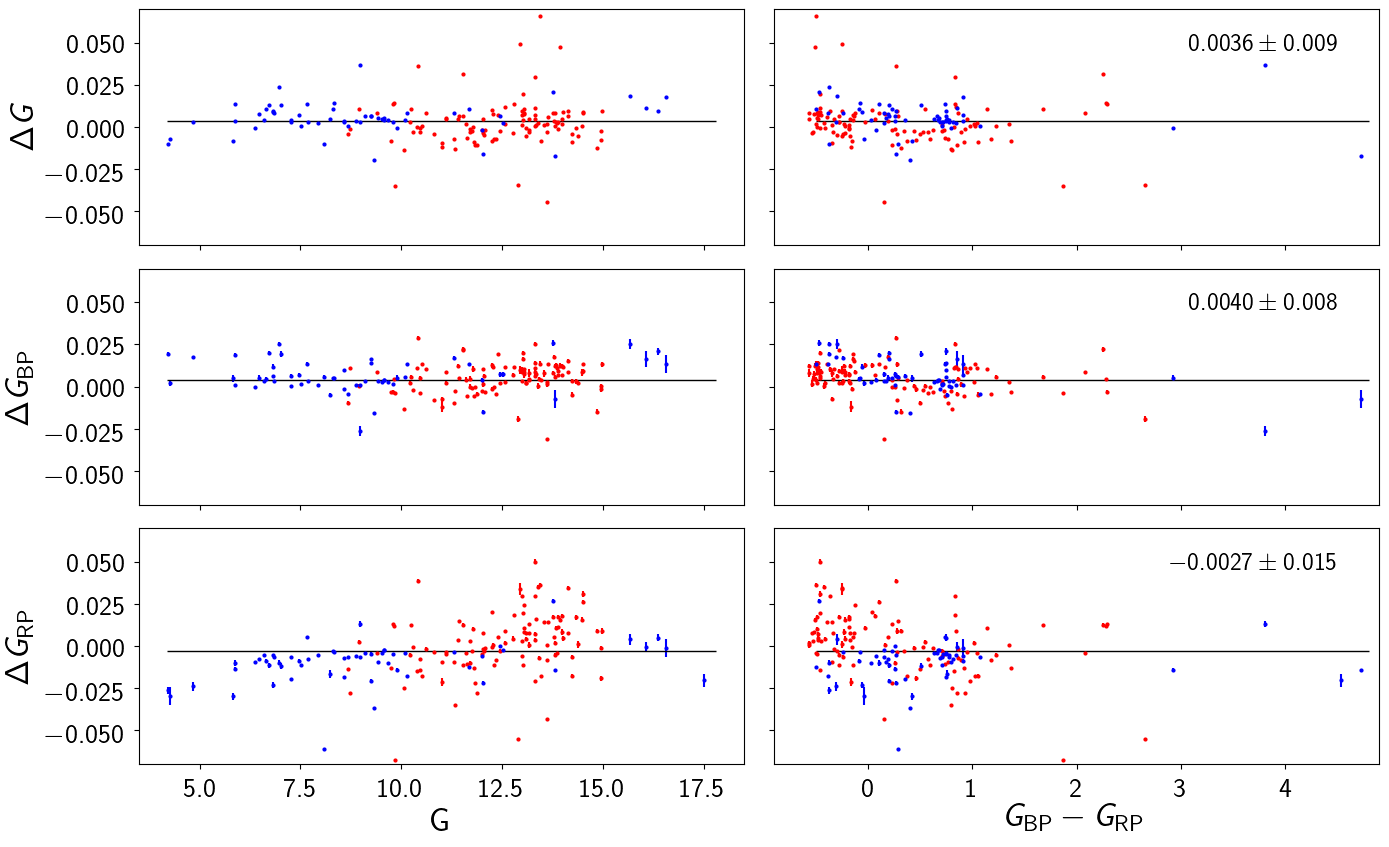}
\caption{Residuals between \edr magnitudes and synthetic ones computed on SEDs from ground and space based observations of SPSS (red symbols) and PVL (blue symbols) sources for $G$ (\textit{top}), \gbp (\textit{middle}), and \grp (\textit{bottom}) as function of $G$ magnitudes (\textit{left}) and \bprp (\textit{right})}
\label{fig:ext_validation}
\end{figure}
Finally, \figref{ext_validation} shows the residuals between the \edr magnitudes of SPSS (red dots) and PVL (blue dots) sources (see Appendix~B for more details) and the synthetic magnitudes computed from the corresponding SEDs obtained from independent ground or space based observations. Residuals are plotted for $G$, \gbp and \grp as function of $G$ and \bprp colour. The horizontal grey lines represent the weighted mean of residuals that in the vast majority of cases amounts to a few mmag. As explained in \secref{sourcephot} at the bright end residual saturation effects dominate the uncertainties.

\section{Known issues with the published photometry}

\subsection{Overestimated mean \gbp flux for faint red sources}\label{sec:fluxlimit}

When computing the weighted mean flux for a source in a given band, epochs with a calibrated flux lower than 1 ${\rm e}^-{\rm s}^{-1}$ were excluded. This lower limit was introduced for \drt \citep{Proc_DR2} to prevent problems caused by extreme outliers in the \gband, however the threshold was applied also to the generation of \gbp and \grp. For the \gband the lower flux limit should not cause any bias because it corresponds to $G\approx25.8$, which is well below the on-board limit used by the VPU to consider a source eligible for observation \citep{VpuDetection}. The on-board detection limit has resulted in epoch observations having $G_{\rm VPU}<20.7$ and even allowing for a generous error on the on-board estimated (and uncalibrated) magnitude $G_{\rm VPU}$, fluxes much lower than the threshold cannot be observed as part of a normal distribution (therefore they can only occur due to problems in the processing). However, this minimum flux threshold can cause an overestimated mean BP flux for faint sources, which tend to have a red colour, and therefore have a much lower flux in \gbp than \grp.

\begin{figure}
    \centering
    \colfig{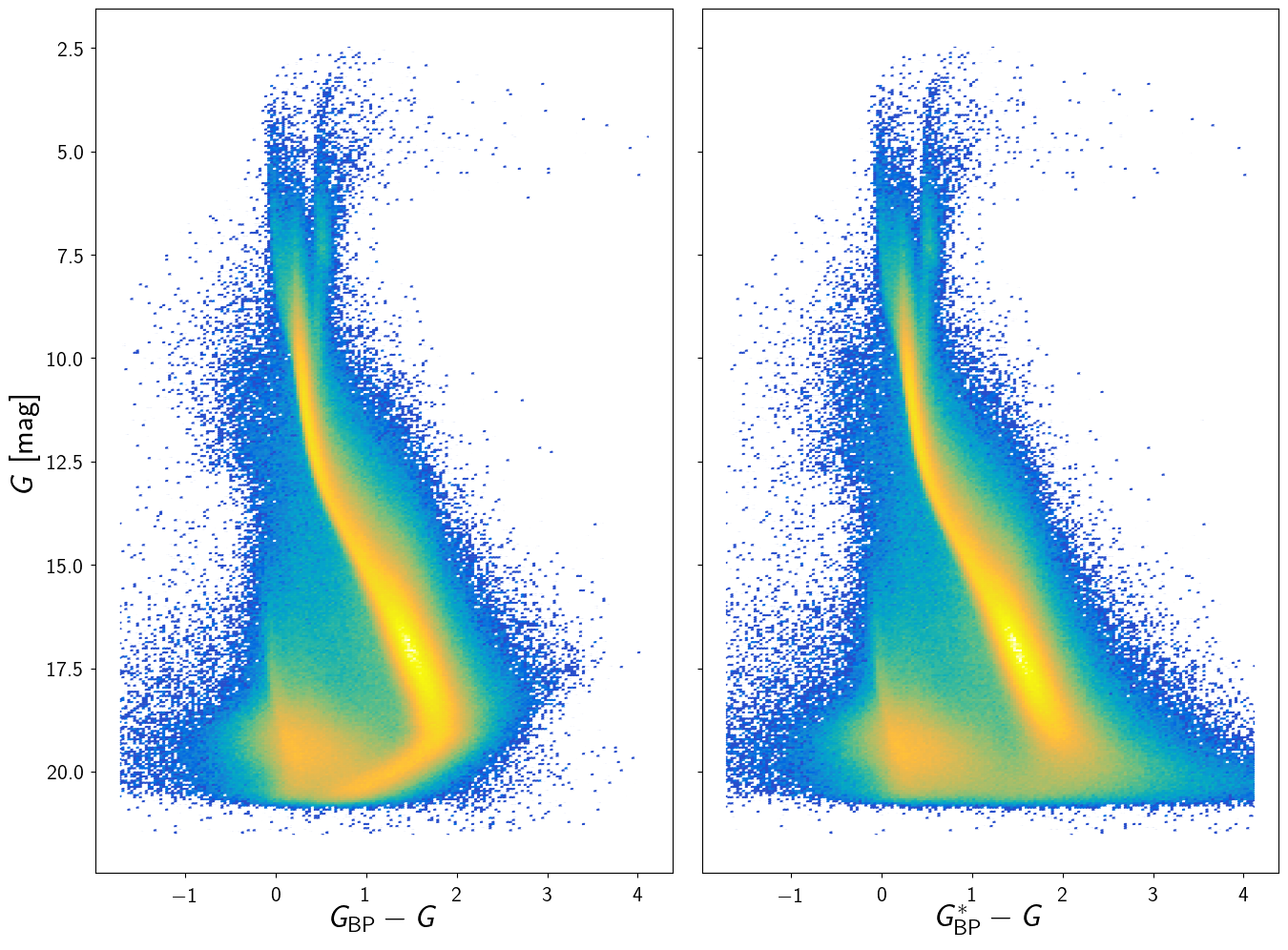}
    \caption{Colour--magnitude diagram for a sample of $\approx3.4$ million nearby sources selected from the \edr archive. The left panel shows the CMD produced using the $G$ and \gbp magnitudes from the \edr archive, which presents a tail like feature  bending progressively towards bluer colours for fainter $G$ magnitudes. The right panel shows the CMD for the same sources but with $\gbp^{\ast}$ recomputed without the low flux threshold.}
    \label{fig:c5ppv79cmd}
\end{figure}

To exemplify the issue, we selected from the \edr archive all gold nearby sources with an error on the parallax smaller than 1 mas providing a sample of $\approx3.4$ million sources. The left panel of \figref{c5ppv79cmd} shows the $G$ vs $\gbp-G$ colour--magnitude diagram (CMD) for these sources. A striking feature of this CMD is the tail at the faint end of the main sequence bending towards bluer colours, which is clearly unexpected. This feature is produced by the fact that the 1 ${\rm e}^-{\rm s}^{-1}$ minimum flux threshold adopted for the \gband was also applied to \gbp: when progressively fainter red sources are observed, the distribution of their epoch photometry will be progressively more clipped at its faint end leading to an overestimated mean flux. To confirm the nature of this feature we performed a simple experiment in which we regenerated the mean \gbp source photometry removing the minimum flux threshold. The result of the experiment is shown in the right panel of \figref{c5ppv79cmd}, which presents the CMD for the same set of sources but using the new $\gbp^{\ast}$ computed without the low flux threshold: the tail feature is no longer visible and the sources previously located there have been redistributed towards redder colours.

\begin{figure}
    \centering
    \colfig{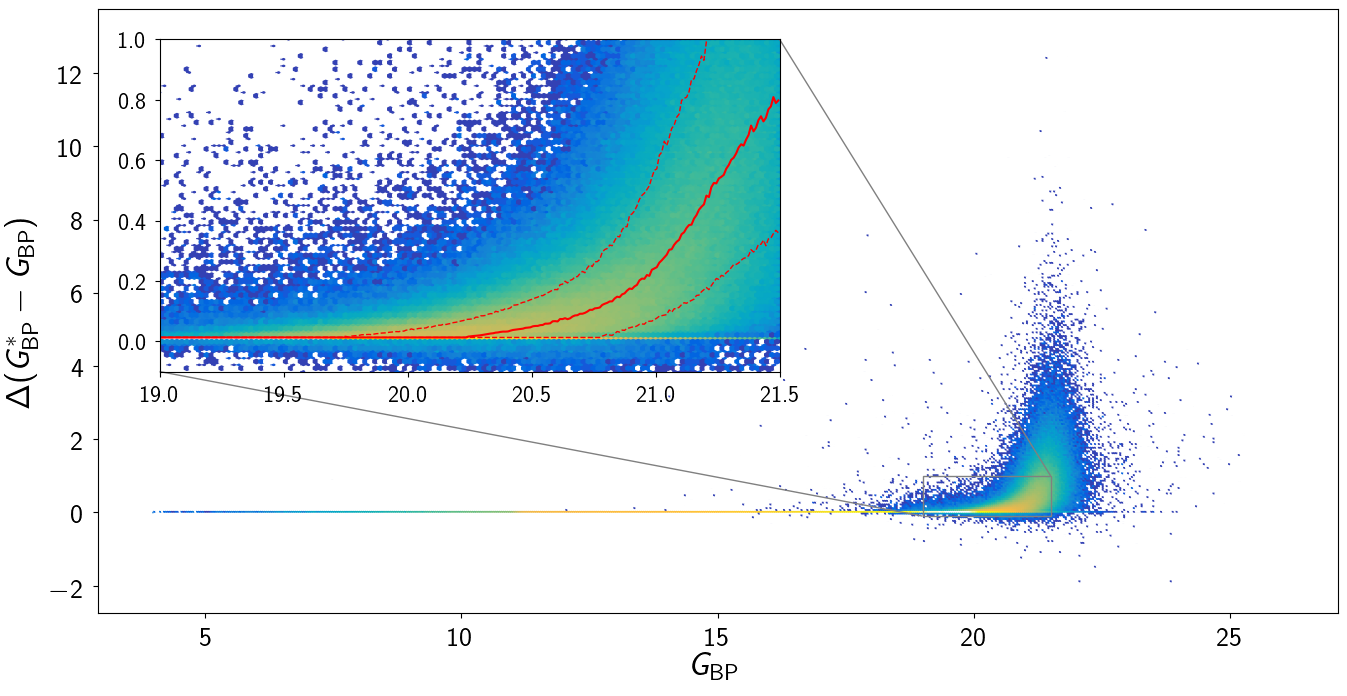}
    \colfig{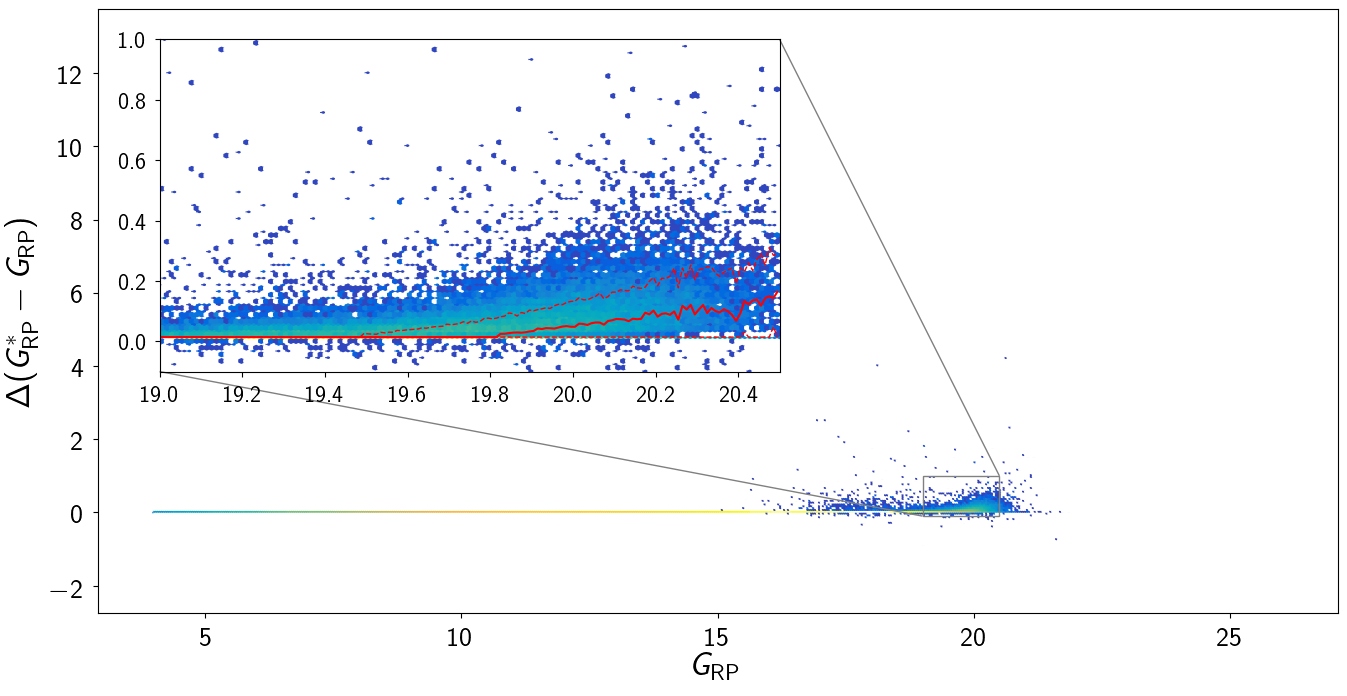}
    \caption{Change in the \gbp  (\textit{top panel}) and \grp (\textit{bottom panel}) magnitude when removing the 1 $e^-s^{-1}$ flux threshold vs. \gbp and \grp magnitude spectively for the sample of $\approx3.4$ million nearby sources.  The insets show a zoom of the transition region where the discrepancy between the two magnitudes becomes significant. The solid red lines show the median of the residual distribution, and the two red dashed lines show the 84th and 16th percentiles.}
    \label{fig:deltaxp}
\end{figure}

The top panel of \figref{deltaxp} shows the difference between the recomputed $\gbp^{\ast}$ magnitude and the \gbp from the \edr archive versus \gbp. The plot shows that the two values are in good agreement until $\gbp\approx20.3$ at which point the discrepancy between the two magnitudes grows progressively larger reaching a size of several magnitudes. The inset panel focuses on the transition region and includes the 16th (dashed), 50th (solid), and 86th (dashed) percentile lines, which confirms that the effect of the low flux threshold is modest for $\gbp<20.3$. 

The low flux threshold has also the effect of reducing the measured scatter in the \gbp mean source photometry. To estimate the size of this effect we selected sources with at least 50 epoch observations in $\gbp^{\ast}$ in a set of 5 mmag slices at $\gbp^{\ast}$=20, 21, 22, 23, 24 and computed the median and MAD of the calibrated epoch fluxes for all the sources in each slice using all epochs or just those with flux larger than 1 ${\rm e}^-{\rm s}^{-1}$. The analysis is done in flux-space because the error distributions in magnitude-space are not symmetric with the discrepancy becoming progressively larger towards fainter fluxes (see \appref{MagFlux}). The scatter measured at these faint magnitudes will have significant contributions from non-poissonian sources of uncertainties, such as background, contamination and blends.
The results are summarised in \tabref{deltabp}: the increase in scatter when using all transits is rather modest for sources close to $\gbp^{\ast}\approx20$ and only increases to $30-40\%$ at fainter $\gbp^{\ast}$ magnitudes. Although the increase in scatter for the $\gbp^{\ast}$=24 slice is smaller than for the $\gbp^{\ast}$=23 slice, it should be noted that these two fainter slices have ten times fewer epochs available than the brighter slices and therefore the discrepancy is probably due to small number statistics. Finally, it should be noted that at $\gbp^{\ast}\approx22$ the scatter is already of the same order of magnitude as the mean flux and since \figref{deltaxp} suggests that the corrected $\gbp^{\ast}$ magnitude could reach values even fainter than 25, it is unlikely that the photometry of these sources would be of much scientific value.

In principle the same problem described for \gbp could also affect \grp. The main reasons are that the same 1 ${\rm e}^-{\rm s}^{-1}$ flux threshold was also applied when generating the \grp source photometry and that the RP spectra undergo the same process as the BP ones and therefore have similar error budgets. We therefore performed the same analysis described above for \grp as well. The bottom panel of \figref{deltaxp} shows the difference between the recomputed $\grp^{\ast}$ magnitude and the \grp and the \grp from the \edr archive versus \grp. As it can be seen, the effect in \grp is considerably smaller than in \gbp; in particular at $\grp=20$  it is $\approx0.05$ magnitudes. Considering the typical uncertainty in the \grp photometry at this magnitude (see \figref{photerrors}) an additional filter on \grp does not seem to be required.

Finally, we used the same dataset to asses the impact of this issue on the \xp flux excess. Although a systematic effect is present, it is also small and generally well within the typical uncertainties on the \xp flux excess (see \secref{quality:cxs}). This is understandable since faint red sources will have the vast majority of their flux in the RP band and therefore even a large change in BP flux will not significantly alter the value of the $C$ ratio. The significance of the systematic differences in $C$ and \cxs can be further reduced by applying the filtering suggested in \secref{filt_gbp}. See also the online \edr documentation for more details.

\begin{table}[htp]
    \caption{Statistics for the five magnitude slices used to characterise the scatter. The first column provides the $\gbp^{\ast}$ magnitude of the 5 mmag slice; the second column is the percentage of epochs that are used when the 1 $e^-s^{-1}$ flux threshold is applied; columns three and four provide the median $\tilde{f}$ and MAD flux when all epochs are used; columns five and six provide the median $\tilde{f}$ and MAD flux when the low flux threshold is applied; column seven provide the percentage increase in scatter when the the low flux threshold is not used.}
    \label{tab:deltabp}
    \centering
    \begin{tabular}{lc|cc|cc|c}
    \hline\hline
    & & \multicolumn{2}{c}{All} & \multicolumn{2}{c}{Threshold} & \\
    $\gbp^{\ast}$ & $N_t/N$ & $\tilde{f}$ & MAD & $\tilde{f}$ & MAD & $\Delta\sigma$ \\
    \hline
    20 & 95.5\% & 139.90 & 32.06 & 140.74 & 31.49 & 1.8\%\\
    21 & 89.1\% &  56.28 & 29.52 &  62.36 & 26.15 & 12.9\%\\
    22 & 70.9\% &  23.34 & 29.51 &  40.02 & 22.86 & 29.1\%\\
    23 & 58.2\% &  10.15 & 29.60 &  34.70 & 21.04 & 40.7\%\\
    24 & 55.5\% &   5.60 & 27.27 &  31.25 & 20.55 & 32.7\%\\
    \hline
    \end{tabular}
\end{table}

\subsection{Sources with poor SSCs}

While validating the photometry for \edr it was realised that the $G$ magnitude distribution has a small tail of very faint sources extending as faint as $G\approx25.5$. Every \gaia transit has an associated uncalibrated magnitude estimated on-board by the VPU. By design \citep{VpuDetection} the VPU detection magnitude does not reach values fainter than $\approx21$ mag since the detection algorithm will not assign a window to fainter detections. Even allowing for a generous error in the on-board estimated magnitude, it is clearly not possible for \gaia to have observed sources much fainter than $G\approx21$. The cause of these unrealistically faint sources was found to be due to unreliable reference SSCs estimated from the mean spectra used in the calibration process. It is indeed possible that if the SSCs fluxes forming the ratios used in the LS calibration model (see \secref{calmodel}) have extreme values the resulting calibration factor could have a value considerably smaller than 1 leading to a much fainter calibrated flux. It should be noted that the calibrations do not have any problem, the issue is caused by unreliable colour information (i.e. the SSCs) being used when applying the calibration to the epoch photometry of some sources.

The unreliable SSCs values were caused mostly by two different issues: 1) sources with mean spectra significantly affected by blending with another source leading to significantly higher flux levels in the boundary SSCs; 2) red sources with extremely low flux in \gbp leading to very low signal-to-noise mean spectrum and hence very unreliable BP SSC values. Sources affected by this issue were identified from having extreme values of the SSC ratios used when applying the LS calibration and their photometry has been removed from the main \edr catalogue \citep[see][for more information]{EDR3_CU9}. 
A separate table with ad hoc photometry produced by calibrating the sources as bronze (i.e. using default SSC values) is available from the \edr known issues web page.

\subsection{Systematics due to use of default colour in the IPD}\label{sec:defaultNuEff}

In determining the \gband fluxes, an appropriate PSF or LSF must be chosen in order to carry out the IPD \citep{EDR3_PSF}. One of the parameters used to select the LSF/PSF is the colour of the source and this is done using the \nueff value determined from the mean \xp spectrum. In some cases, this value is not available, so a default one is used: this will lead to a systematic effect in both astrometry and \gband photometry. In the former case, the handling of the chromaticity effects in the astrometric solution automatically dealt also with this systematic \citep{DPACP-128}. Unfortunately the importance of this effect was not recognised early enough to be included in the photometric calibration model and therefore the only option for \edr is to derive a correction to the internally calibrated mean source photometry. We note that the photometric calibrations (LS and SS) were derived only with data for which the IPD used the appropriate \nueff and therefore are unaffected by this issue.

The correction has been determined using a short period of data for which the IPD was generated twice using the appropriate \nueff values and the default one. The period was chosen such that the scan direction would cross the Small Magellanic Cloud so that a significant number of blue stars were available for the calibration. The analysis of this dataset showed that the systematic generated by the use of a default \nueff is a function of \nueff, AC position, CCD, FoV and magnitude. In principle the correction should be applied to the epoch photometry before generating the mean source photometry, however this approach is impractical since the epoch photometry is not available in \edr. Since there are many observations contributing to the mean \gband photometry in the dataset used, an average correction can be calculated that is only dependent on source properties available in the \edr archive: colour and magnitude. 

\begin{figure}
    \centering
    \colfig{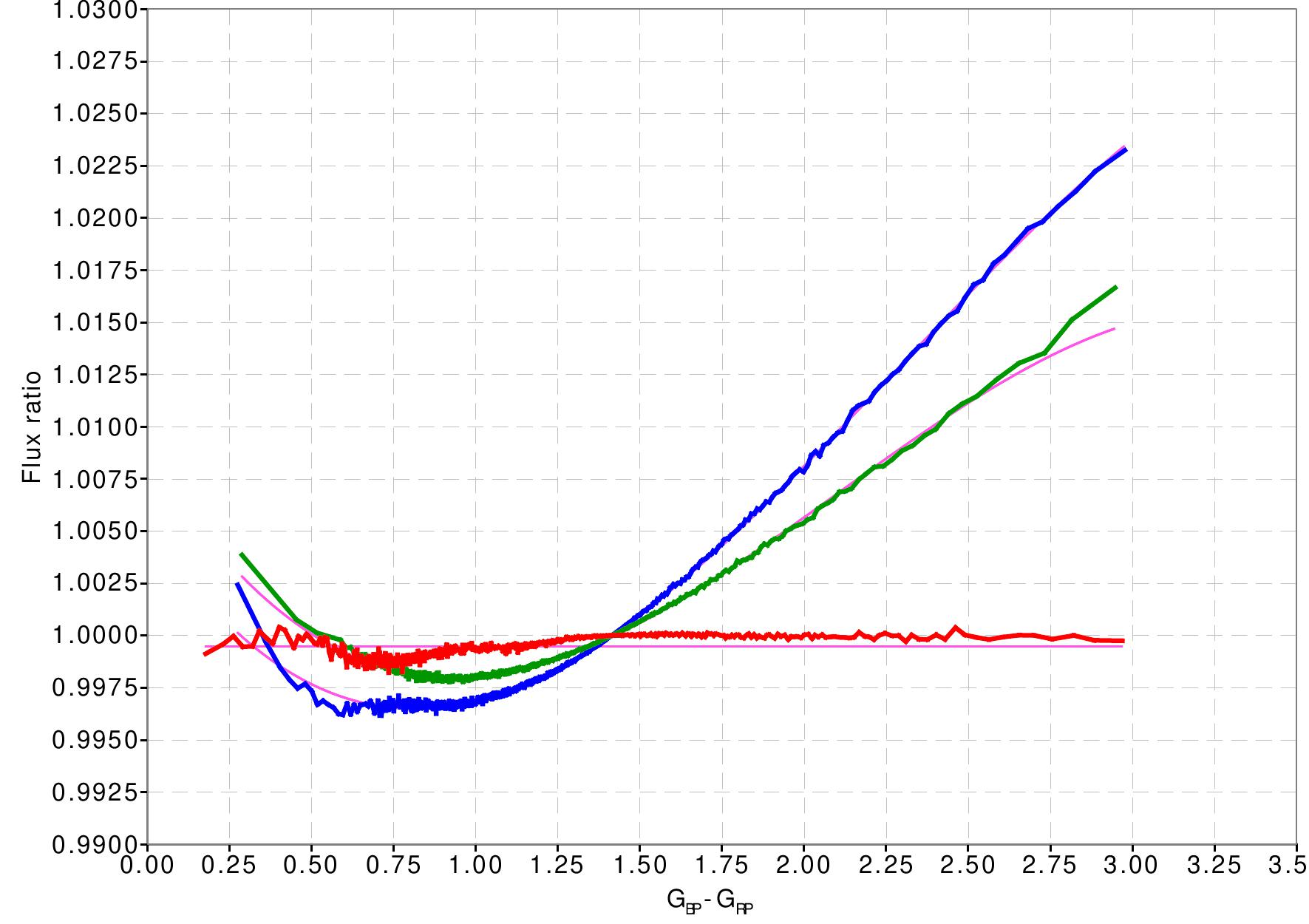}
    \caption{Systematic in $G$ flux caused by using a default \nueff in the IPD process as a function of colour. The analysis is divided into the magnitude ranges $G<13$ (red), $13<G<16$ (green), and $G>16$ (blue). The magenta lines are cubic polynomial fits to these lines.}
    \label{fig:bowtie}
\end{figure}

\begin{figure}
    \centering
    \colfig{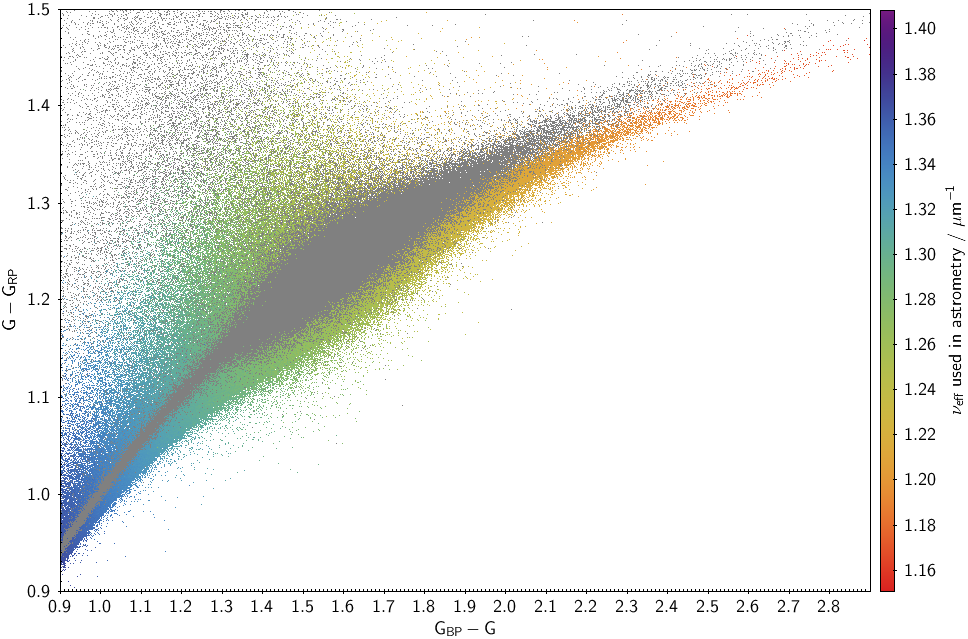}
    \colfig{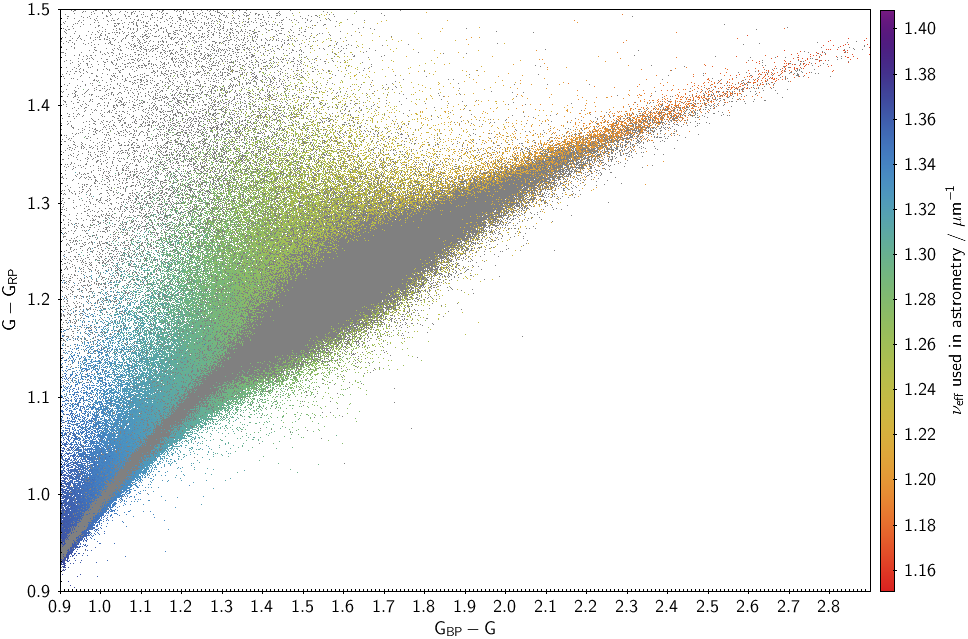}
    \caption{Effect of correcting the photometry of sources for which the IPD used a default colour. \textit{Top panel}: This shows a colour-colour diagram of the reddest main sequence stars in the local neighbourhood with parallax greater than 3 mas. The points are colour coded using the \nueff value used in the processing. If the default \nueff was used, then the points are coloured grey.
    \textit{Bottom panel}: The same sources but with corrected \gband\ photometry.}
    \label{fig:nueffvalidation}
\end{figure}

Investigation as a function of magnitude showed that the systematic was mainly dependent on the window class configuration, thus the analysis was divided into the magnitude ranges $G<13$, $13<G<16$, and $G>16$. There is a further slight magnitude dependence at the faintest magnitudes, but this is not taken into account. \afigref{bowtie} shows that there is no significant correction needed for the brightest range that corresponds to sources observed mainly with 2D observations.
This is probably due to the colour calibration of the PSF not being as good as that of the LSF \citep{EDR3_PSF} and therefore the difference between the normal and default \nueff processing not being as pronounced.
A simple cubic relationship as a function of colour $f(\bprp)$ was fitted to the measured systematic and the coefficients are given in \tabref{nueffcorrection} for the two magnitude ranges for which the correction is required. The corrected $G$ flux can therefore be obtained as $\gflux^\ast=\gflux\times f(\bprp)$. The applicability range in \bprp of these corrections is 0.25 to 3.0. Outside of this range, the correction values at these limits should be used. We note that no correction is needed for $G<13$.
\begin{table}[htp]
    \caption{Coefficients of the correction to the \gband fluxes to be applied to the stars where a default \nueff has been used in the IPD. The correction is a simple cubic polynomial in \bprp. The applicability range in \bprp of these corrections is 0.25 to 3.0. Outside of this range, use the correction values at these limits.
    We note that no correction is needed for $G<13$. The old fluxes are multiplied by this correction to obtain the corrected fluxes.}
    \label{tab:nueffcorrection}
    \centering
    \begin{tabular}{lcccc}
    \hline\hline
    $G$ range & $c_0 $ & $c_1 $ & $c_2 $ & $c_3 $\\
    \hline
    $13<G<16$ & 1.00876 & -0.02540 & 0.01747 & -0.00277\\
    $G>16$      & 1.00525 & -0.02323 & 0.01740 & -0.00253\\
     \hline
    \end{tabular}
\end{table}
To validate this correction, main sequence stars were selected from the HR diagram for stars with parallax greater than 3 mas. The split that can be seen in the $G-\grp$ versus $\gbp-G$ plot, shown in the upper panel of \figref{nueffvalidation}, is due to whether a star has been processed with an actual or default \nueff. The lower panel shows the same plot but with corrected \gband photometry for stars that have had a default \nueff used in the IPD. These can mostly be identified in the \edr archive as those having \texttt{astrometric\_params\_solved}=95. There is also the case of \texttt{astrometric\_params\_solved}=3 for which the value of \nueff used cannot be established from the archive. However, since most of these have used a default \nueff, the correction should be applied  \citep[see][]{DPACP-128}.
As can be seen, the correction removes the split.

Additional checks were carried out for stars redder than \bprp=3.0. Similar to that seen in \figref{nueffvalidation}, this correction removes the split from the data even for the reddest stars.
Due to the nature of the cyclic processing chain, the \nueff value used in the IPD, comes from the previous processing cycle, which contained less data. This means that there are many sources that used a default \nueff in IPD, but have a mean \bprp value in \edr catalogue that can be used to correct for this effect using the procedure described in this section. Since this is a large number, a separate table in the archive is provided with the corrected \gband photometry for the convenience of the user.

\subsection{\gband magnitude term for blue and bright sources}\label{sec:issues_bright}
\begin{figure}
\centering
\colfig{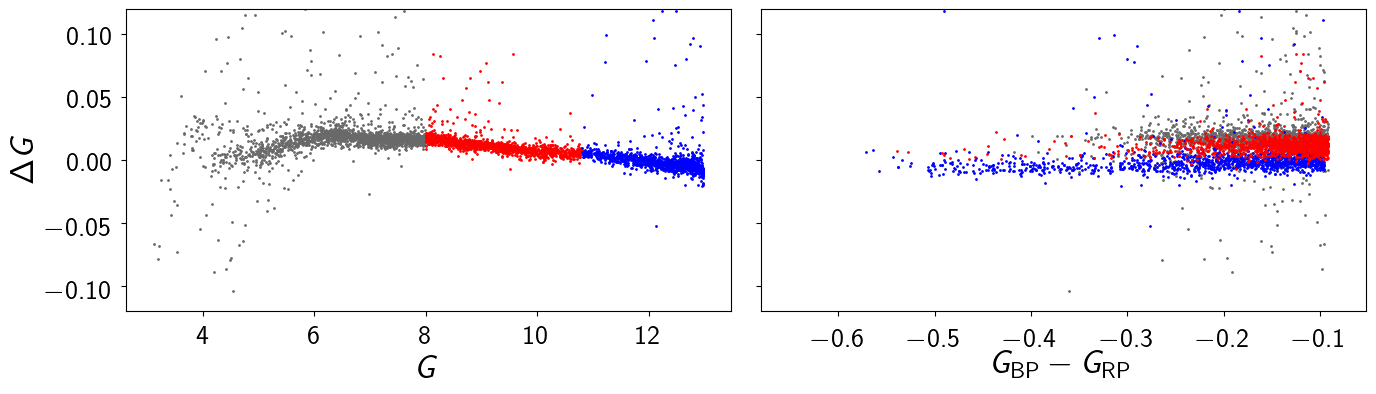}
\caption{Residuals between \edr magnitudes and synthetic ones from \xp spectra computed on all-sky
sources brighter than $G < 13$ and $\bprp < -0.1$ for $G$ as function of $G$ magnitudes (\textit{left}) and \bprp (\textit{right})}
\label{fig:issues_bright}
\end{figure}

An anomaly has been detected in the data while looking at the residuals between \edr and synthetic magnitudes derived from \xp spectra for an all-sky sample of extremely blue and bright sources ($G<13$ and $\bprp < -0.1$): the residuals plotted in  \figref{issues_bright}, show a trend of about 5 mmag/mag for the magnitude range $8<G<13$. More precisely, a linear fit to data in this range results in the relation $\Delta G = 0.054-0.0046\, G$. Brighter than $G=8$, residuals are dominated by saturation effects. Such trends are not seen in the \gbp and \grp cases. The origin of this behaviour is probably due to the PSF/LSF calibration where problems are known to exist with the colour-dependent part of the PSF model \citep{EDR3_PSF}. This magnitude trend is not noticeable in \figref{ext_residuals} and \figref{ext_validation} because very few sources fall in this magnitude and colour interval: the interested reader can find some additional details and plots in the online documentation.

\section{Considerations for the end user}

A major point made by this paper is that there is no silver bullet when it comes to identifying problematic data in a large catalogue like \edr. The best approach is inevitably dependent on the specific scientific goal that the end user is pursuing. However, it is recognised that it can also be valuable to have a set of prescriptions that could be applied when a preliminary exploration of the data is required before committing to a more detailed analysis. In this spirit, this section provides a number of suggestions for possible quality filters that users may want to consider while also pointing out some caveats. This section deals only with the photometric content of \edr, the reader is referred to \cite{DPACP-128} for what concerns the astrometric content of the catalogue.

While it is obviously worthwhile including in the archive query some basic restrictions (e.g. magnitude and/or colour range, minimum number of observations, sky position, basic astrometric parameters, etc.) to benefit from the database indices and restrict the data volume to a manageable size, we suggest applying more detailed filtering as a post-processing operation. This allows the use of quantities that are not available in the \edr archive and to tweak the selection criteria to assess their impact.

\subsection{Photometry release criteria}
In order for the photometry of a given source to be published in \edr the following basic criteria had to be met. When the photometry did not meet these criteria the corresponding values and relevant derived quantities have been nullified. In particular: \texttt{phot\_g\_n\_obs}$\geq10$ is required for the $G$ photometry to be published; \texttt{phot\_bp\_n\_obs}$\geq2$ is required for the \gbp photometry to be published; \texttt{phot\_rp\_n\_obs}$\geq2$ is required for the \grp photometry to be published.
See also \cite{DPACP-128} for a discussion of the criteria applied to the astrometric quantities.

\subsection{Filter on \gbp}\label{sec:filt_gbp}
\asecref{fluxlimit} showed that \gbp tends to be systematically brighter towards the faint end: it would therefore make sense to include a restriction on \gbp in the archive query. From \figref{deltaxp} the restriction $\gbp<m$ could be in the range $20.3\leq m\leq20.9$, which corresponds to the range where $50\%$ of the sources should have a \gbp flux that is unaffected by the systematic to where $50\%$ of the sources are systematically brighter by 0.2 mag. The value chosen for $m$ will have an impact on completeness and on the magnitude and colour range of the selection. Of course if \gbp (or derived quantities) is not required in the analysis, then there is no use in applying the filter. 

To illustrate the effect of this filter we used the set of the nearby sources for which the appropriate \nueff was used in the IPD process\footnote{This required an additional constraint to the criteria defined in \secref{data}: \texttt{astrometric\_params\_solved}>=31}.
\begin{table}[]
    \caption{Effect of filtering on \gbp for the nearby source dataset. \gbp shows the maximum value allowed for \texttt{phot\_bp\_mean\_mag}, $I_{\rm BP}$ shows the corresponding minimum value allowed for \texttt{phot\_bp\_mean\_flux}.} 
    \label{tab:nearby_sel_frac}
    \centering
    \begin{tabular}{c|c|r|r}
    \hline\hline
    \gbp & $I_{\rm BP}$ & \multicolumn{1}{c|}{$N$} & \multicolumn{1}{c}{Fraction}\\
    \hline
    20.3 & 103.61 &  7,575,348 &  $68\%$\\ 
    20.4 &  94.49 &  7,812,148 &  $71\%$\\
    20.5 &  86.18 &  8,047,835 &  $73\%$\\
    20.6 &  78.60 &  8,284,960 &  $75\%$\\
    20.7 &  71.68 &  8,523,503 &  $77\%$\\
    20.8 &  65.37 &  8,769,848 &  $79\%$\\
    20.9 &  59.62 &  9,026,012 &  $82\%$\\
    - &   - & 11,069,066 & $100\%$\\
    \hline
    \end{tabular}
\end{table}
\tabref{nearby_sel_frac} shows the fraction of sources that were retained as a function of the magnitude threshold in the suggested range. The brightest threshold caused nearly one third of the sources to be excluded from the selection, the faintest threshold instead caused $18\%$ of the sources to be excluded.

\subsection{Crowding effects}\label{sec:quality:crowding}

\edr includes two quantities that were not available in \drt and provide the number of BP and RP epoch transits included in the mean source photometry that are likely to be blended\footnote{Columns \texttt{phot\_bp\_n\_blended\_transits} and \texttt{phot\_rp\_n\_blended\_transits} in \texttt{gaia\_source} table in the \edr archive.} with one or more other sources (see \secref{crowding}). A useful metric that can be computed for a source is the blending fraction $\beta$, which is defined as the sum of the number of blended transits in \xp divided by the sum of the number of observations in BP and RP\footnote{In terms of \edr archive columns: $\beta$ =\newline
\noindent(\texttt{phot\_bp\_n\_blended\_transits}+\texttt{phot\_rp\_n\_blended\_transits}) \newline \noindent *1.0\,/\,(\texttt{phot\_bp\_n\_obs} + \texttt{phot\_rp\_n\_obs}) where the term $1.0$ is required to ensure the floating point division.}. To avoid systematic problems caused by crowding, sources could be required to have a low blend fraction, for example $\beta\leq0.1$ allows only $10\%$ of the epochs to be affected by blending.

There are a number of caveats that should be considered when applying this kind of selection. First, the fact that a source has $\beta=0.0$ does not necessarily imply that the source is not affected by crowding. The reason is that the crowding assessment (see \secref{crowding}) is limited to sources that are present in the \gaia catalogue, meaning sources that have been acquired, at least once, by \gaia. Close pairs, namely sources that are close enough on the sky to never be resolved by \gaia as a non-single source, will not be excluded by a filter on $\beta$. The second caveat is that $\beta$ does not take into account the flux ratio between the target source and the blending source(s): for example, if $\beta=0.5$ but the target source is $\gbp=14.0$ and the blending source is $\gbp=19.0$, then the effect of the blending source on the target source is probably negligible. In principle, users can assess this effect, at least in the case where the blending is from a source that is close to the target source on the sky (an occasional blend could be due to a source coming from the other FoV). First all sources affected by blending can be detected using a filter on $\beta$, then for each of these sources a cone search query could be performed to find other sources in the \edr archive that are close enough to the target source to result in a blend. Since the size of a \xp window is approximately $3.5\times2.1$ arcsec (AL$\times$AC), sources that are closer than $\approx1.05$ arcsec should always be blended whereas sources that are at a distance larger than $\approx1.05$ arcsec but closer than $\approx1.75$ arcsec will occasionally be blended depending on the satellite scan direction. Once the blending source(s) have been identified it will become possible to make a more informed decision on whether the blend is likely to have a significant effect on the photometry or not.

\subsection{Filter on \xp corrected flux excess}\label{sec:quality:cxs}

\asecref{xpxs} introduced the corrected \xp flux excess, \cxs, which is obtained for a given source by subtracting from the \xp flux excess $C$ \citep[see][]{Phot_DR2} the expected excess at the source colour produced by the polynomial defined in \tabref{cxs_fit}. We remind the user that \cxs is not available as a column in the \edr archive and therefore has to be computed as described in \secref{xpxs}. \cxs provides a measure of consistency between the \gband, BP and RP photometry and therefore can be used to exclude sources showing inconsistencies. \asecref{xpxs} analysed in detail different possible causes of the inconsistency, showing that it could originate in any of the bands. This is the major limitation with \cxs: it only indicates the presence of an inconsistency, without an indication to where it originates. Filtering on \cxs can also be problematic when completeness is important since it will have the effect of excluding variable and extended sources (see \figref{xsVarExt}).

To devise a selection criteria for \cxs we made use of the Stetson and Ivezic standards \citep{2000PASP..112..925S, Ive07} to determine the \cxs scatter versus $G$ magnitude using all the sources in the sample with $G>9$ (to avoid problems with low number statistics at the bright end). The scatter was measured in bins of 0.01 mag and the resulting dataset was then fitted with a simple power law in $G$ magnitude:
\begin{equation}
    \sigma_{C^{\ast}}(G)=c_0 + c_1G^m,\label{eq:cxss_fit}
\end{equation}
with $c_0=0.0059898$, $c_1=8.817481\cdot10^{-12}$, and $m=7.618399$. This fit is considered to represent the $1\sigma$ scatter for a sample of well behaved isolated stellar sources with good quality \gaia photometry.

\begin{figure}
    \centering
    \colfig{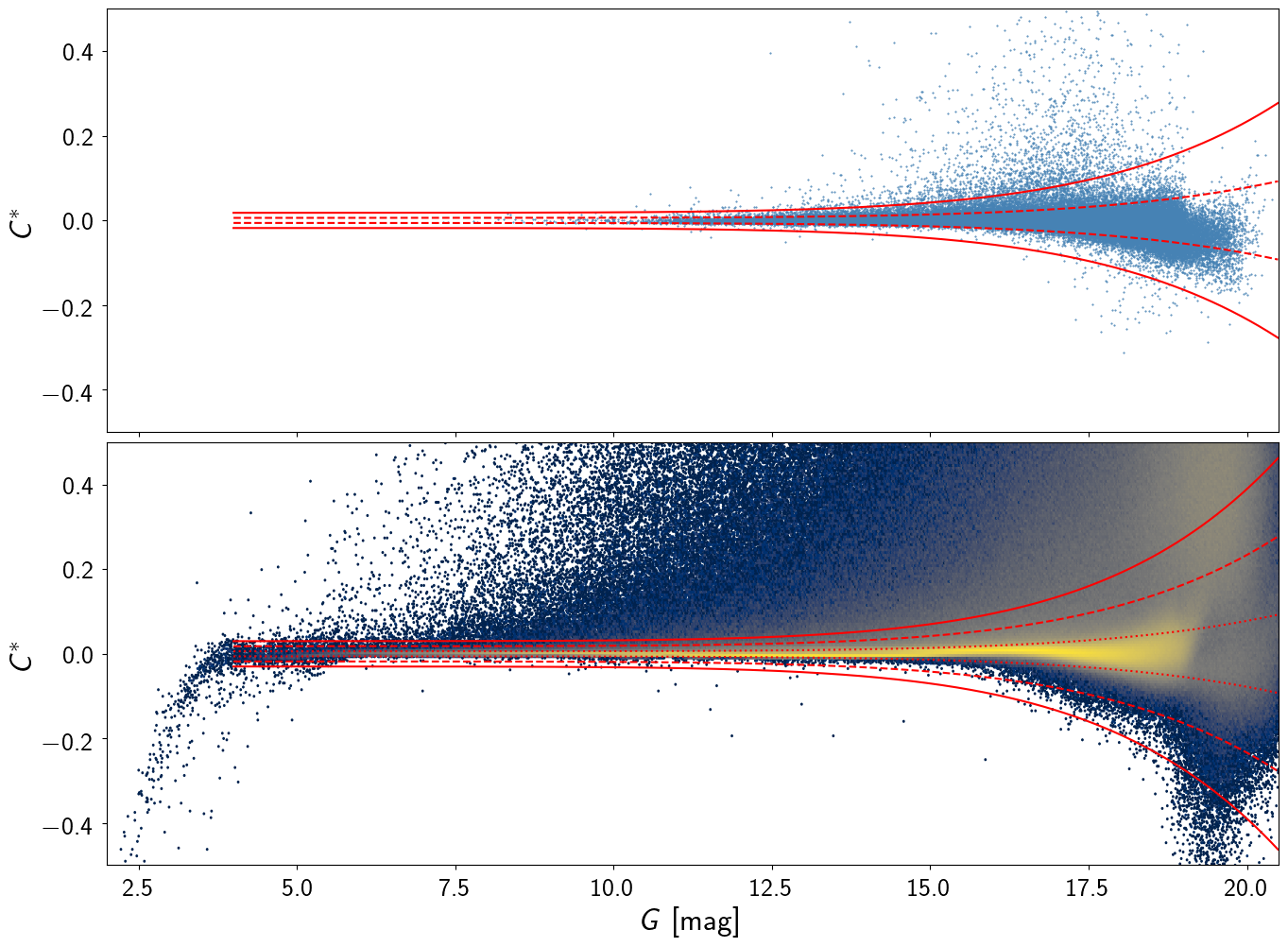}
    \caption{Corrected \xp flux excess vs. magnitude for the Stetson and Ivezic dataset (\textit{top panel}) including the $\pm\sigma$ (solid) and $\pm3\sigma$ (dashed) scatter lines and for the nearby source dataset (\textit{bottom panel}) including the $\pm\sigma$ (solid), $\pm3\sigma$ (dashed) and $\pm5\sigma$ (dotted) scatter lines. The scatter lines are defined by \equref{cxss_fit} with the fit coefficients provided in the text.}
    \label{fig:cxs_vs_mag}
\end{figure}

The top panel of \figref{cxs_vs_mag} shows the \cxs dependence on $G$ magnitude with the $\sigma$ and $3 \sigma$ lines represented by the fit described above.
The  bottom panel of \figref{cxs_vs_mag} shows the \cxs dependence on $G$ magnitude for a sample of nearby sources (limited to $\gbp<20.75$) showing the $\pm \sigma$, $\pm 3 \sigma$ and $\pm5 \sigma$ lines. A possible filter on the corrected \xp flux excess can be defined in terms of the fitted scatter line as $|\cxs|<N\sigma_{\cxs}$. The filter should only be applied for $G>4$ mag as for brighter magnitudes the effects of saturation are still too large (see \secref{saturationCorrection}).
\begin{figure}
    \centering
    \colfig{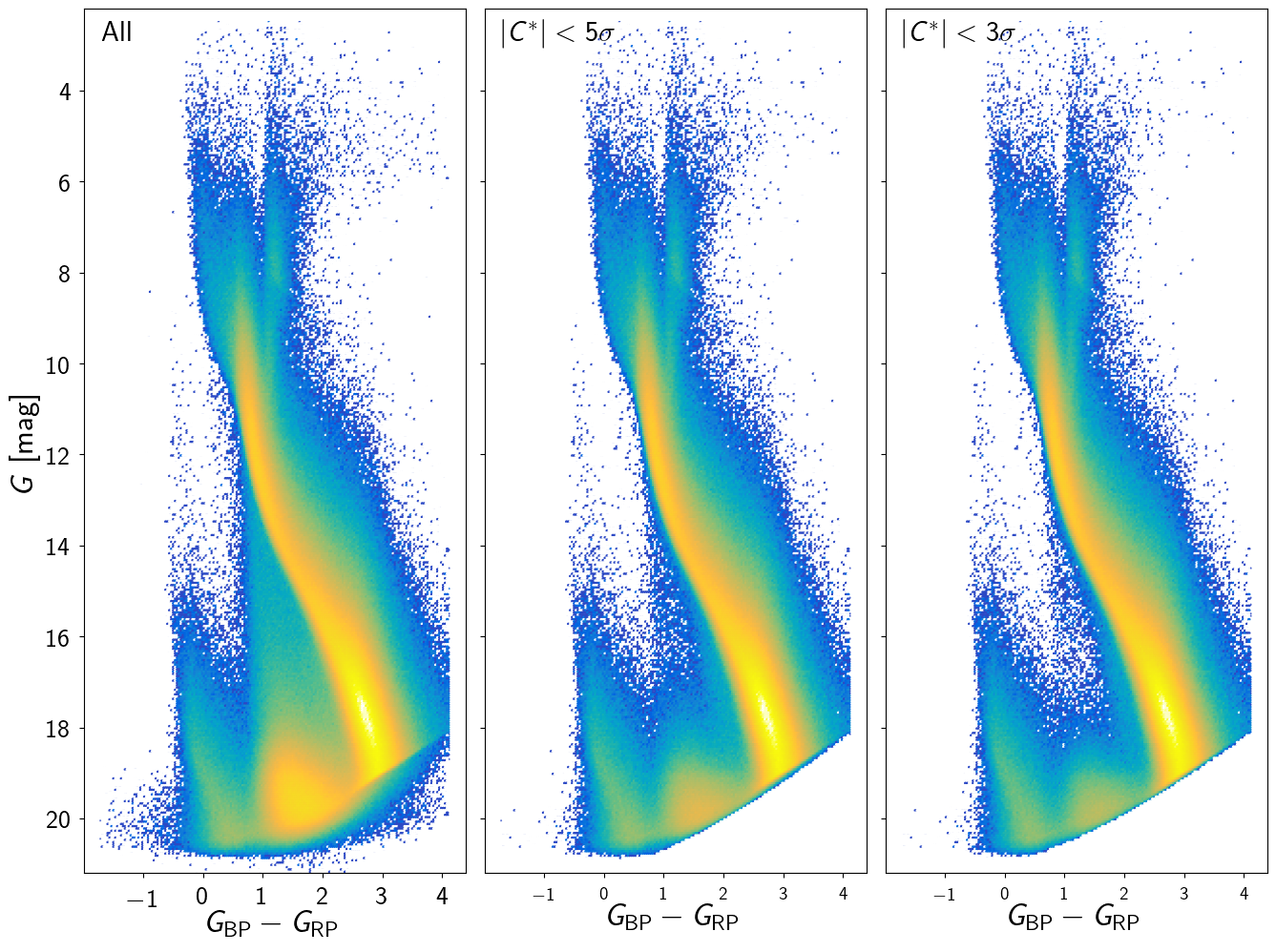}
    \colfig{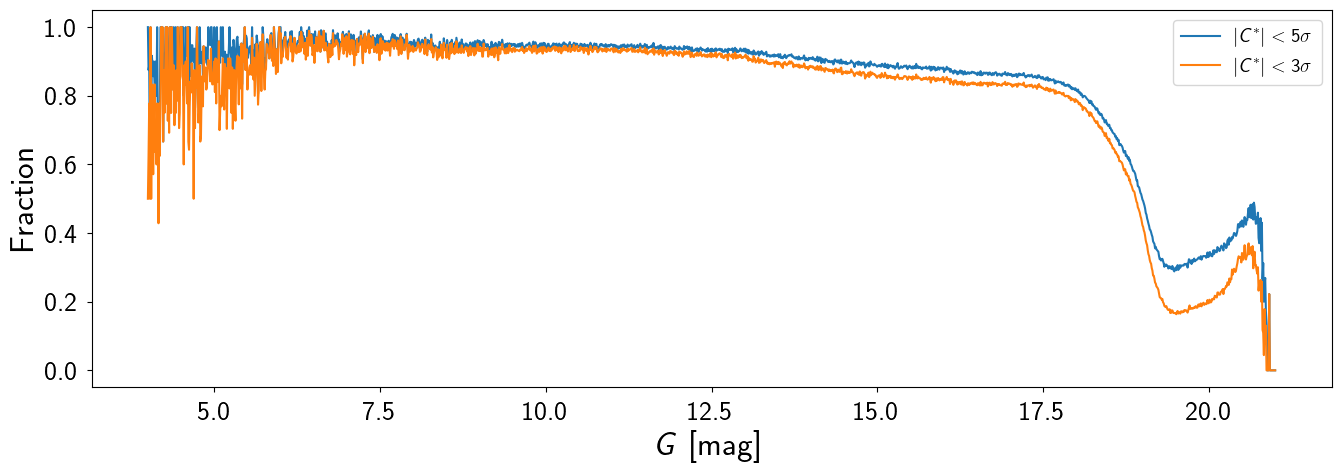}
    \caption{Colour--magnitude diagram for the nearby source sample for all sources with $\gbp<20.75$ (\textit{left top panel}), the subset of sources with $|\cxs|$ smaller than $5\sigma$ (\textit{central top panel}) and smaller than $3\sigma$ (\textit{right top panel}). The bottom panel shows the fraction of sources selected using the two thresholds.}
    \label{fig:cxs_cmd}
\end{figure}
To illustrate the effect of the \cxs filter, we use the set of nearby sources with $\gbp<20.75$ (see \secref{filt_gbp}). \afigref{cxs_cmd} shows the CMDs for the full dataset and then for two selections that applied a $5\sigma$ and a $3\sigma$ cut on \cxs. The bottom panel of \figref{cxs_cmd} shows the fraction of sources as a function of $G$ magnitude for the two filtered datasets.  The effect of the \gbp magnitude filter (see \secref{filt_gbp}) is also clearly visible as a progressive brighter faint-magnitude limit towards red colours visible in the CMDs. This also explains the fact that the fraction of selected sources has a minimum at $G\approx19.5$ to then increase again for fainter magnitudes where the sources in the sample have bluer colours and are less likely to have a large \cxs (see \figref{xpfit}). The \cxs filter seems to be mostly reducing the population of sources between the white dwarf and main sequence.

\subsection{Caveat on comparisons with previous releases}\label{sec:dontmix}

There are many reasons why the data from \gaia DR2 should not be used in conjunction with or compared to that in EDR3.
Due to the way that the photometric systems have been set up in the two releases, the passbands are different and cannot be compared directly. In general there will be colour terms between them. Although the difference between the \edr and DR2 passbands are smaller than between DR2 and DR1 (\gband only), it will still amount to a few percent.
Specifically for the \gband, the PSF and LSF calibrations \citep{EDR3_PSF} have improved greatly for data release and a number of systematic effects have been corrected because of this, for example the linearity of the magnitude scale.
Also, since the PSF and LSF fits are now much better for point sources, the difference in the photometry between extended and point sources will be amplified. In some cases this difference will amount to 0.5 mag or more. This issue is further discussed in \appref{galaxies}.

The source IDs are in the majority unchanged between \drt and EDR3, but it is still a significant number \citep[see Sect. 7 of][for a detailed discussion]{EDR3_XM}. Moreover, the list of transits associated with a given source ID may have changed significantly following improvements in the cross match process.
Finally, comparisons between data releases are not recommended in general since they mainly show issues that are present in the old data that are no longer relevant. Interpretation of the differences are also difficult to make for the reasons given above.

\section{Conclusions}\label{sec:end}

In this paper we have presented the photometric content of \edr and described the process of producing calibrated photometry for \gband, BP and RP. A few issues that have been discovered during the validation of the photometry in preparation for the data release have been discussed and possible mitigation strategies have been suggested. Although it has been stressed that selecting good quality data from the \edr catalogue must be tailored to the specific scientific goal of the end user, a number of quality metrics have been presented: the recommendation is to use them only in the preliminary exploratory analysis while a better ad hoc approach is being devised.
Finally, we conclude providing a summary of the major improvements in the \edr photometry.
\begin{itemize}
    \item Apart from saturation effects for very bright sources (see \secref{saturationCorrection}), the significant magnitude term found in the \drt photometry \citep{CasagrandeVandenBerg, Weiler2017} is not visible anymore \figref{ext_validation}: over all there is no trend larger than 1 mmag/mag. 
    However for blue sources (\gbp-\grp $<-0.1$) there is an indication of a differential colour term between sources brighter and fainter than G=11 at a level of $1\%$ in magnitude (see \secref{issues_bright}).
    \item Using only one passband for the entire magnitude and colour range, does not leave systematics above the $1\%$ level in magnitude in all bands; this increases to $\sim2\%$ for the 4148 sources with $G<13$ and $\gbp-\grp<-0.1$.
    \item Better background estimation has been carried out in all three passbands. For the \gbp and \grp bands this new processing is described in \secref{overproc:bkg}. The validation of this is shown in \figref{skyCxs} where no trace of zodiacal light is present unlike in \drt.
    \item The consistency between $G$, \gbp, and \grp has improved as is shown in \secref{xpxs}.
    \item Better saturation handling for the \gband has been carried out for \edr. This is seen in \figref{photerrors}, where the bump at $G=11.2$ has been reduced with respect to \dro and DR2. This can also be seen for $G<6$ in this plot and in the analysis of \secref{saturationCorrection} where the correction is close to zero.
    \item In the processing that lead to \edr, the handling of bad data was considerably improved. This involved pre-filtering data that had been identified as problematic and excluding some periods where the photometry was of poorer quality and could not be calibrated well enough. This is evidenced in Figs.~\ref{fig:skewnesssky} and \ref{fig:veryfaintgsky} where no great circles (sometimes referred to as `cat scratches') are visible.
    \item In comparison between \dro and DR2, no discontinuities are seen at $G=13$ and 16 in the comparison with external photometry (see \figref{extcatcomp}). This reflects the better stability that has been achieved in establishing a consistent photometric system between the window class configurations.
    \item
    Since the number of observations available for a given source is, on average, larger in \edr than in DR2 (thanks to the longer time range covered) the uncertainties on the mean source photometry are also smaller. However, an even larger overall improvement was achieved with the addition of more terms to the calibration models as described in \secref{calmodel}. This is clearly visible in \figref{photerrors} since the comparison is based on sources with the same number of observations in all three data releases and therefore the differences cannot be ascribed to the different time ranges covered by the releases.
\end{itemize}

All the improvements listed above are part of the ongoing huge effort to extract the most homogeneous, precise and accurate photometry from \gaia data. This requires the calibration of hundreds of instrumental characteristics and inhomogeneities in space and time, in the attempt to standardise the overall system. 
As extensively discussed by \citet{clem13}, with current technology, this goal may be achieved up to typical uncertainties of $\la 1\%$. This can be seen as the current state-of-the-art limit in photometry \citep[see also][for recent, large-scale examples]{Ive07,APASS2020}. The current sub-$1\%$ precision we achieved for an enormous number of stars, covering the full range of astrophysical parameters, allows us to reveal tiny effects, that we progressively try to understand and correct. Through this continuous process, the precision and accuracy of the photometric data increases with the subsequent data releases, allowing for more sophisticated treatment and understanding of the system complexity.


\begin{acknowledgements}

This work presents results from the European Space Agency (ESA) space mission \textit{Gaia}. 
\textit{Gaia} data are being processed by the \textit{Gaia} Data Processing and Analysis Consortium (DPAC). 
Funding for the DPAC is provided by national institutions, in particular the institutions participating in the \textit{Gaia} MultiLateral Agreement (MLA). 
The \textit{Gaia} mission website is \url{https://www.cosmos.esa.int/gaia}. 
The \textit{Gaia} Archive website is \url{https://archives.esac.esa.int/gaia}.

The \gaia\ photometric data processing has been financially supported by (in alphabetical order by country):
the Tenure Track Pilot Programme of the Croatian Science Foundation and the Ecole Polytechnique Fédérale de Lausanne and the Project TTP-2018-07-1171 Mining the Variable Sky, with funds of the Croatian-Swiss Research Programme;
the Agenzia Spaziale Italiana (ASI) through contracts I/037/08/0, I/058/10/0, 2014-025-R.0,
2014-025-R.1.2015  and 2018-24-HH.0 to the Italian Istituto Nazionale di Astrofisica (INAF), 
and INAF;
the Spanish Ministry of Science, Innovation and University (MICIU/FEDER, UE) through grants RTI2018-095076-B-C21, ESP2016-80079-C2-1-R, and the Institute of Cosmos Sciences University of Barcelona (ICCUB, Unidad de Excelencia ’Mar\'{\i}a de Maeztu’) through grants MDM-2014-0369 and CEX2019-000918-M;
the United Kingdom Particle Physics and Astronomy Research Council (PPARC), the United Kingdom Science and Technology Facilities Council (STFC), and the United Kingdom Space Agency (UKSA) through the following grants to the University of Bristol, the University of Cambridge, the University of Edinburgh, the University of Leicester, the Mullard Space Sciences Laboratory of University College London, and the United Kingdom Rutherford Appleton Laboratory (RAL): PP/D006511/1, PP/D006546/1, PP/D006570/1, ST/I000852/1, ST/J005045/1, ST/K00056X/1, ST/K000209/1, ST/K000756/1, ST/L006561/1, ST/N000595/1, ST/N000641/1, ST/N000978/1, ST/N001117/1, ST/S000089/1, ST/S000976/1, ST/S001123/1, ST/S001948/1, ST/S002103/1, and ST/V000969/1.

We would like to thank J.~Martin-Fleitas for providing the response loss data that was used to produce \afigref{calperiods}; C.~Ducourant and L.~Galluccio for providing SDSS parameters for a selection of galaxies; L.~Rimoldini for providing the catalogue of variable and extragalactic sources that was used to produce \afigref{xsVarExt}; M.~J.~Irwin for useful discussions on calibrating broad band photometry and statistical methods; and C.~Bailer-Jones for kindly reviewing an earlier version of this manuscript.

We additionally made use of TOPCAT \citep[\url{http://www.starlink.ac.uk/topcat/}]{TOPCAT}, Astropy, a community-developed core Python package in Astronomy \citep{ASTROPY}, IPython \citep{IPYTHON}, and Matplotlib \citep{MATPLOTLIB}.

This research has made use of the SIMBAD database, operated at CDS, Strasbourg, France.

\end{acknowledgements}

\bibliographystyle{aa} 
\bibliography{refs} 

\begin{appendix} 

\section{External calibration of \xp spectra}\label{sec:extxpcal}

Since externally calibrated \xp spectra have been used to reconstruct the passbands of the \edr photometric systems (\secref{extcal}), we give here a brief description of the models and the strategy implemented to perform the external calibration of the \xp spectra, provided that a full description will be presented in \citeip{Montegriffo}
The general scheme for the calibration of \xp spectra is similar to the one implemented for the photometric data, being split into an internal calibration - aimed at bringing all observations onto a common reference system (often referred to as the `mean instrument') - and an external calibration describing an instrument model capable of reproducing the observational mean spectrum of a source given its SED. Once the instrument model is defined, it can be used to reconstruct the SED of each source starting from the corresponding \xp mean spectra. 

The process defining the instrument model follows a forward modelling approach.
Given a number of calibrators it is possible to derive the optimal set of model parameters that minimises in a least squares sense the difference between predicted and observable spectra.
If we specify with $u$ the location of a sample in the AL reference system, the relationship between the observational mean \xp spectrum $n_e(u)$ and the corresponding spectral photon distribution $n_p(\lambda)$ is given by:
\begin{equation}
\label{eq:extXpInstrumentModel}
n_{e}(u)=  \int_0^\infty n_p(\lambda) \, L(u - D(\lambda), \lambda) \, R(\lambda) \, {\rm d}\lambda,
\end{equation}
where the instrument model is a combination of the effective monochromatic LSF $L(u, \lambda)$, the dispersion relation $D(\lambda)$ and the overall instrument response function $R(\lambda)$, which includes the contribution of mirrors and prisms transmissivities, CCDs quantum efficiency etc.
Two distinct models are built for the BP and RP instruments.
The traditional approach to derive a simple response of the instrument as a function of wavelength by computing the ratio between the observational spectrum and the SED for a limited set of (possibly featureless) calibrators does not work very well for \gaia because the width of the LSF is rather large compared to the wavelength scale of the response variations: as a consequence, the derived response changes with the spectral type of the calibrator. The LSF cannot be ignored and must be taken into account.

If the previous relation is discretised to a finite wavelength grid, it can be conveniently expressed as a matrix multiplication:
\begin{equation}
\label{eq:extXpInstrumentModelMatrix}
\overrightarrow{n_e} = \mathbf{K} \cdot \overrightarrow{n_p},
\end{equation}
where the matrix $\mathbf{K}$ represents the instrument model sampled on the same wavelength grid of $\overrightarrow{n_p}$ and on the same  sample grid of $\overrightarrow{n_e}$. This alternative formulation points out clearly the linear nature of the instrument model.
In principle, given the availability of an arbitrarily large set of calibrators spanning the widest variety of SEDs, it should be possible to derive the instrument matrix of \equref{extXpInstrumentModelMatrix} at once from a least squares fit: unfortunately the pool of astrophysical sources that satisfy the requirements for a reliable flux calibrator (isolated and point-like source with stable flux and high S/N ratio) shrinks to a set of stars spanning a limited range of astrophysical parameters, and SPSS represent such a selection of sources. Moreover it can be demonstrated through principal component analysis that a set of calibrators such as the SPSS can only constrain a limited number of instrument components (roughly one fifth of the number of required components). We note that a similar problem holds also for the passband determination as discussed in \secref{extcal}. To mitigate these limitations we have followed two parallel strategies:
1) we developed the current model starting from a nominal instrument model (built on our pre-launch knowledge of the instrument hardware) so that only constrained deviations are allowed between the fitted model and the starting one; 
2) we extended the calibrator set (SPSS) with a wide variety of sources featuring strong emission lines over the entire wavelength range (mostly QSO and WR stars) and several sources taken by the PVL set (especially the reddest sources). 
Since these additional calibrators often exhibit flux variability (see \secref{extds}), we implemented an iterative scheme where each update of the instrument state is followed by a grey flux calibration of all non-SPSS sources (the flux level of each source is scaled to minimise the difference between predicted and observed flux distributions). The model is bootstrapped on the nominal model and, since its formulation (\equref{extXpInstrumentModel}) is not linear with respect its parameters, the optimal model is then evaluated employing the Differential Evolution algorithm \citep{storn}.

Once the instrument model is defined, one could use the integral equation \ref{eq:extXpInstrumentModel} to solve for the unknown SED, expressed in some convenient parametric form, starting from the observational spectrum. However, this equation is known as a Fredholm Integral Equation of the first kind and it is generally difficult to solve because the problem is essentially ill-conditioned: being the observed spectrum $n_p(u)$ affected by noise, there are many solutions that satisfy exactly an integral solution slightly perturbed from the original. To mitigate this problem we adopted a different strategy: since \xp mean spectra are represented as a linear combination of a proper set of basis functions $B_i(u)$ \citeipp{Carrasco}, for each basis function we set up an integral equation as \equref{extXpInstrumentModel}, where $n_p(u)$ is substituted with the basis $B_i(u)$. In this way  we can solve these equations to build up a new set of basis functions (called `inverse basis functions') whose images through the instrument model are the basis functions that represent the observed mean spectra. In other words, by solving the integral equation for analytic functions that are by definition noise free  we can achieve a numerically stable representation of externally calibrated SEDs. BP and RP spectra are reconstructed separately and then merged to form a unique SED.

\section{External datasets}\label{sec:extds}

\subsection{Spectro-photometric standard stars}

The SPSS were selected to be good flux calibrators for {\em Gaia}, according to the criteria detailed by \citet{SPSS}. The original requirement was to provide a final precision of $\simeq$1\% and an accuracy of 1--3\% of the {\em Gaia} flux calibration with respect to Vega \citep{bohlin14}. Additionally, because no existing dataset satisfied simultaneously all the requirements, a dedicated observing campaign was carried out, including: (i) a strict constancy monitoring of all candidate SPSS within $\pm$0.5\% \citep{SPSS3}; (ii) a photometric campaign to provide BVR magnitudes for all the candidate SPSS (Altavilla et al., submitted to MNRAS); and (iii) a spectro-photometric campaign to derive accurate flux tables \citep[][Pancino et al., in preparation]{SPSS2}\footnote{The SPSS flux tables and ancillary data can be obtained in advance of publication by contacting the authors.}.

The preliminary SPSS version used to calibrate EDR3 contains the best 113 SPSS out of a total of about 200 surviving SPSS candidates after constancy assessment (of the $\approx300$ initial candidates, about 100 were discarded because of variability). Therefore, the current SPSS release contains about half of the full SPSS set and only $\simeq$1500 spectra on a total of $\simeq$6500 (roughly 25\%). As a result, the extreme blue and red portions of the SPSS flux tables are expected to significantly improve in the future. The SPSS reference system is tied to the CALSPEC reference system\footnote{\url{https://www.stsci.edu/hst/instrumentation/reference-data-for-calibration-and-tools/astronomical-catalogs/calspec}} \citep{Bohlin2019} by means of the three pure-hydrogen white dwarfs G191-B2B, GD 71, and GD 153. More specifically, the SPSS flux reference system is tied to the 2013 version of CALSPEC, which is about 0.6\% brighter than the current version \citep{bohlin14}. A comparison of the 36 stars in common with the latest version of CALSPEC confirms that the SPSS are indeed 1\% brighter. On the other hand, the synthetic Johnson-Cousins magnitudes obtained from the SPSS are on average 1\% fainter than the reference magnitudes by \citep{landolt92}, for the 37 stars in common. The $\sim$1\% flux difference between different standard systems appears to be the ultimate accuracy limit of current observing technology \citep[see the discussion by][]{clem13}.

\subsection{Stetson secondary standard stars}\label{app:stetson}

The publicly available Stetson secondary standards\footnote{\url{https://www.canfar.net/storage/list/STETSON/Standards}} are a set of more than 200\,000 stars in different wide fields such as stellar clusters, standard fields, supernova remnants, planetary nebulae, deep galactic fields and the like. Their Johnson-Cousins UBVRI magnitudes are accurately calibrated (within 1\%) on the \citet{landolt92} system with the method described by \citet{2000PASP..112..925S} and updated as in \citet{stetson19}. The database is constantly evolving, with daily updates, but the reference system remains stably anchored to the \citet{landolt92} one, only the number of stars and their individual uncertainties change with time. The version of the database that we used here was downloaded in 2019. From a sample of 177\,915 UBVRI secondary standard stars assembled by merging the catalogues of Stetson’s standard fields, we selected a subsample of $\sim$100\,000 (a) having an unambiguous counterpart in Gaia DR3 and (b) being strictly clustered along the locus of bona fide stars in the $G-\gbp$ versus $G-\grp$ colour diagram. The latter criterion ensures that the selected stars have precise and consistent photometry in the three \gaia bands. We stress again that, in the present context, we use this catalogue only as a convenient sample of bona fide, well-measured stars, as all the data that we used to refine the external calibration come entirely from \gaia, namely original or synthetic $G$, \gbp and \grp photometry. The photometry available in the Stetson archive was not used.

\subsection{Passband validation library}

The PVL was created as a validation set, covering a larger variety of spectral types with respect to the flux calibrators, the SPSS. In the O and B spectral type range there are luminous blue variables, $\beta$\,Cephei and $\alpha$\,Cygni stars, slowly pulsating B stars, hot subdwarfs, and a large fraction of binaries \citep{2019A&A...623A.110G,2017PASA...34....1D}. Similar considerations hold for very red stars, which can vary because of chromospheric and magnetic activity, rotation and spots \citep{2019ApJ...879..114I}, or in the case of brown dwarfs, cloud coverage variations \citep{2018haex.bookE..94A}. Therefore, without a careful constancy monitoring on a large range of timescales, it is very difficult to pre-select suitable flux standards of extreme spectral types. For this reason the SPSS list does not include stars of types later than M2 and contains very few bright O and B stars. The PVL was thus built using space-based HST (Hubble Space Telescope) spectra, by including: (i) 39 CALSPEC hot stars that were not already included in the SPSS set for the above reasons; (ii) 17 stars from the HOTSTAR set \citep{khan18}, excluding stars in planetary nebulae; (iii) the red star Proxima Centauri \citep{ribas17}; (iv) the two brown dwarfs in the CALSPEC set (2M0036+18 and 2M0559-14)\footnote{These two faint and extremely red stars have negligible flux bluewards of 600--800~nm, and thus they have no HST blue spectra. To cover the entire {\em Gaia} wavelength range, the spectra were extended to 330~nm, by adjusting the appropriate theoretical spectra from the \citet{2012RSPTA.370.2765A} library.} and (v) the red CALSPEC stars GJ\,555 and VB\,8. We did not consider stars that showed known variability above $\pm$5\% and the typical variability in the PVL set is $\pm$2\% (for comparison it is $\pm$0.5\% for the SPSS). For this reason, we have preferred to rely on the SPSS set as flux calibrators to ensure we can achieve our goal of 1\% accuracy. All PVL stars that were calibrated on the latest CALSPEC reference system were re-aligned with the SPSS system, namely the 2013 CALSPEC system, by increasing their flux by 0.6\% \citep{bohlin14}. In this way, the combined SPSS + PVL dataset is as homogeneous as possible and can be used to perform a range of validation tests. The PVL flux tables will be published together with the SPSS ones (Pancino et al., in preparation).

\section{Colour--colour transformations}\label{sec:cctran}
\newcommand{\gminr}{\ensuremath{G-G_{\rm RP}}\xspace}
\newcommand{\bming}{\ensuremath{G_{\rm BP}-G}\xspace}
\def\gmag{\ensuremath{G}\xspace}
\newcommand{\hip}{Hipparcos\xspace}
\newcommand{\tyctwo}{Tycho-2\xspace}

\begin{figure*}
\centerline{	
	\includegraphics[width=0.67\columnwidth]{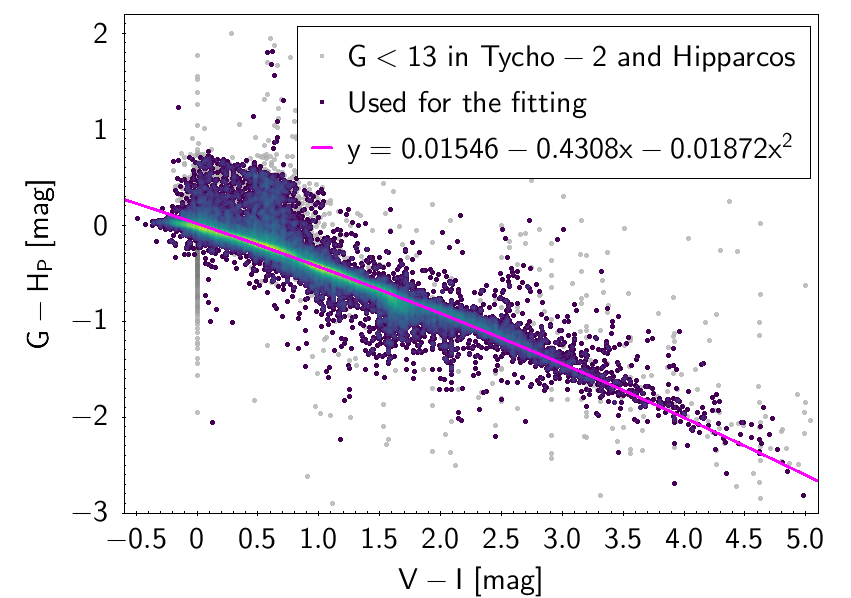}
	\includegraphics[width=0.67\columnwidth]{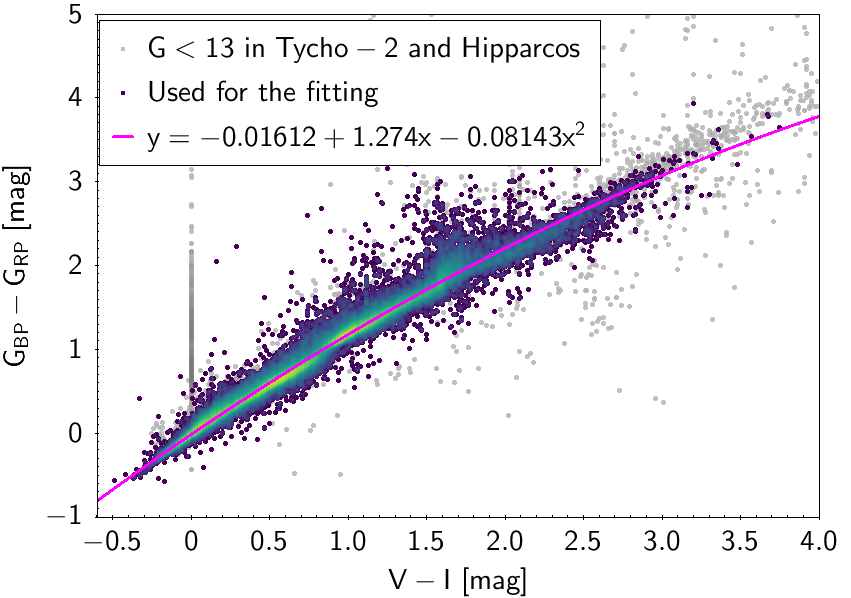}
	\includegraphics[width=0.67\columnwidth]{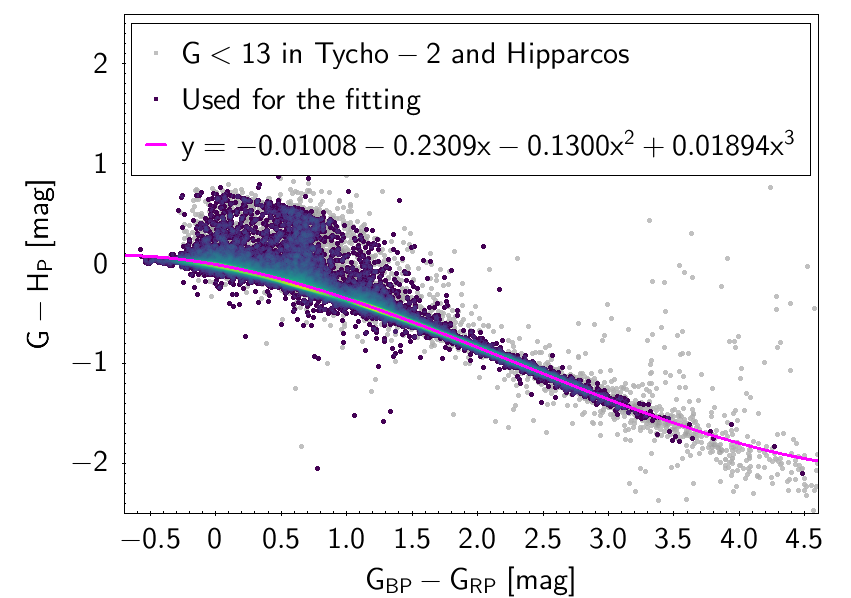}}
\centerline{	
	\includegraphics[width=0.67\columnwidth]{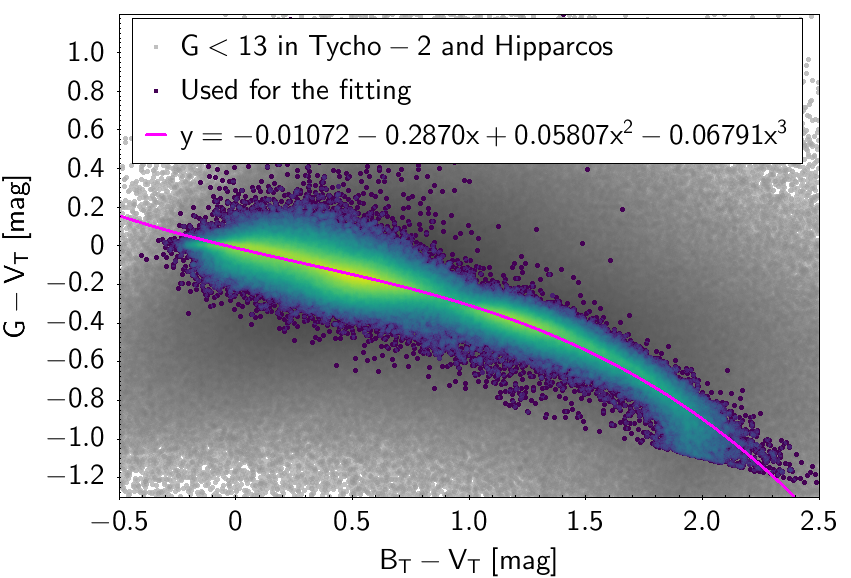}
	\includegraphics[width=0.67\columnwidth]{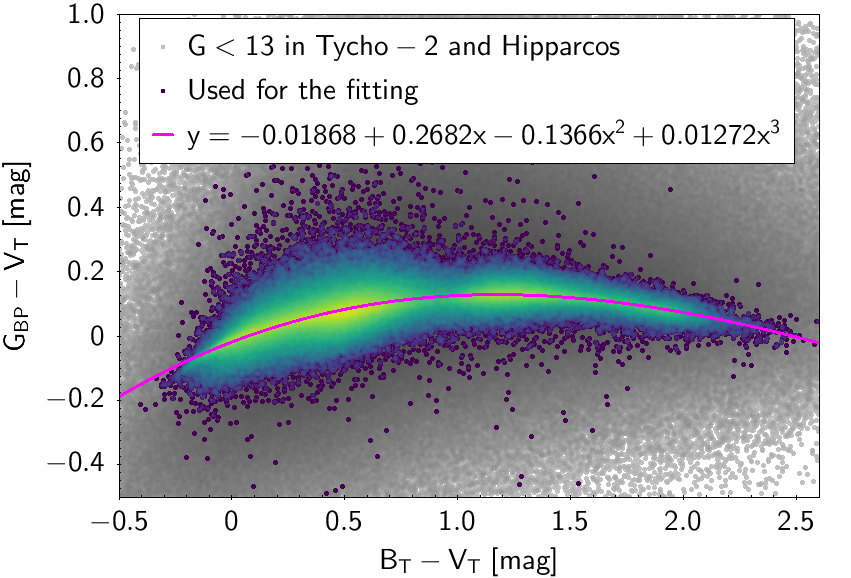}
	\includegraphics[width=0.67\columnwidth]{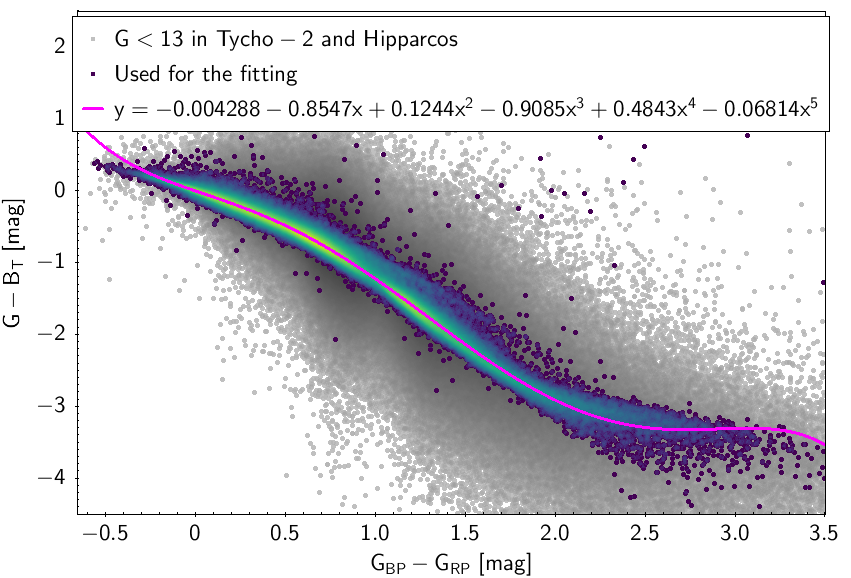}}
\centerline{
	\includegraphics[width=0.67\columnwidth]{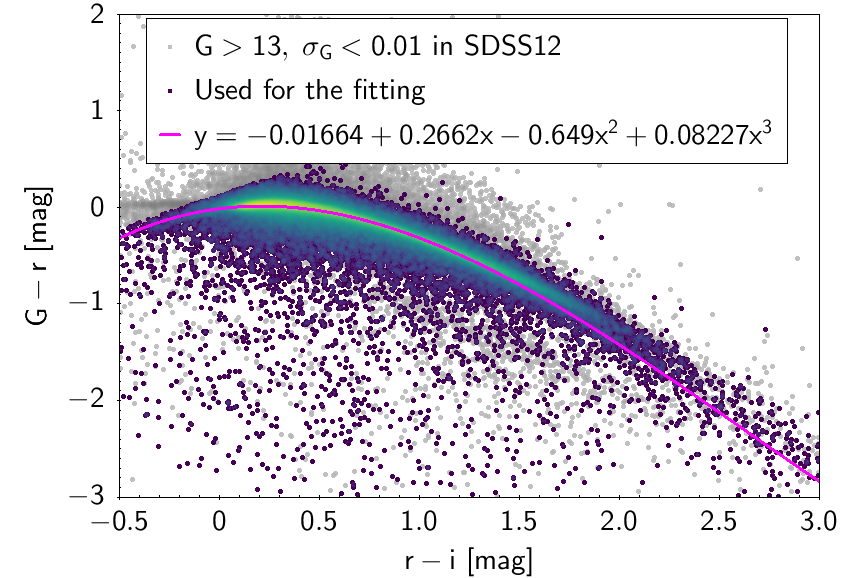}
	\includegraphics[width=0.67\columnwidth]{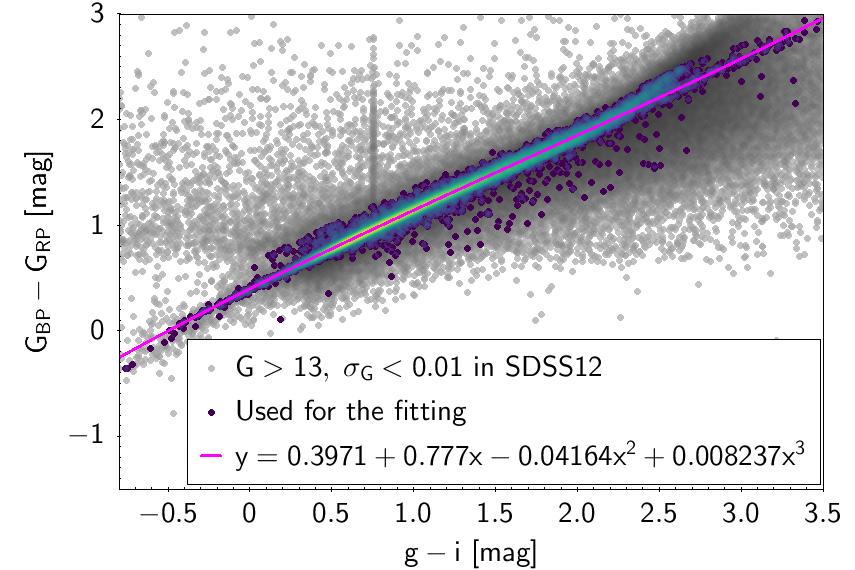}
	\includegraphics[width=0.67\columnwidth]{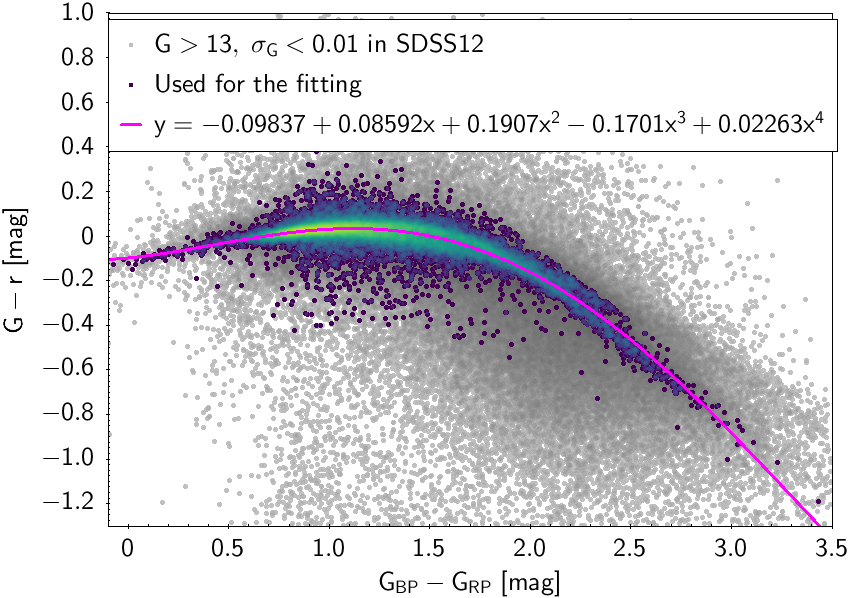}}
\centerline{	
	\includegraphics[width=0.67\columnwidth]{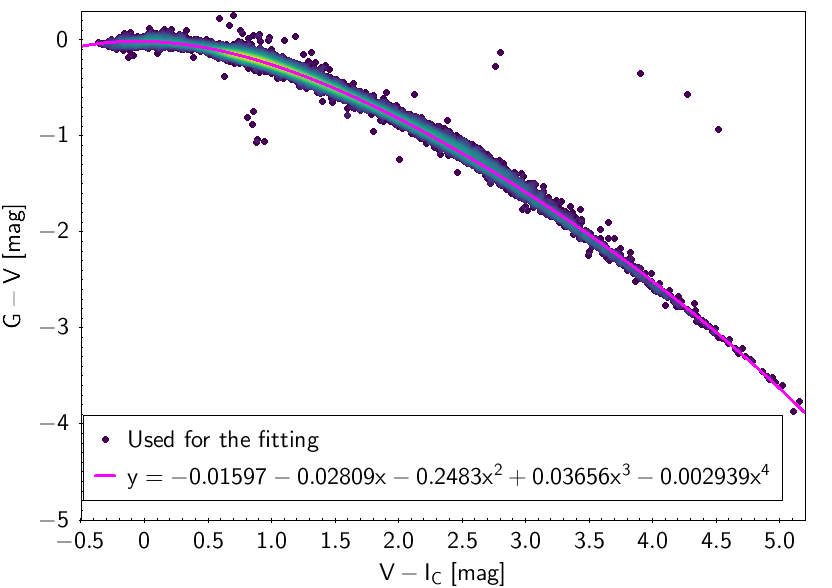}
	\includegraphics[width=0.67\columnwidth]{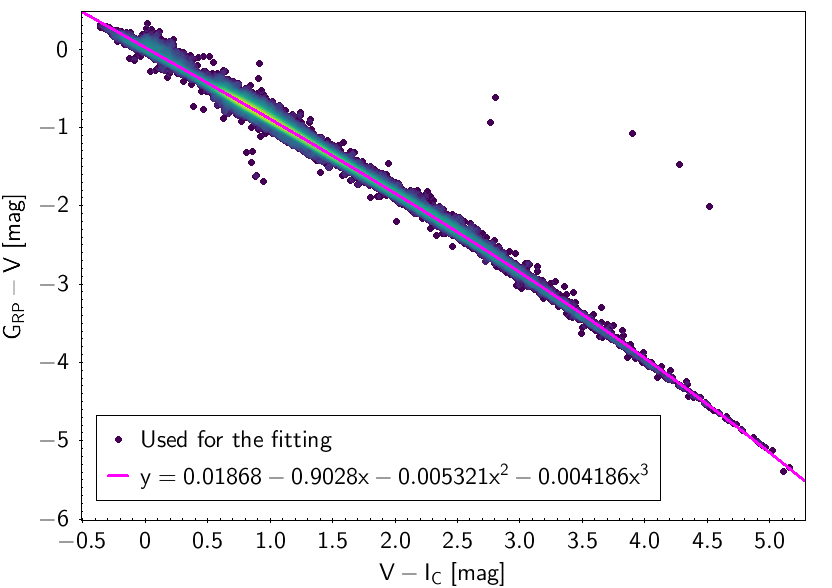}
	\includegraphics[width=0.67\columnwidth]{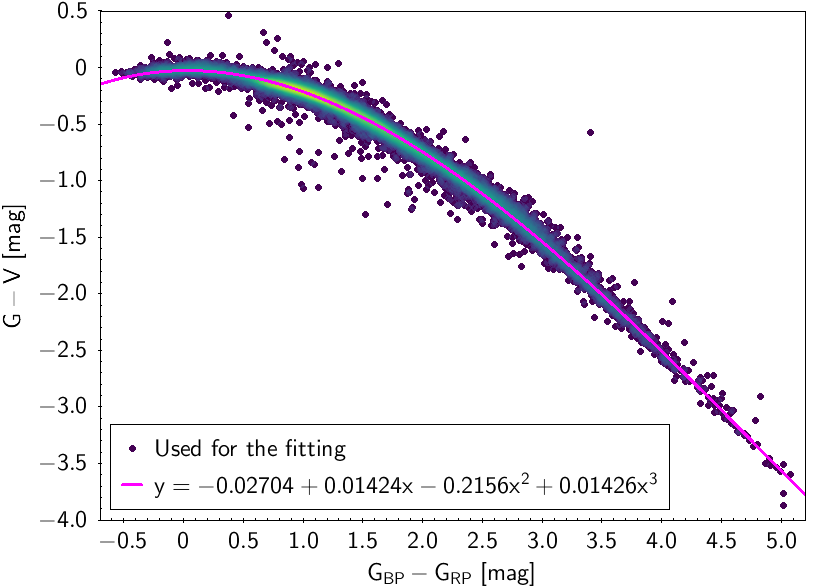}}
\centerline{
	\includegraphics[width=0.67\columnwidth]{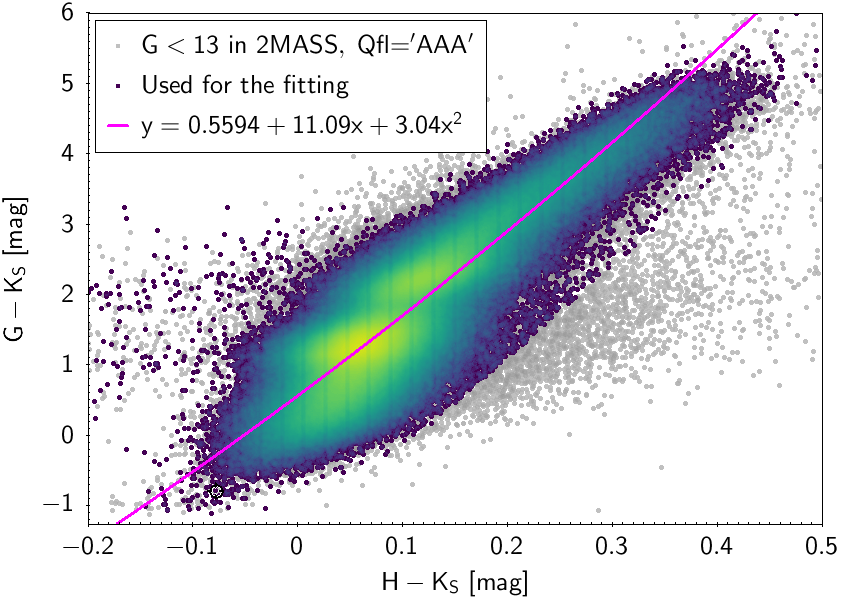}
	\includegraphics[width=0.67\columnwidth]{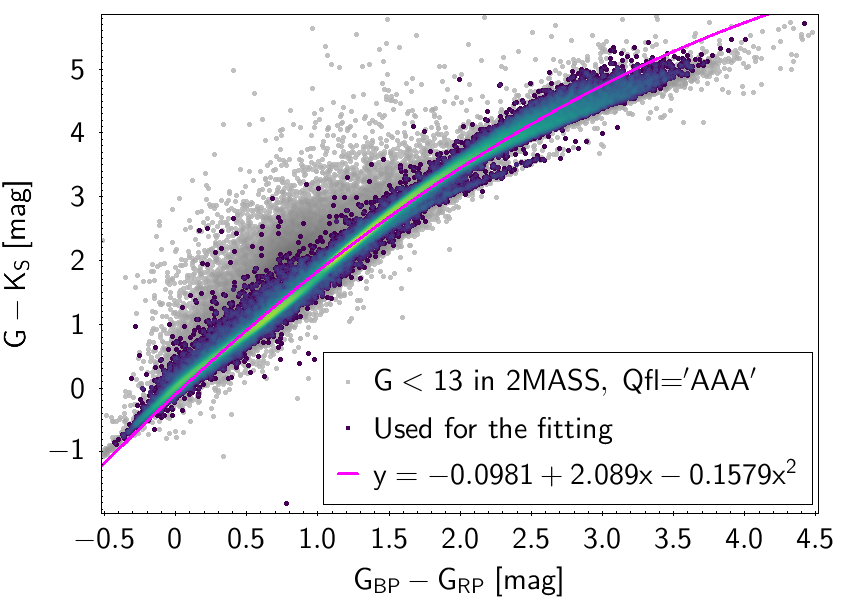}
	\includegraphics[width=0.67\columnwidth]{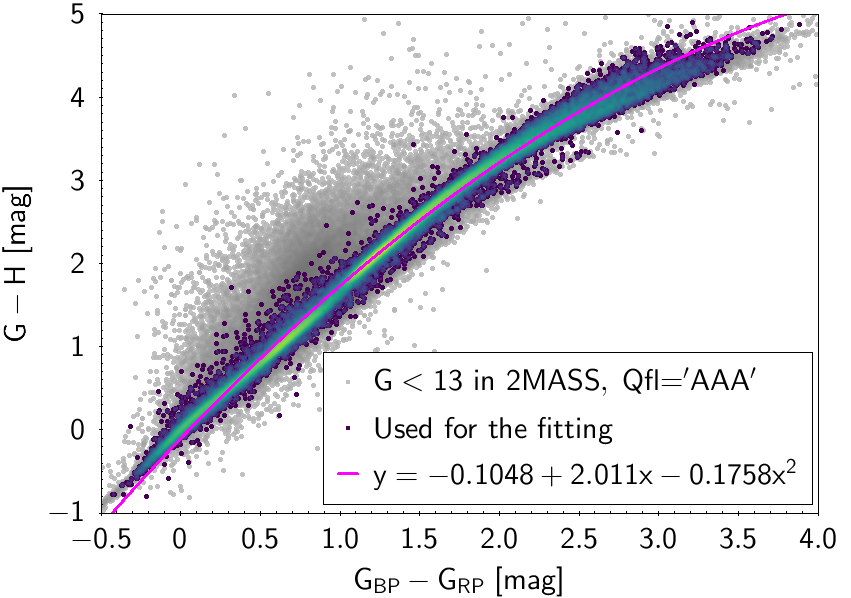}}
\caption{A selection of photometric relationships between \edr\ and  \hip, (top) \tyctwo, SDSS12, Johnson-Cousins and 2MASS (bottom).}
\label{fig:colcoltrans} 
\end{figure*}

This section gives colour-colour transformations that relate the \edr photometric systems to other systems. Relationships for
\hip \citep{1997ESASP1200.....E},
\tyctwo \citep{2000A&A...355L..27H},
SDSS12 \citep{2015ApJS..219...12A},
Johnson-Cousins \citep{2000PASP..112..925S} and
2MASS \citep{2006AJ....131.1163S}
are provided here. For all fits, except Johnson-Cousins, only those sources with small magnitude error and small \xp excess flux were used. In the case of Johnson-Cousins, all available sources where used due to the high quality of these standards.
\edr\ sources with $G<13$~mag, photometry in the three \gaia passbands and in the external photometric systems were cross-matched to these external catalogues. The magnitude limit was used in order to limit the influence of photometric noise on the derived relationships. However, this magnitude range is not appropriate for the SDSS12 transformations since SDSS12 sources brighter than 14~mag are saturated. Thus, for the SDSS12 transformations \edr sources with $\sigma_G<0.01$ and SDSS12 magnitudes fainter than 15 were used. 
In order to obtain good quality fits, filtering on data quality was applied to the data (see  the \edr online documentation for more details). The validity of these fits is only applicable in the colour ranges used for the fits (see \tabref{colcolapplicability}). The coefficients of the polynomials representing the transformations derived between \gaia and \hip, \tyctwo, SDSS12, Johnson-Cousins and 2MASS can be found in \tabref{colcoltrans}. A selection of these photometric relationships can be seen in \figref{colcoltrans}. A complete set of figures can be found in the \edr online documentation. The relationships shown here were derived using an early internal version of the release. Thus, some sources used in the fit could have been filtered out in the final publication.
The purpose of these relationships is to provide a general transformation valid for the widest possible set of stellar populations and types. This will provide a reasonable estimate of their photometry when transforming from one system to another. There are cases in which different types, particularly M giants and dwarfs, that show different behaviour in the colour-colour diagram. In such cases, a single fit was carried out for the most populous type (usually M giants), covering the widest range of colours. Thus, for many of the relationships shown here the red end is only valid for M giants and not  M dwarfs. 
\begin{table}
\caption[Applicable range for the relationships]{Applicable range for the relationships between the \edr system and the other photometric systems considered.
\label{tab:colcolapplicability}
}
\begin{center}
\begin{tabular}{c c}
\hline 
\multicolumn{2}{c}{\textbf{\hip relationships}}  \\
$\gmag-H_P=f(B-V)$  		 		 & $-0.25<B-V<1.9^a$\\
$\gmag-H_P=f(V-I)$  		 		 & $-0.25<V-I<5.0$	\\
$\gmag-H_P=f(\bprp)$  	 		 & $-0.5<\bprp<4.0$\\
$\gbp-H_P=f(V-I)$  		 		 & $-0.2<V-I<3.0$\\
$\grp-H_P=f(V-I)$  		 		 & $-0.4<V-I<3.5$\\
$\bprp=f(V-I)$  	 		 	 & $-0.5<V-I<3.5$\\
\hline
\multicolumn{2}{c}{\textbf{\tyctwo relationships}}\\
$\gmag-V_T=f(B_T-V_T)$		 		 & $-0.2<B_T-V_T<2.0^b$\\
$\gmag-V_T=f(\bprp)$	 		 	 & $-0.35<\bprp<4.0$\\
$\gmag-B_T=f(\bprp)$	 			 & $-0.3<\bprp<3.0$\\
$\gbp-V_T=f(B_T-V_T)$	 			 & $-0.2<B_T-V_T<2.5$\\
$\grp-V_T=f(B_T-V_T)$	 			 & $-0.3<B_T-V_T<2.0^c$\\
$\bprp=f(B_T-V_T)$  	 			 & $-0.3<B_T-V_T<2.0^d$\\
\hline
\multicolumn{2}{c}{\textbf{SDSS12 relationships}} \\
$\gmag-g=f(g-i)$        				& $-1.0<g-i<9.0$\\ 
$\gmag-r=f(r-i)$        				& $-0.5<r-i<2.0$ \\
$\gmag-i=f(r-i)$        				& $-0.35<r-i<2.0$  \\
$\gbp-g=f(g-i)$         				& $-0.6<g-i<3.5$  \\
$\grp-r=f(r-i)$         				& $-0.9<g-i<8.0$  \\ 
$\bprp=f(g-i)$       				& $-0.5<g-i<3.5^e$  \\
$\gmag-r=f(\bprp)$   				& $0.0<\bprp<3.0^f$ \\
$\gmag-i=f(\bprp)$   				& $0.5<\bprp<2.0$\\
$\gmag-g=f(\bprp)$   				& $0.3<\bprp<3.0^g$\\
\hline
\multicolumn{2}{c}{\textbf{Johnson-Cousins relationships}}\\
$\gmag-V=f(V-I_C)$			 	 & $-0.4<V-I_C<5.0$\\ 
$\gmag-V=f(V-R)$			 	 & $-0.15<V-R<2.3^h$\\ 
$\gmag-V=f(B-V)$			 	 & $-0.4<B-V<3.3^i$\\ 
$\gbp-V=f(V-I_C)$				 & $0.0<V-I_C<4.0$\\ 
$\grp-V=f(V-I_C)$		 		 & $-0.4<V-I_C<5.0$\\ 
$\bprp=f(V-I_C)$	 			 & $-0.4<V-I_C<5.0$\\ 
$\gmag-V=f(\bprp)$	 			 & $-0.5<\bprp<5.0$\\
$\gmag-R=f(\bprp)$	 			 & $0.0<\bprp<4.0^j$\\
$\gmag-I_C=f(\bprp)$	 			 & $-0.5<\bprp<4.5$\\
\hline
\multicolumn{2}{c}{\textbf{2MASS relationships}}\\
$\gmag-K_S=f(H-K_S)$			  	  & $-0.1<H-K_S<0.4$\\
 $\gbp-K_S=f(H-K_S)$			  	  & $-0.1<H-K_S<0.4$\\
 $\grp-K_S=f(H-K_S)$			  	  & $-0.1<H-K_S<0.4$\\
  $\bprp=f(H-K_S)$			  	  & $-0.1<H-K_S<0.4$\\
  $\gmag-K_S=f(\bprp)$			  & $-0.5<\bprp<2.5$\\
  $\gmag-H=f(\bprp)$			  	  & $-0.5<\bprp<2.5$\\
  $\gmag-J=f(\bprp)$			  	  & $-0.5<\bprp<2.5$\\
  $\gmag-K_S=f(J-K_S)$			  	  & $-0.2<H-K_S<1.1$\\
  $\gbp-K_S=f(J-K_S)$			  	  & $-0.2<H-K_S<1.1$\\
  $\grp-K_S=f(J-K_S)$			  	  & $-0.2<H-K_S<1.1$\\
  $\bprp=f(J-K_S)$			  	  & $-0.1<H-K_S<1.1$\\
\hline
\multicolumn{2}{l}{\footnotesize{$^a$ For $B-V>1.4$ this is only valid for M giants.}} \\
\multicolumn{2}{l}{\footnotesize{$^b$ For $B_T-V_T>1.7$ this is only valid for M giants.}}\\
\multicolumn{2}{l}{\footnotesize{$^c$ For $B_T-V_T>1.7$ this is only valid for M giants.}} \\
\multicolumn{2}{l}{\footnotesize{$^d$ For $B_T-V_T>1.7$ this is only valid for M giants.}}\\
\multicolumn{2}{l}{\footnotesize{$^e$ For $\bprp>2.25$ this is only valid for M giants.}} \\
\multicolumn{2}{l}{\footnotesize{$^f$ For $\bprp>2.0$ this is only valid for M giants.}}\\
\multicolumn{2}{l}{\footnotesize{$^g$ For $\bprp>2.0$ this is only valid for M giants.}} \\
\multicolumn{2}{l}{\footnotesize{$^h$ For $V-R>0.9$ this is only valid for M giants.}} \\
\multicolumn{2}{l}{\footnotesize{$^i$ For $B-V>1.3$ this is only valid for M giants.}}\\
\multicolumn{2}{l}{\footnotesize{$^j$ For $\bprp>2.0$ this is only valid for M giants.}}\\
\hline
 \end{tabular}
 \end{center}
 \end{table}

\begin{table*}
\caption[Coefficients of the photometric transformations]{Coefficients of the transformation polynomials derived between the
 \hip, \tyctwo, SDSS12, Johnson-Cousins and 2MASS systems and that of \edr.}
\label{tab:colcoltrans}
\begin{center}
\begin{tabular}{c c c c c c c c}
\hline 
\multicolumn{7}{c}{\textbf{\hip relationships}}\\
& & $\mathbf{B-V}$ & $\mathbf{(B-V)^{2}}$ & $\mathbf{(B-V)^{3}}$ &&&  $\mathbf{\sigma}$ \\   
$\gmag-Hp$          & -0.02392 &-0.4069 & 0.04569 &-0.0452 &&&  0.02417 \\     
& & $\mathbf{V-I}$ & $\mathbf{(V-I)^{2}}$ & $\mathbf{(V-I)^{3}}$   &&& $\mathbf{\sigma}$ \\   
$\gmag-Hp$          & 0.01546 &-0.4308 &-0.01872 &   &&& 0.08191 \\
$\gbp-Hp$    & -0.02696 & 0.1086 & -0.009148 & 0.004715 &&&  0.06\\
$\grp-Hp$    & -0.006437 &-1.194 & 0.09962 &  &&&  0.1024     \\
$\bprp$   & -0.01612 & 1.274 &-0.08143 &  &&&  0.082 \\
& & $\mathbf{\bprp}$ & $\mathbf{(\bprp)^{2}}$ & $\mathbf{(\bprp)^{3}}$   &&& $\mathbf{\sigma}$ \\   
$\gmag-Hp$          & -0.01008 &-0.2309 &-0.1300 & 0.01894  &&& 0.06066 \\
\hline
\multicolumn{7}{c}{\textbf{\tyctwo relationships}}\\
& & $\mathbf{B_T-V_T}$ & $\mathbf{(B_T-V_T)^{2}}$ & $\mathbf{(B_T-V_T)^{3}}$   &&& $\mathbf{\sigma}$ \\   
$\gmag-V_T$  & -0.01072 &-0.2870 & 0.05807 &-0.06791   &&& 0.06084     \\
$\gbp-V_T$   & -0.01868 & 0.2682 &-0.1366 & 0.01272   &&& 0.04127   \\
$\grp-V_T$   & -0.04424 &-1.197 & 0.4948 &-0.1757   &&& 0.09359 \\
$\bprp$   &  0.02621 & 1.458 &-0.6176 & 0.1817   &&& 0.06834 \\
& & $\mathbf{\bprp}$ & $\mathbf{(\bprp)^{2}}$ & $\mathbf{(\bprp)^{3}}$ & $\mathbf{(\bprp)^{4}}$ &$\mathbf{(\bprp)^{5}}$  & $\mathbf{\sigma}$ \\   
$\gmag-V_T$          & -0.01077 &-0.0682 &-0.2387 & 0.02342 &  &  & 0.05350    \\
$\gmag-B_T$          & -0.004288 &-0.8547 & 0.1244 &-0.9085 & 0.4843 &-0.06814 & 0.07063 \\
\hline
\multicolumn{7}{c}{\textbf{SDSS12 relationships}}\\
                   &           & $\mathbf{g-i}$      & $\mathbf{(g-i)^{2}}$      & $\mathbf{(g-i)^{3}}$         &&& $\mathbf{\sigma}$\\	
$\gmag-g$   	   & -0.1064  &-0.4964 &-0.09339 & 0.004444  &&&0.0872 \\ 
$\gbp-g$          &  0.06213 &-0.2059 &-0.06478 & 0.007264 &&& 0.02944\\
$\grp-g$          & -0.3306 &-0.9847 &-0.02874 & 0.002112 &&& 0.04958\\
$\bprp$         &  0.3971 & 0.777 &-0.04164 & 0.008237 &&& 0.03846\\

            	   &	   & $\mathbf{r-i}$	 & $\mathbf{(r-i)^{2}}$      & $\mathbf{(r-i)^{3}}$	    &&& $\mathbf{\sigma}$ \\   
$\gmag-r$   	   & -0.01664 & 0.2662 &-0.649  & 0.08227   &&& 0.123 \\
$\gmag-i$   	   & -0.01066  & 1.298  &-0.7595  & 0.1492   &&& 0.07112 \\
            	   &	   & $\mathbf{\bprp}$ & $\mathbf{(\bprp)^{2}}$ & $\mathbf{(\bprp)^{3}}$ & $\mathbf{(\bprp)^{4}}$  && $\mathbf{\sigma}$ \\   
$\gmag-r$   	   & -0.09837	& 0.08592 &0.1907 &-0.1701 & 0.02263 && 0.03776 \\
$\gmag-i$   	   & -0.293	& 0.6404 &-0.09609  & -0.002104	& && 0.04092 \\
$\gmag-g$   	   &  0.2199	&-0.6365 &-0.1548  & 0.0064 & && 0.0745  \\
\hline
\multicolumn{7}{c}{\textbf{Johnson-Cousins relationships}}\\
& & $\mathbf{V-I_C}$ & $\mathbf{(V-I_C)^{2}}$ & $\mathbf{(V-I_C)^{3}}$ & $\mathbf{(V-I_C)^{4}}$ &&    $\mathbf{\sigma}$ \\  
$\gmag-V$           & -0.01597 & -0.02809 &-0.2483 & 0.03656 &-0.002939 && 0.0272 \\
$\gbp-V$       & -0.0143 & 0.3564 &-0.1332 & 0.01212 & && 0.0371 \\
$\grp-V$       &  0.01868 &-0.9028 &-0.005321 &-0.004186 & && 0.03784\\
$\bprp$   & -0.03298 & 1.259 &-0.1279 & 0.01631 & && 0.04459\\
& & $\mathbf{V-R}$ & $\mathbf{(V-R)^{2}}$ & $\mathbf{(V-R)^{3}}$ &&&     $\mathbf{\sigma}$ \\  
$\gmag-V$           & -0.03088 &-0.04653 &-0.8794 & 0.1733 &&&  0.0352 \\
& & $\mathbf{B-V}$ & $\mathbf{(B-V)^{2}}$ & $\mathbf{(B-V)^{3}}$ &&&     $\mathbf{\sigma}$ \\  
$\gmag-V$           & -0.04749 &-0.0124 &-0.2901 & 0.02008 &&&  0.04772\\
& & $\mathbf{\bprp}$ & $\mathbf{(\bprp)^{2}}$ & $\mathbf{(\bprp)^{3}}$ & $\mathbf{(\bprp)^{4}}$  &&  $\mathbf{\sigma}$ \\  
$\gmag-V$           & -0.02704 & 0.01424 &-0.2156 & 0.01426 & && 0.03017\\
$\gmag-R$           & -0.02275 & 0.3961 &-0.1243 &-0.01396 & 0.003775 && 0.03167\\
$G-I_C$           &  0.01753 & 0.76 &-0.0991 & & && 0.03765\\
\hline
\multicolumn{7}{c}{\textbf{2MASS relationships}}\\
& & $\mathbf{H-K_S}$ & $\mathbf{(H-K_S)^{2}}$ &&&&   $\mathbf{\sigma}$  \\
$\gmag-K_S$           &  0.5594 & 11.09 &3.040 &&&&  0.3743 \\
$\gbp-K_S$          &  0.5922 & 15.36 & 1.691 &&&&  0.499\\
$\grp-K_S$          &  0.1882 & 10.3 &-3.976 &&&&  0.2956 \\
$\bprp$         &  0.1836 & 8.456 & -3.781 &&&&  0.2361 \\
& & $\mathbf{\bprp}$ & $\mathbf{(\bprp)^2}$     &&&& $\mathbf{\sigma}$ \\  
$\gmag-K_S$           & -0.0981 & 2.089 &-0.1579   &&&& 0.08553 \\
$\gmag-H$           & -0.1048 & 2.011 &-0.1758   &&&& 0.07805 \\
$\gmag-J$           & 0.01798 & 1.389 &-0.09338   &&&& 0.04762\\
& & $\mathbf{J-K_S}$ & $(\mathbf{J-K_S})^{2}$ & $(\mathbf{J-K_S})^{3}$ & $(\mathbf{J-K_S})^{4}$  && $\mathbf{\sigma}$  \\
$\gmag-K_S$         &  0.1683 & 3.803 &-1.45 & 0.7867 & && 0.1309 \\
$\gbp-K_S$        &  0.1777 & 5.28 &-4.384 & 4.451 &-1.273 && 0.174 \\
$\grp-K_S$        &  0.08089 & 2.655 &-1.488 & 1.618 &-0.5068 && 0.07997 \\
$\bprp$         &  0.09396 & 2.581 &-2.782 & 2.788 & -0.8027&& 0.09668 \\
\hline
 \end{tabular}
 \end{center}
 \end{table*}

\subsection{Saturation correction}\label{sec:saturationCorrection}

\begin{figure*}
\centerline{
\includegraphics[width=0.67\columnwidth]{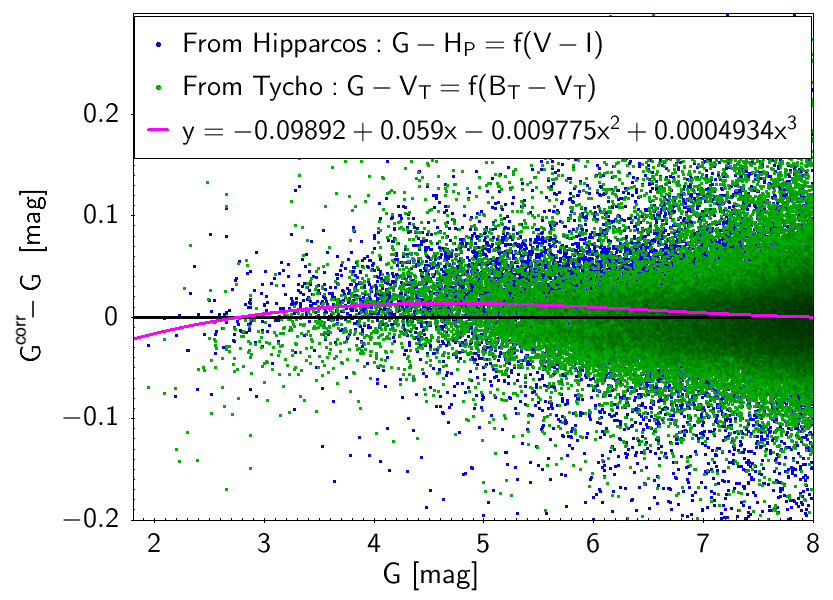}
\includegraphics[width=0.67\columnwidth]{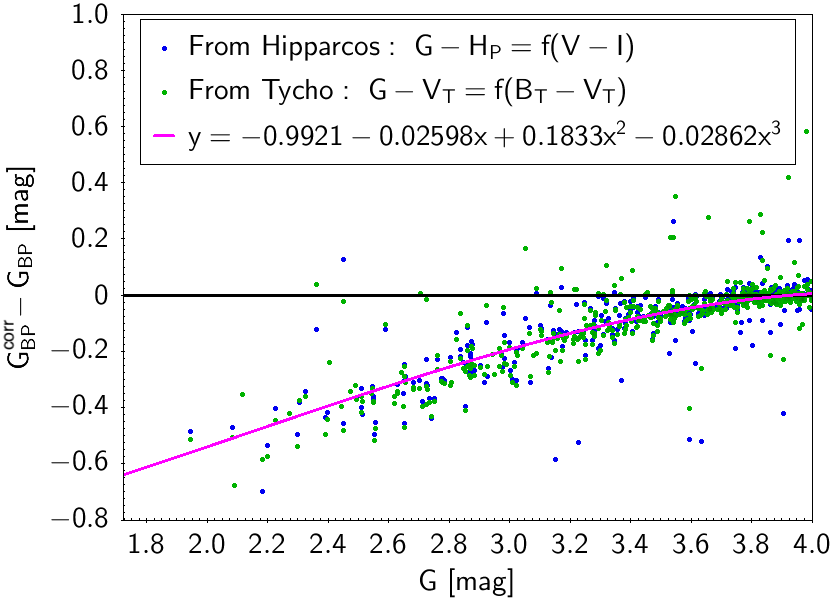}
\includegraphics[width=0.67\columnwidth]{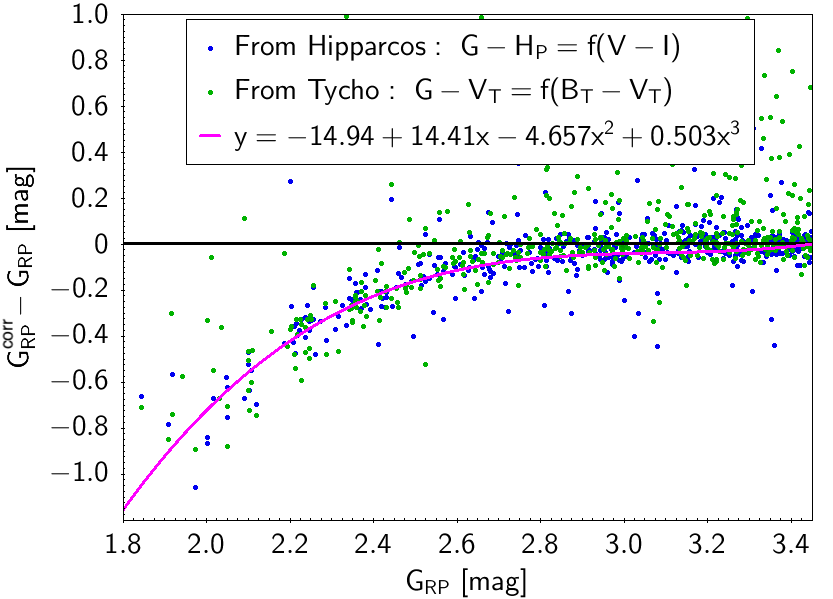}}
\caption[Saturation corrections]{Saturation corrections for $G$ (left), \gbp (centre) and \grp (right) passbands.}
\label{fig:saturationcorrection} 
\end{figure*}

The effect of saturation on the photometry of bright stars is shown in \figref{saturationcorrection}. The impact of saturation on the results of the \gband photometry has decreased with respect to \drt because of improvements in the handling of saturation in the PSF fitting \citep{EDR3_PSF}. The figure shows the residuals when \hip or \tyctwo photometry is transformed into the \edr system, using the transformations in \tabref{colcoltrans}, and compared with the \edr photometry. The \tyctwo and \hip data are combined to derive empirical corrections. The corrected magnitudes from the mean magnitudes in \edr, $G^{\rm corr}_{\rm XP}$ can be obtained with the following equations:
\begin{eqnarray}
\label{eq:satcorr}
G^{\rm corr}-G &=& -0.09892+0.059G-0.009775G^2\nonumber\\
&& +0.0004934G^3\label{eq:gsat}\\
\gbp^{\rm corr}-\gbp &=& -0.9921-0.02598G+0.1833G^2\nonumber\\
&&-0.02862G^3\label{eq:bpsat}\\
\grp^{\rm corr}-\grp &=&  -14.94+14.41\grp-4.657\grp^2\nonumber\\
&&+0.503\grp^3\label{eq:rpsat}
\end{eqnarray}
We note that the relationship for the corrected \grp is in terms of \grp rather than in $G$. This is because the analysis in \grp had a much smaller dispersion than in $G$.
The relationships should only be used in the following ranges:
\begin{itemize}
\item[]
$2.0 < G < 8$ for \equref{gsat}
\item[]
$2.0 < G < 3.94$ for \equref{bpsat}
\item[]
$2.0 < \grp < 3.45$ for \equref{rpsat}
\end{itemize}

\section{Gaps}\label{sec:gaps}

In the period covered by \edr there are a number of gaps in the photometric coverage. There are several factors that could cause a gap, the most common are: decontamination and refocussing events, gaps in the reconstructed attitude \citep[closely related to gaps in the crossmatch,][]{EDR3_XM}, satellite outages, gaps in the \xp calibration libraries, quality filtering applied during the processing and for the selection of the \edr content \citep{EDR3_CU9}. Some gaps affect only certain instruments. \tabref{gaps} provides a list of the known gaps in the \edr photometry.

\longtab[1]{
\begin{center}
\begin{longtable}{cccl}
\caption{Caption}List of known time gaps in the data contributing to the \edr' photometry.\label{tab:gaps}\\
\hline\hline
Start & End & Duration & Cause\\
\hline
\endfirsthead
\hline
Start & End & Duration & Cause\\
\hline
\endhead
\hline
\endfoot
\hline
\endlastfoot
1105.086491 & 1105.397602 & 0.311111 & Attitude\\
1185.162879 & 1185.354545 & 0.191667 & Attitude\\
1189.165656 & 1189.353156 & 0.187500 & Attitude\\
1193.017045 & 1193.035101 & 0.018056 & Attitude\\
1241.843433 & 1241.889267 & 0.045833 & Attitude\\
1261.364266 & 1261.537878 & 0.173611 & Attitude\\
1297.893433 & 1297.936488 & 0.043056 & Attitude\\
1316.490655 & 1316.491631 & 0.000976 & Attitude\\
1316.491631 & 1324.101353 & 7.609722 & Decontamination\\
1324.101353 & 1326.797599 & 2.696246 & Attitude\\
1336.678154 & 1336.786488 & 0.108333 & Attitude\\
1380.717043 & 1380.897598 & 0.180556 & Attitude\\
1401.753153 & 1401.951764 & 0.198611 & Attitude\\
1436.318431 & 1436.330931 & 0.012500 & Attitude\\
1443.949918 & 1443.974918 & 0.025000 & Refocussing\\
1471.942041 & 1472.237875 & 0.295833 & Attitude\\
1498.048985 & 1498.240652 & 0.191667 & Attitude\\
1623.572595 & 1623.694817 & 0.122222 & Attitude\\
1649.117039 & 1649.139261 & 0.022222 & Attitude\\
1649.650372 & 1649.672594 & 0.022222 & Attitude\\
1650.165650 & 1650.183705 & 0.018056 & Attitude\\
1650.205927 & 1650.221205 & 0.015278 & Attitude\\
1650.576761 & 1650.589261 & 0.012500 & Attitude\\
1651.092039 & 1651.104539 & 0.012500 & Attitude\\
1651.189261 & 1651.835094 & 0.645833 & Attitude\\
1652.335094 & 1652.453150 & 0.118056 & Attitude\\
1655.672594 & 1655.694816 & 0.022222 & Attitude\\
1770.014259 & 1770.214259 & 0.200000 & Attitude\\
1773.710092 & 1773.835092 & 0.125000 & Attitude\\
1788.114259 & 1788.168425 & 0.054167 & Attitude\\
1849.126758 & 1849.144813 & 0.018056 & Attitude\\
1919.019812 & 1919.125368 & 0.105556 & Attitude\\
1943.455923 & 1943.553145 & 0.097222 & Attitude\\
1951.342034 & 1951.465645 & 0.123611 & Attitude\\
1962.371201 & 1962.478145 & 0.106944 & Attitude\\
2094.003143 & 2095.568421 & 1.565278 & Attitude\\
2099.223977 & 2099.412865 & 0.188889 & Attitude\\
2111.240643 & 2111.457310 & 0.216667 & Attitude\\
2139.417031 & 2139.523976 & 0.106944 & Attitude\\
2142.289254 & 2142.394809 & 0.105556 & Attitude\\
2147.971198 & 2147.994809 & 0.023611 & Attitude\\
2150.432309 & 2150.535087 & 0.102778 & Attitude\\
2154.119809 & 2154.223976 & 0.104167 & Attitude\\
2165.160087 & 2165.178142 & 0.018056 & Attitude\\
2172.835087 & 2173.064253 & 0.229167 & Attitude\\
2178.246198 & 2178.261475 & 0.015278 & Attitude\\
2179.636475 & 2179.764253 & 0.127778 & Attitude\\
2192.251753 & 2195.218420 & 2.966667 & Attitude\\
2233.850363 & 2233.878141 & 0.027778 & Attitude\\
2233.898975 & 2233.921197 & 0.022222 & Attitude\\
2233.967030 & 2234.010086 & 0.043056 & Attitude\\
2235.662863 & 2235.676752 & 0.013889 & Attitude\\
2246.647586 & 2246.842030 & 0.194444 & Attitude\\
2330.615640 & 2330.615706 & 0.000066 & Attitude\\
2330.615706 & 2338.962373 & 8.346667 & Decontamination\\
2354.546195 & 2355.447584 & 0.901389 & Attitude\\
2386.614250 & 2386.643417 & 0.029167 & Attitude\\
2405.967028 & 2408.643417 & 2.676389 & Attitude\\
2408.935083 & 2409.968417 & 1.033333 & Attitude\\
2499.493415 & 2499.680915 & 0.187500 & Attitude\\
2574.644410 & 2574.727743 & 0.083333 & Refocussing\\
2574.727743 & 2574.828137 & 0.100394 & Attitude\\
2651.929524 & 2651.962858 & 0.033333 & Attitude\\
2751.340634 & 2751.530912 & 0.190278 & Attitude\\
3045.130908 & 3048.183685 & 3.052778 & Attitude\\
3205.136461 & 3205.172572 & 0.036111 & Attitude\\
3254.087849 & 3254.286460 & 0.198611 & Attitude\\
3269.454515 & 3269.469793 & 0.015278 & Attitude\\
3271.503127 & 3271.532293 & 0.029167 & Attitude\\
3314.864237 & 3314.882293 & 0.018056 & Attitude\\
3317.535070 & 3317.562848 & 0.027778 & Attitude\\
3542.114234 & 3542.311456 & 0.197222 & Attitude\\
3603.251733 & 3605.226733 & 1.975000 & Attitude\\
4009.661450 & 4009.855894 & 0.194444 & Attitude\\
4074.210060 & 4076.062838 & 1.852778 & Attitude\\
4112.768393 & 4112.769320 & 0.000927 & Attitude\\
4112.769320 & 4121.385986 & 8.616666 & Decontamination\\
4182.015614 & 4182.029503 & 0.013889 & Attitude\\
4263.518390 & 4263.718390 & 0.200000 & Attitude\\
4399.057277 & 4399.251722 & 0.194444 & Attitude\\
4477.440610 & 4477.737832 & 0.297222 & Attitude\\
4478.018387 & 4478.083665 & 0.065278 & Attitude\\
4545.371164 & 4545.383664 & 0.012500 & Attitude\\
4626.857274 & 4627.057274 & 0.200000 & Attitude\\
4729.669773 & 4729.683662 & 0.013889 & Attitude\\
4795.372550 & 4795.805883 & 0.433333 & Attitude\\
4845.776715 & 4845.819771 & 0.043056 & Attitude\\
4873.516993 & 4873.558660 & 0.041667 & Attitude\\
4965.730880 & 4965.753103 & 0.022222 & Attitude\\
5056.171157 & 5056.215601 & 0.044444 & Attitude\\
5078.630879 & 5078.835046 & 0.204167 & Attitude\\
5203.610044 & 5203.651711 & 0.041667 & Attitude\\
\end{longtable}
\end{center}
}

Decontamination campaigns involved actively heating different parts of the focal plane assembly to allow the water-based contamination to sublimate and being vented out. A decontamination campaign terminates when the active heating is disabled, however at that stage the satellite is not yet in thermal equilibrium, which is slowly reached over the course of several revolutions. Using the photometric calibrations from the period after the decontamination campaigns it was possible to detect time ranges during which the system photometric response was changing significantly with each OBMT revolution. The LS calibrations for these time ranges are not capable of tracking the fast time evolution of the system and therefore the calibrated AF epochs have been excluded from contributing to the source photometry. The PSF/LSF modelling for these periods was also problematic with the running solution unable to keep up with the very rapid changes in the instrument. This resulted also in lower quality raw fluxes produced by the IPD process \citep[see Figure 10 in][]{EDR3_PSF}.
The LS calibration for BP and RP did not show the same problems and therefore the corresponding epochs were not excluded. \tabref{af_exclusions} provides the time ranges for which epoch observations were excluded from the source photometry.

\begin{table}[hp]
    \caption{List of time ranges for which certain epochs were excluded from the mean source photometry because the calibration could not track the fast changes in system response. The first and second columns provide the start and stop OBMT revolution of the exclusion period; the third column provides the set of excluded epoch CCD observations.}
    \label{tab:af_exclusions}
    \centering
    \begin{tabular}{c|c|l}
    \hline\hline
    From & To & CCDs\\
    \hline
    1324.10135 & 1336.08621 & All AFs, both FoV\\
    1336.08621 & 1344.07611 & AF1 only, both FoV\\
    2328.61877 & 2330.61571 & AF1 preceding FoV only\\
    2338.96237 & 2350.94621 & All AFs, both FoV\\
    2350.94621 & 2358.93543 & AF1 only, both FoV\\
    4121.38599 & 4133.39359 & All AFs, both FoV\\
    4133.39359 & 4141.39866 & AF1 only, both FoV\\
    \hline
    \end{tabular}
\end{table}

\section{Mean and predicted SSCs for silver and bronze photometry}\label{sec:sscs}

A fraction of \gaia sources has incomplete colour information: a source may be missing either or both BP and RP spectrum shape coefficient (SSC) sets. A `silver' source has incomplete or missing either BP or RP SSCs. A source can be classified as `bronze' if it has incomplete or missing both BP and RP SSCs or if silver processing has failed. A `gold' source has complete SSC information.

\begin{figure}
\centerline{	
	\includegraphics[width=0.5\columnwidth]{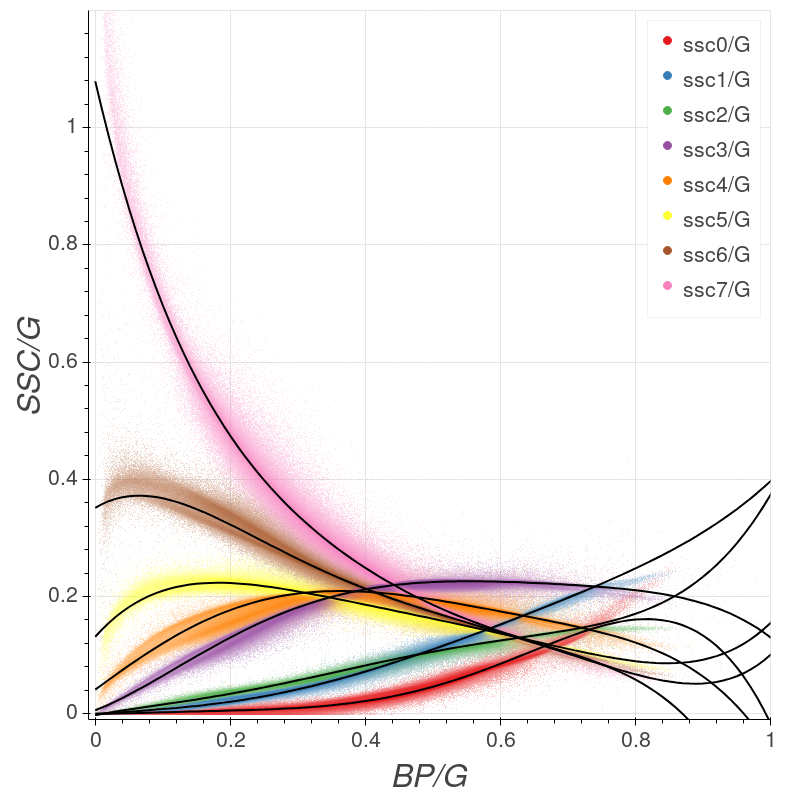}
	\includegraphics[width=0.5\columnwidth]{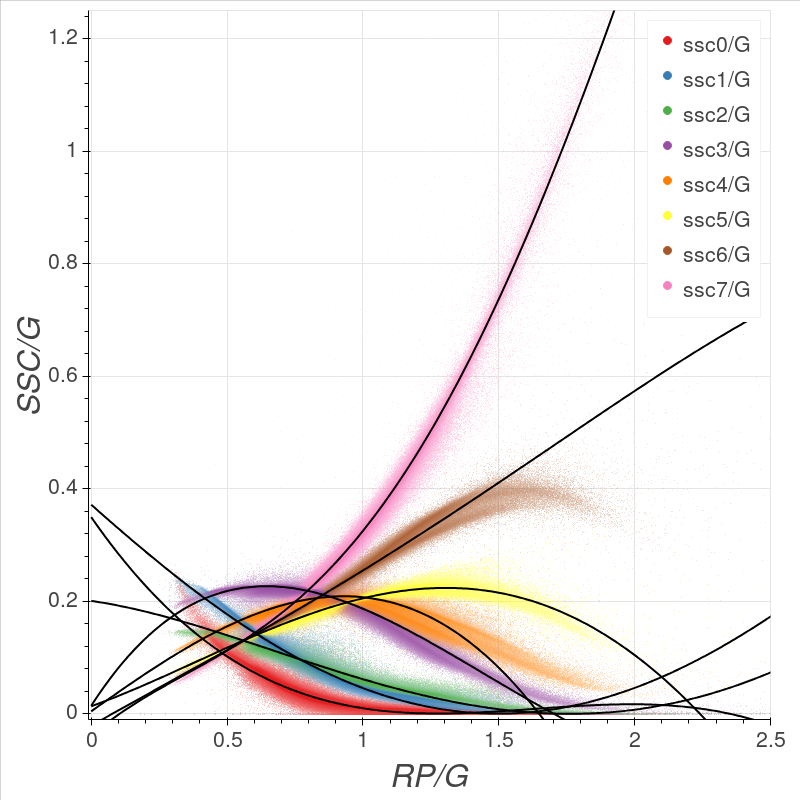}}
\centerline{	
	\includegraphics[width=0.5\columnwidth]{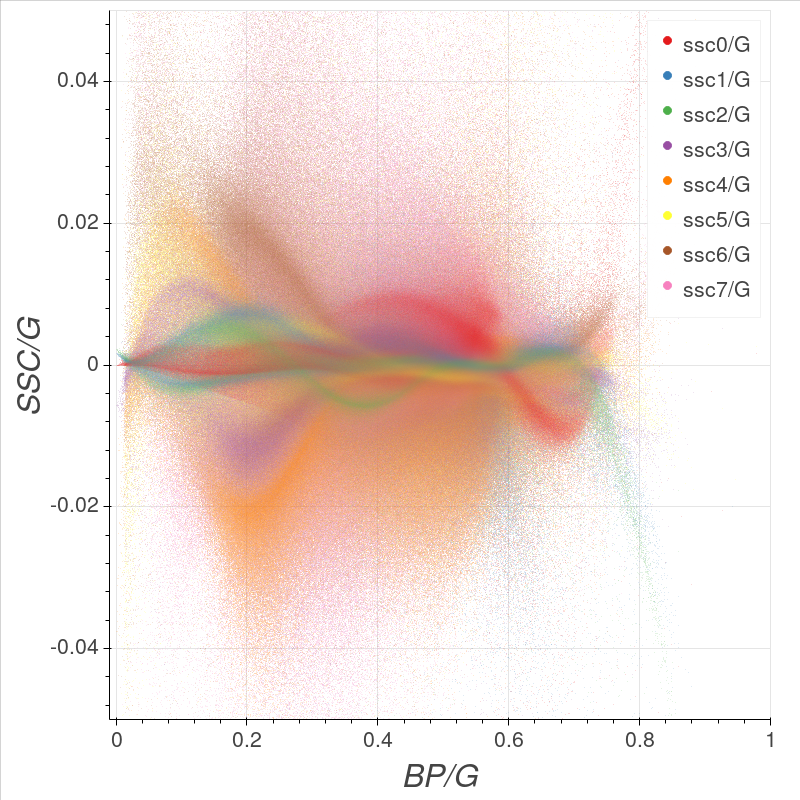}
	\includegraphics[width=0.5\columnwidth]{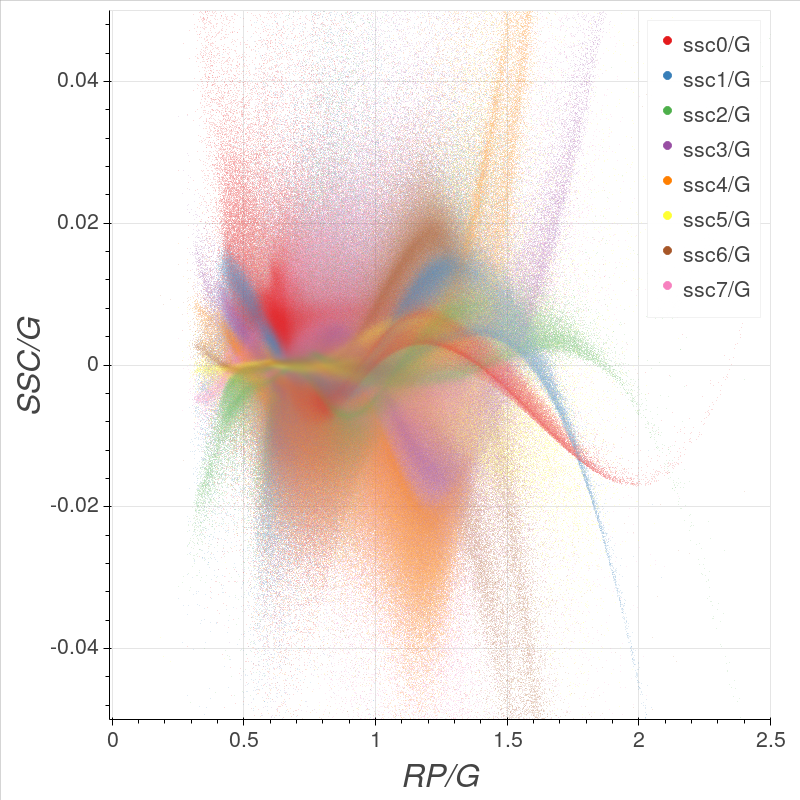}}
\caption[Results of the estimation procedure]{Results of the estimation procedure. Top: Distribution of sources in the SSC/G vs. XP/G space (colour-coded according to the legend) and the results of the corresponding fits (black lines). Bottom: Residuals.}
\label{fig:SilverSSC} 
\end{figure}

In order to calibrate the silver and bronze sources a statistical approach has been adopted to estimate the missing SSCs. In both cases a subset of $\approx3$~million sources was used for the calibrations. This subset was a selection that further flattened the distributions in colour, magnitude and sky position and as such is not dominated by the central colours or a particular region in the sky.
In the case of the bronze sources, a set of default colours (SSC values) is used. This has been derived from the median SSC values of the subset above. In order to estimate the missing SSC values of the silver sources, we assume that the colour-SSC space distribution of carefully selected $2.9\times10^6$ gold sources in the sample above and with more than five valid transits ($G$, BP and RP) is representative of the overall distribution of the sources observed by Gaia. A fifth-degree polynomial is then fitted to the SSC/$G$ versus XP/$G$ flux ratio distributions in order to determine empirical relationships from which the missing values are estimated. The distribution of the sources and the results of the procedure are shown in \figref{SilverSSC}.

\section{Photometry of extragalactic sources}\label{sec:galaxies}

\asecref{xpxs} showed that galaxies tend to have large values of the corrected \xp flux excess \cxs (see the top-left panel of \figref{xsVarExt}), this has already been used to select galaxies in \drt \citep[see e.g.][]{liu2020}. This behaviour can be explained considering the \gaia acquisition system and processing for \gband and \xp data. Fainter sources are acquired in the AF CCDs using the window configuration that corresponds to a viewing size of $0.35\times2.1$ (AL$\times$AC) arcsec. \xp spectra are instead acquired with a window configuration providing a AL$\times$AC viewing size of $3.5\times2.1$ arcsec. Since the \xp uncalibrated epoch flux is derived by integrating the pre-processed epoch spectra (see \secref{bprp}), this is equivalent to aperture photometry with a rectangular aperture. The \gband photometry instead is the result of an LSF fit, where the LSF \citep{EDR3_PSF} is optimised for point sources and therefore it is likely to produce an underestimated flux as the observed sources become progressively less stellar-like. The net result is that the \xp flux will tend to be significantly larger than the \gband one with the colour of the source unlikely to play a noticeable role since its effect will be much smaller (see \figref{xpfit}, top panel).

\begin{figure}[!htp]
    \centering
    \colfig{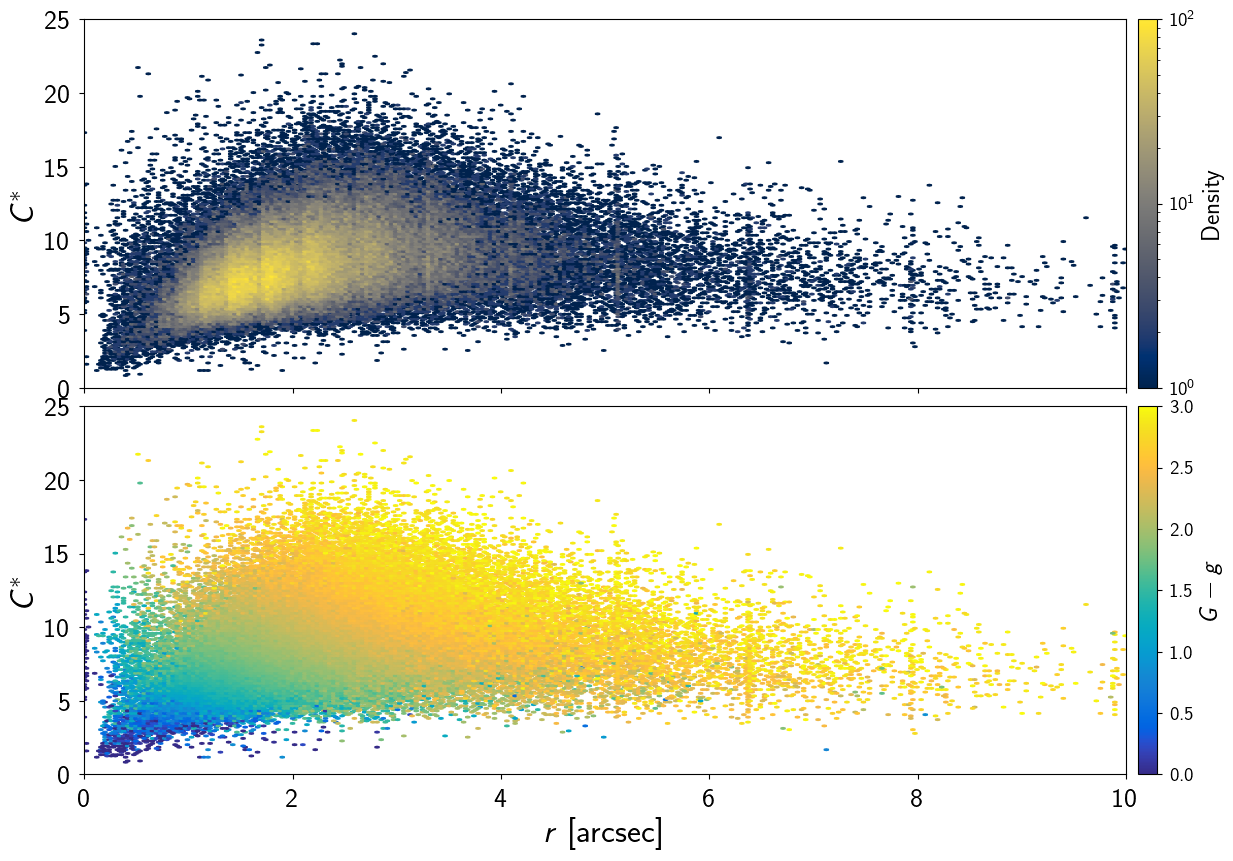}
    \caption{Dependence of the corrected \xp flux excess factor \cxs with the apparent size of the galaxy provided in terms of its de Vaucouleurs radius for a sample of $\approx 146$ thousand galaxies selected from SDSS DR12. The top panel shows the density, the bottom panel the colour scale shows the difference between the \gaia \gband magnitude and the SDSS $g$ magnitude.}
    \label{fig:gal_xcs_size}
\end{figure}

This explanation was verified using a sample of 146,605 galaxies selected from the SDSS DR12 release \citep{sdssdr12} and extracted from CDS \citep{cds}. \afigref{gal_xcs_size} shows the \xp corrected flux excess versus the apparent size of the galaxies as measured by the De Vaucouleurs radius. The top panel shows the distribution of the galaxies confirming that \cxs increases with the angular size of the galaxy up to the point when the galaxy becomes larger than the size of the \xp window becoming flat for larger sizes. The bottom panel of \figref{gal_xcs_size} shows the same dependency but with the colour scale showing the difference between the \gaia $G$ and the SDSS $g$ magnitude. Although the two bands are not the same, this difference can be used as a first order approximation of the discrepancy between the \gaia photometry and the SDSS photometry: as expected, the discrepancy is smaller for galaxies of small apparent size and then increases significantly for progressively larger objects.

\begin{figure}
    \centering
    \colfig{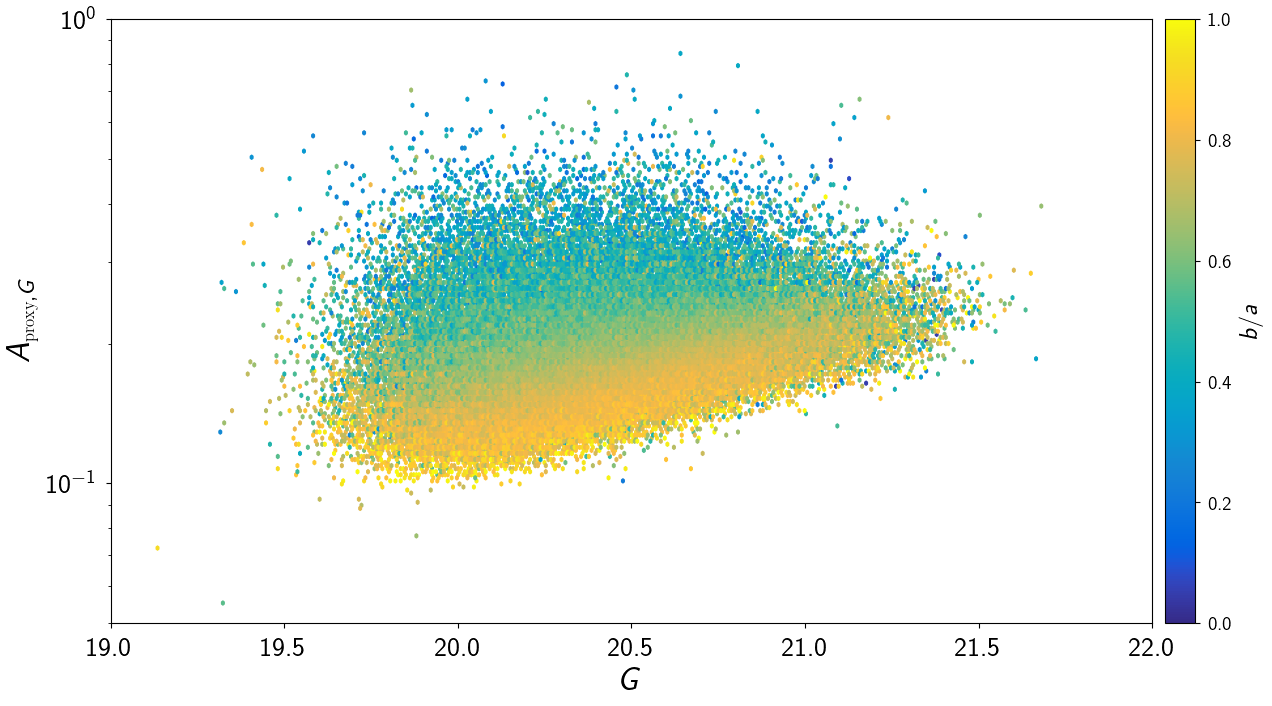}
    \caption{Dependence of the variability proxy $A_{\textrm{proxy},G}$ for the \gband with the apparent size of the galaxy provided in terms of its de Vaucouleurs radius for a sample of $\approx 146$ thousand galaxies selected from SDSS DR12. The colour scale shows the ratio between semi-minor and semi-major axes of the galaxies.}
    \label{fig:gal_var_ell}
\end{figure}

Because of the very small size of the AF windows, it is likely that the \gband photometry of extended sources will show an excess of scatter mimicking variability. The more elongated the galaxy the larger the excess scatter is expected to be since the measured epoch flux will be affected by the scanning direction of the satellite. This effect can be seen in \figref{gal_var_ell} which shows the variability proxy\footnote{The variability proxy corresponds to the estimated fractional error on a single AF CCD observation assuming all observations have equal weight.} for the \gband versus the $G$ magnitude \citep[see e.g.][]{NAMI} with the colour scale showing the ratio between the semi-minor and semi-major axes of the galaxy as available from the SDSS archive. As expected, the more elongated the galaxy, the larger the excess scatter of the \gband photometry. Because of the much larger size of the windows, this pseudo-variability is not observed in the \xp photometry.
Variability studies using \gaia data should take this into account to avoid polluting their sample with galaxies.

Finally, it should be noted that the PSF and LSF modelling and IPD determination has been considerably improved \citep{EDR3_PSF} in \edr for point sources: significant differences with respect to the \drt photometry are therefore to be expected for extended sources. This is yet another example of the limitations in comparing the \edr and DR2 photometry.

\section{RP zooming}\label{sec:rpzooming}

The precision reached in the photometric calibrations is such that even small effects can be analysed in detail. An example of this is offered by the signature of RP zooming in the SS calibrations for the RP CCDs.

\begin{figure}[!ht]
    \centering
    \colfig{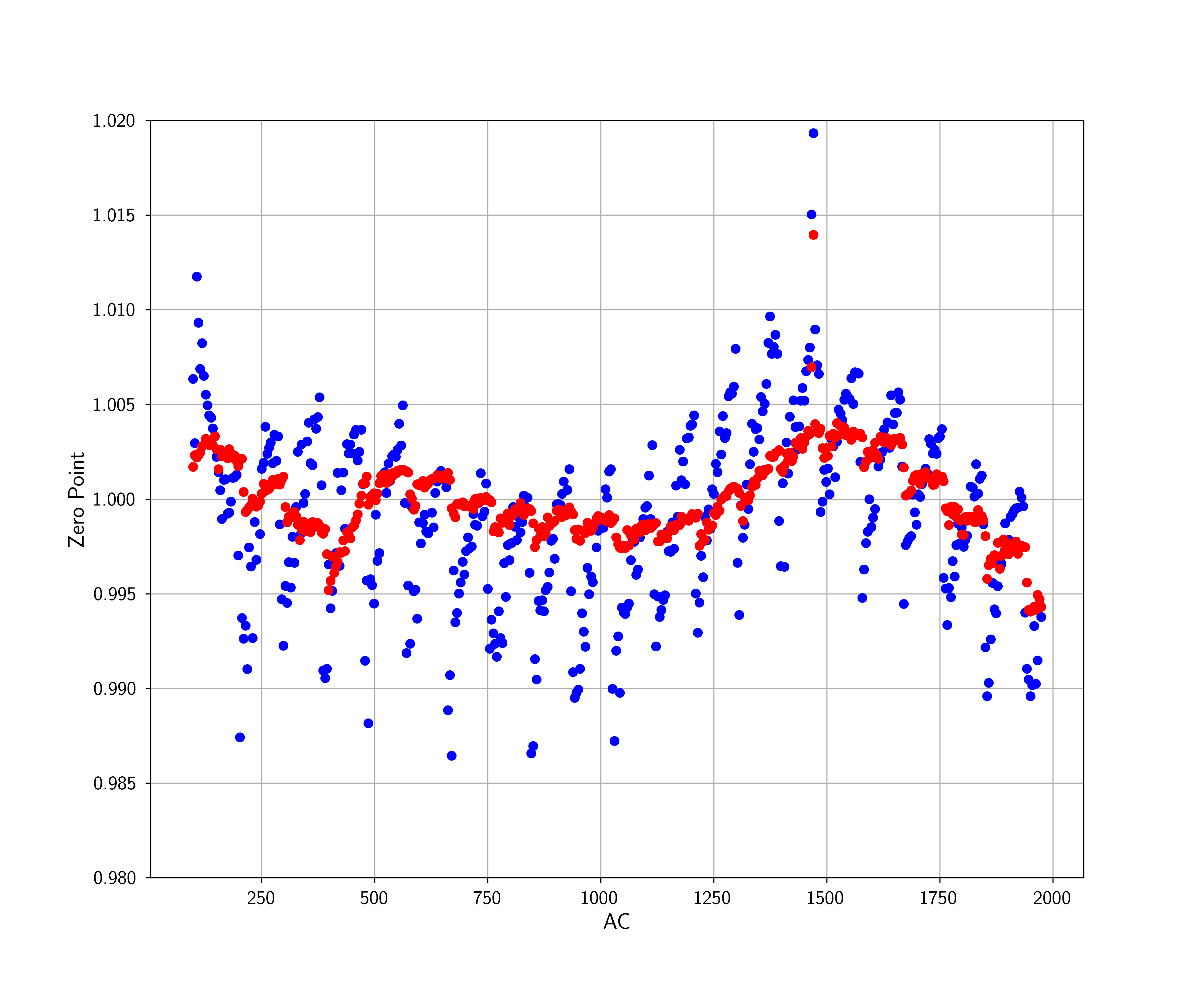}\\
    \colfig{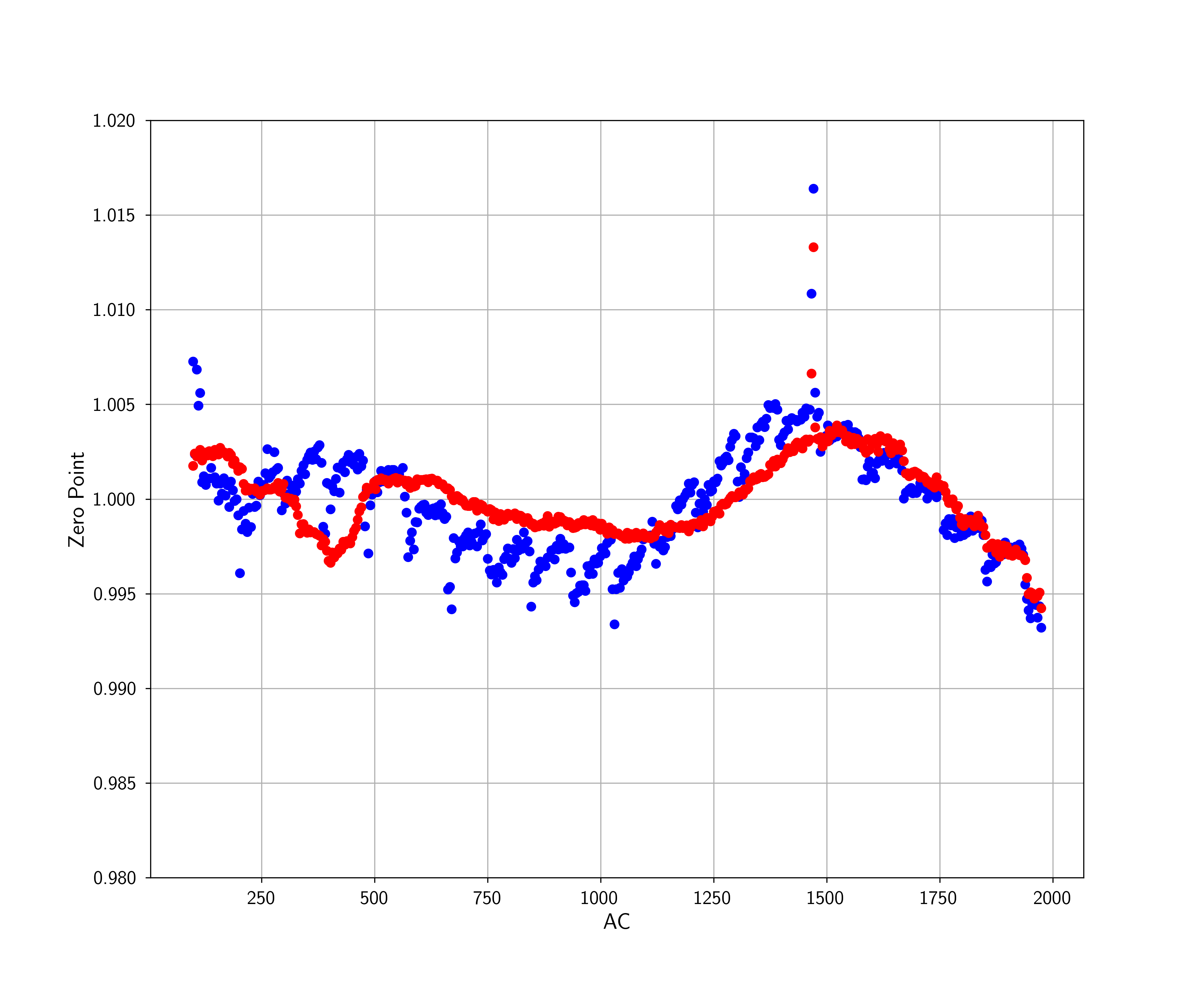}
    \caption{CCD response as a function of AC position as derived from the SS calibration. The top panel shows the response when the flux loss has been modelled and corrected only in the range $\pm2$ pix; the bottom panel shows the response when the flux loss has been modelled and corrected in the range $\pm4$ pix. The blue dots show the preceding FoV; the red dots show the following FoV.}
    \label{fig:rpzooming}
\end{figure}

The RP prism has a very low convergence (about $1\%$), which reduces the total telescope focal length but only in the across-scan direction. It leads to the loss of samples located at the extreme AC edges of the CCD: up to 70 pixels may be lost in the RP FoV. Of course, the most affected CCDs are the extreme ones (row 1 and 7 for RP), and only a minor effect is expected on the central CCD (row 4). To mitigate for this effect additional optical elements are used to effectively magnify the RP optical path to fully cover the RP CCDs. The VPU allocates the window positions based on a set of on-board lookup tables that take into account the AC motion of the source on the focal plane: for a given AC position of the window, as assigned by the VPU, there will be a distribution of how well centred the source will be within the window. Although this is true also for BP, the RP zooming has the effect of widening the distribution of this centring error. This means that to effectively model flux loss in RP, it is necessary to adopt a wider range for the centring error. This is shown in the top panel of \figref{rpzooming} which shows the CCD response determined by the SS calibration for an initial test run which was using the $\pm2$ pixel clamped range for the centring error. The saw-tooth pattern is caused by residual flux-loss that was not corrected due to the restriction in centring error. When the correction range for the flux loss is expanded to $\pm4$ pixel, as shown in the bottom panel of \figref{rpzooming}, the pattern fully disappears in the 
preceding FoV (red) and is considerably reduced in the following FoV (blue), although a systematic effect is still present at the $\approx2.5$ mmag level. The probable reason for the remaining error is that the flux loss terms in the LS calibrations is only a quadratic in centring error and that for RP more terms are needed.

\section{On the use of fluxes and magnitudes}\label{sec:MagFlux}
The error distribution of the fluxes is reasonably symmetric and close to being Gaussian for most magnitudes (but see also \secref{fluxlimit}). This is the reason why all the photometric calibrations are carried out in flux-space. The transformation from fluxes to magnitudes is non-linear and would cause the error distribution to become asymmetric. If the calibrations were to be done in magnitude-space a bias would be created. While this would be small, at the 1\% level, the aim of the \gaia project is to push the photometric accuracy to the mmag level and beyond.
The use of fluxes in the photometric processing, with flux errors being Gaussian-distributed, has the additional advantage of supporting the use of a maximum likelihood estimator for the generation of mean photometry. Furthermore, using inverse variance weighting ensures maximum signal to noise for the mean \citep[see e.g.][]{lupton1993statistics}.
%
In general, the asymmetry caused by the flux-magnitude transformation is small, but since the photometry is being published close to the magnitude limit it is important to consider. The error asymmetry caused by this transformation between plus and minus magnitudes for the epoch G photometry is 5\%, 10\%, and 20\% for G magnitudes of 19, 20, and 21, respectively. For BP, the fluxes can get lower and have larger asymmetries. This is the reason why magnitude errors are not given in the \gaia archive -- a single magnitude error is not sufficient. If working in magnitude space is required, then lower and upper bounds of the magnitude error should be computed from the $I-\sigma_I$ and $I+\sigma_I$ values converted to magnitudes using the zeropoints given in \tabref{ext_zptab}.

It is recommended that users work in flux space at the faint end,  that is to adopt a forward modelling approach and compare the model and data in flux space not in magnitude space.

\section{\gaia-related acronyms}\label{sec:acronyms}

\begin{table}[hp]
    \caption{\gaia-related acronyms used in the paper. Each acronym is also defined at its first occurrence in the paper.}
    \label{tab:acronyms}
    \centering
    \begin{tabular}{l|l|l}\hline\hline
Acronym & Description & See \\\hline
AC & ACross scan & \secref{data} \\
AF & Astrometric Field (in Astro) & \secref{data}\\
AL & ALong scan (direction) & \secref{data} \\
BP & Blue Photometer & \secref{data}\\
CCD & Charge-Coupled Device & \secref{data}\\
DPAC & Data Processing and Analysis Consortium & \secref{intro}\\
FWHM & Full Width at Half-Maximum & \secref{extcal} \\
FoV & Field of View & \secref{crowding} \\
GPS & Galactic Plane Scan & \secref{overproc:bkg} \\
HEALPix & Hierarchical Equal-Area iso-Latitude Pixelisation & \secref{standards} \\
IDU & Intermediate Data Update & \secref{data}\\
IPD & Image Parameter Determination & \secref{data}\\
LS & Large Scale & \secref{algorithm}\\
LSF & Line Spread Function & \secref{data}\\
OBM T &On-Board Mission Timeline & \secref{data}\\
OBMT-Rev & On-Board Mission Timeline in units of satellite revolutions &\secref{data} \\
PVL & Passband Validation Library & \secref{standards} \\
PSF & Point Spread Function & \secref{data}\\
RP & Red Photometer & \secref{data}\\
SM & Sky Mapper & \secref{data}\\
SPSS & Spectro Photometric Standard Stars & \secref{standards}\\
SS & Small Scale & \secref{algorithm}\\
SSC & Spectrum Shape Coefficients & \secref{calmodel} \\
TDI & Time-Delayed Integration (CCD) & \secref{crowding} \\
VPU & Video Processing Unit & \secref{sourcephot} \\\hline
\end{tabular}
\end{table}

\end{appendix}

\end{document}